\documentclass[10pt,a4paper]{article}
\pdfoutput=1
\usepackage{setspace}
\usepackage{amsfonts,amssymb,amsmath,stmaryrd,bm}
\usepackage[title,titletoc]{appendix}
\usepackage{graphics,epsfig,epstopdf}

\newcommand{\foot}{\footnote}
\newcommand{\bi}{\bibitem}
\newcommand{\ci}{\cite}
\newcommand{\rf}{\eqref}

\newcommand{\fr}{\frac}

\newcommand{\x}{\times}

\newcommand{\sech}{\textrm{\,sech\,}}
\newcommand{\cosech}{\textrm{\,cosech\,}}

\newcommand{\del}{\partial}
\newcommand{\dpl}{\partial_+}
\newcommand{\dm}{\partial_-}

\newcommand{\ra}{\rightarrow}
\newcommand{\Ra}{\Rightarrow}
\newcommand{\lra}{\leftrightarrow}

\newcommand{\Z}{\mathbb{Z}}

\newcommand{\id}{\tbf{1}}
\newcommand{\inv}[1]{#1^{-1}}

\newcommand{\td}{\tilde}

\newcommand{\al}{\alpha}
\newcommand{\bet}{\beta}
\newcommand{\g}{\gamma}
\newcommand{\de}{\delta}
\newcommand{\De}{\Delta}
\newcommand{\e}{\epsilon}
\newcommand{\z}{\zeta}
\newcommand{\ze}{\zeta}

\newcommand{\q}{\theta}

\newcommand{\s}{\sigma}

\newcommand{\ch}{\chi}

\newcommand{\vt}{\vartheta}
\newcommand{\vp}{\varphi}

\newcommand{\be}{\begin{equation}}
\newcommand{\ee}{\end{equation}}
\newcommand{\ba}{\begin{array}}
\newcommand{\ea}{\end{array}}
\newcommand{\bea}{\begin{eqnarray}}
\newcommand{\eea}{\end{eqnarray}}
\newcommand{\bp}{\begin{pmatrix}} 
\newcommand{\emp}{\end{pmatrix}}
\newcommand{\bit}{\begin{itemize}}
\newcommand{\eit}{\end{itemize}}
\newcommand{\no}{\nonumber}
\newcommand{\la}{\label}

\newcommand{\nin}{\noindent}
\newcommand{\hs}{\hspace}
\newcommand{\vs}{\vspace}

\newcommand{\mc}{\mathcal}
\newcommand{\mf}{\mathfrak}
\newcommand{\mbb}{\mathbb}
\newcommand{\trm}{\textrm}
\newcommand{\tbf}{\textbf}
\newcommand{\mbf}{\mathbf}

\newcommand{\Tr}{\trm{Tr}}

\newcommand{\STr}{\trm{STr}}

\newcommand{\com}[2]{[ #1,\,#2 ]}
\newcommand{\acom}[2]{\{#1 ,\,#2\}}

\newcommand{\ord}[1]{\mathcal{O}(#1)}
\newcommand{\sds}{\inplus}

\newcommand{\brr}{\big]}
\newcommand{\bll}{\big[} 

\newcommand{\Lag}{\mathcal{L}}

\newcommand{\psl}{\Psi_{_L}}
\newcommand{\psr}{\Psi_{_R}}
\newcommand{\zel}{{\zeta_{_L}}}
\newcommand{\zer}{{\zeta_{_R}}}
\newcommand{\chl}{{\chi_{_L}}}
\newcommand{\chr}{{\chi_{_R}}}

\newcommand{\ppn}{\bar{\mf{P}}_+}
\newcommand{\pmn}{\bar{\mf{P}}_-}
\newcommand{\ppmn}{\bar{\mf{P}}_\pm}
\newcommand{\ppne}{{\cal P}_+}
\newcommand{\pmne}{{\cal P}_-}
\newcommand{\ppmne}{{\cal P}_\pm}

\newcommand{\ket}[1]{\left|#1\right>}

\newcommand{\su}{\mf{su}}
\newcommand{\so}{\mf{so}}

\newcommand{\N}{\mc{N}}
\newcommand{\SA}{{\mf{s}}}
\newcommand{\JA}{{\mf{t}}}

\newcommand{\kt}{\tilde{k}}
\newcommand{\kh}{\hat{k}}
\newcommand{\2}{\mbf{2}}
\newcommand{\Ph}{P}
\newcommand{\po}{p_0}
\newcommand{\pot}{\td{p}_0}

\newcommand{\Sq}{S_q{}}

\newcommand{\adss}[1]{$AdS_{#1} \x S^{#1}$}

\newcommand{\up}{{\pmb{+}}}
\newcommand{\um}{{\pmb{-}}}
\newcommand{\upm}{{\pmb{\pm}}}
\newcommand{\ump}{{\pmb{\mp}}}

\newcommand{\Sc}{\mbb{S}}

\newcommand{\adt}{\dot{a}}
\newcommand{\bdt}{\dot{b}}

\newcommand{\agdt}{\dot{\alpha}}
\newcommand{\bgdt}{\dot{\beta}}

\newcommand{\Adt}{\dot{A}}
\newcommand{\Bdt}{\dot{B}}
\newcommand{\Cdt}{\dot{C}}
\newcommand{\Ddt}{\dot{D}}

\begin{document}

\thispagestyle{empty}

\vs{-1cm}
\rightline{Imperial-TP-BH-2011-01}

\begin{center}
\vspace{2cm}
{\Large \bf    Towards the quantum S-matrix of the Pohlmeyer reduced version \\
\vspace{0.2cm} of \adss{5}  superstring theory
\vspace{1cm}}
\vspace{.2cm}

{B. Hoare\,\footnote{benjamin.hoare08@imperial.ac.uk} and 
A.A. Tseytlin\,\footnote{Also at Lebedev Institute, Moscow.
tseytlin@imperial.ac.uk}}
\\
\vskip 0.6cm
{\em 
 Theoretical Physics Group \\
 Blackett Laboratory, Imperial College\\
 London SW7 2AZ, U.K.}
\end{center}

\setcounter{footnote}{0}

\begin{abstract}
\noindent
We investigate the structure of the  quantum S-matrix for  perturbative
excitations of the Pohlmeyer reduced version of the \adss{5} superstring
following arXiv:0912.2958. The reduced theory is a  fermionic extension of a 
gauged WZW model with an integrable potential. We  use as an  input  the result
of the  one-loop perturbative scattering amplitude computation and an analogy
with simpler reduced  \adss{n} theories  with  $n=2,3$. The reduced \adss{2}
theory is equivalent to the $\N=2$ 2-d supersymmetric sine-Gordon model for
which the exact quantum S-matrix is known. In the reduced \adss{3} case the
one-loop perturbative S-matrix, improved by a contribution of a local
counterterm, satisfies the group factorization property and  the Yang-Baxter
equation, and reveals the existence of a novel quantum-deformed  2-d
supersymmetry  which is not manifest in the  action. The   one-loop 
perturbative S-matrix  of the reduced   \adss{5}  theory has the  group
factorisation property  but  does not satisfy the Yang-Baxter equation 
suggesting some  subtlety with the realisation of quantum integrability. As a
possible resolution, we  propose  that the S-matrix  of  this theory may be
identified with the  quantum-deformed $[\mf{psu}(2|2)]^2 \ltimes \mbb{R}^2$ 
symmetric  R-matrix  constructed in arXiv:1002.1097. We  conjecture the exact
all-order   form of  this  S-matrix and discuss its possible relation to the 
perturbative  S-matrix  defined by the path integral. As in the   \adss{3} case
the   symmetry of the S-matrix  may be interpreted as an extended 
quantum-deformed  2-d supersymmetry. 
\end{abstract}

\newpage

\vspace{2cm}
\tableofcontents

\setcounter{footnote}{0}


\renewcommand{\theequation}{1.\arabic{equation}}
\setcounter{equation}{0}
\section{Introduction\la{introduction}}

The aim of this paper is to continue the investigation \ci{ht1,ht2} into the
S-matrix of the Pohlmeyer reduced version  of superstring theory on \adss{5}.
One motivation is to shed light on an eventual  first-principles solution 
of the  \adss{5}  superstring based on quantum integrability. 

We  shall  view   the reduced  \adss{5} theory  as a member of a class of  
\adss{n}  ($n=2,3,5$) theories which are Pohlmeyer reductions of Green-Schwarz
superstring  sigma models  based on  \adss{n}  supercosets. These reduced
theories \ci{gt1,ms,gt2}  are fermionic extensions of generalised sine-Gordon
models. Various examples of such bosonic models (called also ``symmetric space
sine-Gordon models'') based  on  a  $G/H$ gauged WZW  theory with an integrable 
potential term were considered in, e.g., \ci{pol,bakas,park,mist,bakps,fpgm}.
Due to their relation  via Pohlmeyer reduction to classical GS superstring
theory  on \adss{n} (and, more generally, of their  bosonic truncations to
classical string theory on symmetric spaces) there has been recent interest in
these models \ci{mi1,mirp,hm1,hm2,schm,hm3,gevic}. 

Let us briefly recall the basic setup. The fields of the reduced theory are all
valued in certain subspaces of a particular representation of the superalgebra
$\hat{\mf{f}}$, whose corresponding supergroup $\hat{F}$ is the global symmetry
of the original superstring  sigma model. The latter is based  on the 
supercoset ${\hat{F}}/{G} $, where $G$ is a bosonic subgroup of $\hat{F}$ (fixed
by a $\mbb{Z}_4$) and is the  gauge group of the superstring sigma model. For
the \adss{5} superstring  \ci{mt}  the supercoset is 
\be\no
\fr{PSU(2,2|4)}{Sp(2,2) \x Sp(4)}\ .
\ee
The  gauge group $H$ of the reduced theory  is a subgroup of $G$ that appears 
upon solving the Virasoro constraints. The reduced theory action is a
fermionic extension of  the $G/H$ gauged WZW theory with an integrable
potential \ci{gt1}\,\foot{We choose Minkowski signature in 2 dimensions with
$d^2 x = dx^0 dx^1$,\  $\dpl \equiv \del_0 \pm \del_1$. For algebras $[\mf{a}]^2
= \mf{a}\oplus \mf{a}$, i.e. the direct sum. We also use the notation $\sds$ for
a semi-direct sum of algebras and $\ltimes$ for a central extension.  For 
example, for the semi-direct sum $\mf{a}\sds \mf{b}$ we have the commutation 
relations: \ $\com{\mf{a}}{\mf{a}}\subset \mf{a}$, 
             $\com{\mf{b}}{\mf{b}}\subset \mf{b}$ 
and          $\com{\mf{a}}{\mf{b}}\subset \mf{b}$. \la{foot1}}
\be\la{gwzw}\begin{split}
\mc{S} & 
 =  \ \fr{k}{8\pi\nu}  \trm{STr} \Big[ \fr{1}{2}\,\int d^2x  \; 
                                       \ \inv{g}\dpl g\ \inv{g}\dm g\  
                                       - \fr{1}{3}\,\int d^3x \;
    \ \epsilon^{mnl} \ \inv{g} \del_m g\ \inv{g}\del_n g \ \inv{g}\del_l g
\\  & \qquad\quad +  \,\int d^2x \;
                   \ \Big( A_+\dm g\inv{g} - A_-\inv{g}\dpl g 
                           - \inv{g} A_+g A_-  +  \tau(A_+)A_- 
                           +  \mu^2\,(\inv{g} T g T - T^2)\Big) 
\\  & \qquad\quad + \int d^2x\; \big( \psl T D_+ \psl + \psr T D_- \psr
                                     + \mu\;g^{-1}\psl g \psr\big)\Big]\,.
\end{split}\ee
Here $g \in G$, $A_\pm\in \mf{h}$=alg$(H)$ and the fermionic fields
$\psl,\,\psr$ take values in fermionic subspaces of $\hat{\mf{f}}$. $k$ is a
coupling constant (level) and  $\nu$ is the index of the representation of 
$G$ in which $g$ is taken as  a matrix.\,\foot{In all the theories we consider
in the main body of this paper  the fundamental representations of $SU$ and $Sp$
groups are  used. Therefore, we have $\nu=\fr{1}{2}$. In particular, as in
\ci{gt2}, the reduced \adss{3} theory is written in terms of a single copy of
the fundamental representation of the supergroup $PSU(1,1|2)$. In appendix
\ref{bosonic} bosonic theories with $G=SO(N+1)$ are considered. For these
theories with  the fundamental representation one has $\nu=1$.} $\mu$ is a
parameter defining the mass of perturbative excitations near $g=\id$. The
standard  symmetric gauging corresponds to  $\tau=\id$\  ($\tau$ is an
automorphism of  $\mf{h}$); for an  abelian gauge group $H$ there is an option
of axial gauging corresponding to $\tau(h)=-h$. The constant matrix $T$ defining
the potential commutes with $H$ (see, e.g., \ci{gt1,mirp} for  details).

In the case of the \adss{5} superstring the Pohlmeyer reduced theory has certain
unique features; in particular, it is UV-finite \ci{rt} and its one-loop
semiclassical partition function is equivalent to that of the original \adss{5}
superstring \ci{hiti}. This suggests \ci{gt1} that it may be quantum-equivalent
to the \adss{5} superstring. If this were the case,  the Pohlmeyer reduced
theory could be used as a starting point for a 2-d Lorentz covariant
``first-principles'' solution of the \adss{5} superstring. The Lorentz
invariance of \eqref{gwzw} is a desirable feature as lack of 2-d Lorentz
symmetry in the light-cone gauge \adss{5} superstring S-matrix  leads, e.g.,  to
a complicated structure for the corresponding thermodynamic Bethe Ansatz for the
full quantum superstring spectrum (see, e.g., \ci{tba} and references therein).

The form  of the light-cone gauge \adss{5} superstring S-matrix (corresponding
to the spin-chain magnon S-matrix on the gauge theory side \ci{stau}) is fixed,
up to a phase,  by the residual global  $PSU(2|2)\x PSU(2|2)$ symmetry of the 
light-cone gauge Hamiltonian \ci{bsmat,afz}. This S-matrix is the starting
point for the conjectured Bethe Ansatz solution for the superstring energy
spectrum based on its integrability (see \ci{review}). Just as for the  standard
2-d sigma models \ci{zz} or other similar massive theories \ci{dh,hm1,smatrev} 
the starting point for solving the Pohlmeyer reduced theory is to find its 
exact S-matrix. Any proposal for the exact quantum S-matrix  should be, of
course, consistent with the perturbative S-matrix computed from the path
integral defined by the classical action. 

This motivates the study of the perturbative S-matrix of the Pohlmeyer reduced
\adss{n} superstring. In \ci{ht1} we computed the tree-level two-particle
S-matrix for the $8+8$ massive excitations of the reduced \adss{5} theory
employing the light-cone $A_+=0$ gauge.\,\foot{The generalised sine-Gordon
models have been  mostly studied  in the case of abelian gauge groups
\ci{mirpa,mirfp,cam,mirat}, for which there is an option of axial gauging. In
this case  the vacuum is unique up to gauge transformations. In the case of a
non-abelian  $H$  there is a  non-trivial vacuum moduli space, no global
symmetry and upon integrating out the gauge fields $A_\pm$  one is left
with a Lagrangian that has no perturbative expansion about the trivial vacuum.
This problem   is an artifact of the gauge fixing procedure  on $g$. If instead
one chooses the light-cone gauge  $A_+=0$  \ci{ht1} one is  able  to construct a
perturbative Lagrangian for the asymptotic excitations and compute the S-matrix.
In the case of a non-abelian $H$ \ci{ht2} this procedure leads to an S-matrix
that does not satisfy YBE (already at the tree level).} 

Remarkably, the resulting S-matrix factorises in the same way as the 
non-Lorentz invariant $[\mf{psu}(2|2)]^2\ltimes \mbb{R}^3$  symmetric light-cone
gauge S-matrix \ci{kmrz,af,review} of the \adss{5} superstring. The factorised
S-matrix has an intriguing similarity with a particular limit of the
quantum-deformed $ \mf{psu}(2|2) \ltimes \mbb{R}^3$ invariant R-matrix of
\ci{bk,bcl}.

In \ci{ht2} the perturbative computation was extended to the one-loop level 
for the bosonic part of the $G/H= SO(N+1)/SO(N)$ theories defined by \rf{gwzw}.
It was argued that the Lagrangian describing the physical fields constructed 
from the gauged WZW model \rf{gwzw} should be supplemented by a particular
one-loop counterterm  coming from  the path integral. For the $G/H=SU(2)/U(1)$
theory\,\foot{This theory is classically equivalent to the complex sine-Gordon
theory, as seen by fixing a gauge on the group field $g$ and integrating out
$A_\pm$.} the one-loop  counterterm contributions to the S-matrix were computed
in three different ways, all giving the same result. These contributions  were
precisely those needed to restore the validity of the Yang-Baxter equation (YBE)
at the  one-loop level \ci{dvm1,dvm2,dvm3} and to match the exact quantum
soliton S-matrix  proposed in \ci{dh}.

An alternative approach to exploring these models is based on constructing
soliton solutions \ci{hm2,hm3} and semi-classically quantising them. Assuming 
quantum integrability \ci{hm1} one can then conjecture an exact soliton
S-matrix.

\

In this paper we investigate the quantum S-matrices for the perturbative massive
excitations of the models \rf{gwzw} which are the Pohlmeyer reductions of the
\adss{2}, \adss{3} and \adss{5} superstring models.\,\foot{In this paper by the
\adss{3} and \adss{2} superstring theories  we mean the formal supercoset
truncations of the full 10-d superstring theories in $AdS_3 \times S^3 \times
T^4$  \ci{bab} and in  $AdS_2 \times S^2 \times T^6$ \ci{sor}.} Studying the
three  models together  turns out to be useful as it reveals certain universal 
features of their symmetries and S-matrices and thus helps to shed light on 
the structure of the most non-trivial case of the  \adss{5} theory. 

The reduced \adss{2} theory is equivalent \ci{gt1} to the $\mc{N}=2$
supersymmetric sine-Gordon model \ci{ku1,ku2}.  Below we will review the
construction of the exact S-matrix for its perturbative excitations
\ci{ku1,ku2,sw,ahn} and demonstrate the agreement with the direct one-loop
computation of scattering amplitudes.

In the reduced \adss{3} theory the one-loop S-matrix computed starting  from
the classical Lagrangian does not satisfy the  Yang-Baxter equation, but 
we will show that one can find a  local counterterm  that  restores  the  YBE
and thus integrability, much  like in the bosonic  complex sine-Gordon theory 
case discussed in \ci{ht2}. The addition of the counterterm not only restores 
the validity of the YBE but also ensures the group factorisation property and 
leads to a novel {\it quantum-deformed  supersymmetry} of the S-matrix. The 
existence of a  hidden 2-d supersymmetry in the classical reduced \adss{3} and
\adss{5} theories was conjectured, by analogy with the \adss{2} case, in
\ci{gt1} and  was recently discussed in \ci{schm} and demonstrated, at least
on-shell, in \ci{iv} and also off-shell for the Lagrangian \eqref{gwzw} in
\ci{hms}.

Assuming that the group-factorisation and the quantum-deformed supersymmetry
are true symmetries of the theory we conjecture an exact 2-d Lorentz invariant 
quantum S-matrix for the perturbative excitations of the reduced \adss{3}
theory. The phase factor is fixed by the unitarity and crossing constraints 
(and is similar to that in the reduced \adss{2} theory). We check that the
resulting exact S-matrix expanded in $1/k$  agrees with our one-loop
computation. 

\

In the  \adss{5} case we observe that the one-loop S-matrix group-factorises in
the same way as at the tree level in \ci{ht1}. However, there is a tree-level
anomaly in the YBE  \ci{ht1,ht2} which is a general feature of the models
\rf{gwzw} with a non-abelian gauge group $H$. As in the bosonic case \ci{gt2}, 
this anomaly cannot be cancelled by adding a local two-derivative counterterm
without breaking the manifest non-abelian symmetry, indicating some subtlety
with a realisation of integrability. 

Motivated by the quantum-deformed supersymmetry we discovered in the S-matrix
of the reduced \adss{3} theory and the quantum-deformed non-abelian symmetry
expected  in soliton S-matrices of the  bosonic theories \ci{hm1,hm2,hm3} here
we conjecture that the  S-matrix for the perturbative massive excitations of the
reduced \adss{5} theory may be related to a trigonometric relativistic limit of
the quantum-deformed $\mf{psu}(2|2)\ltimes \mbb{R}^3$ invariant R-matrix of 
\ci{bk,bcl} which satisfies the YBE  by construction.\,\foot{This is the limit
\ci{bcl} that has a similarity to the factorised tree-level S-matrix of
\ci{ht1}. For similar Lorentz invariant R-matrices see references in \ci{bk}. In
particular, a relativistic quantum-deformed $\mf{sl}(2|2)$ invariant R-matrix
was constructed in \ci{glzt}. One may expect \ci{bcl} this R-matrix to be
related to that discussed in section \ref{bkcomp} of this paper.} The phase
factor can be again fixed by unitarity and crossing and is the same as in the
reduced \adss{2} and \adss{3} theories. The one-loop expansion of the resulting
R-matrix indeed has a similar structure to the one-loop S-matrix that we found
by direct computation.

One possibility is that the S-matrix for the perturbative excitations in a
gauge-fixed Lagrangian is given by a certain non-unitary rotation of the
quantum-deformed R-matrix. The violation of the YBE  by the S-matrix computed 
directly from the Lagrangian may be related to some tension between 
gauge-fixing of the non-abelian $H$ symmetry and the  conservation of hidden
charges. It is possible also that the  physical excitations whose S-matrix is
the quantum-deformed R-matrix may be some non-trivial gauge-invariant
combinations of the Lagrangian fields.\,\foot{In the reduced \adss{3} and
\adss{5} theories the quantum deformation parameter is, respectively, 
$q = e^{-2i\pi/k}$ and $ e^{-i\pi/k}$, i.e.  it is the coupling constant $k$
that controls the deformation.} At the moment, these are speculations and it
is an open question as to what is the origin of the quantum deformation. An
alternative approach based on semi-classical quantization of solitons
\ci{hm1,hm2,hm3,hms} may shed more light on this issue.

\

The structure of the rest  of this paper is as follows. In section \ref{intro}
we review the construction \ci{ht1} of the Lagrangian for the perturbative 
excitations of the  theory \rf{gwzw} near $g=\id$. We then present the result
of the standard Feynman diagram computation of the one-loop S-matrix for the
special cases corresponding to the reduced \adss{2}, \adss{3} and \adss{5}
theories.

In \ci{gt1} it was shown that the reduced \adss{2} theory is equivalent to the
$\mc{N}=2$ supersymmetric sine-Gordon model. In section \ref{seclag22} we review
the known construction of the exact S-matrix for the perturbative excitations 
of the $\mc{N}=2$ supersymmetric sine-Gordon model and demonstrate its
consistency with the direct one-loop computation. We also discuss various
aspects of this S-matrix, such as the phase factor and the symmetries that will
be useful for understanding the more complicated  cases of the reduced \adss{n} 
models with $n =3,5$. 

In section \ref{pr33} we consider the reduced \adss{3} theory. It is shown that 
as in the complex sine-Gordon theory \ci{dvm1,dh,ht2} one can add a local  
counterterm  to restore the satisfaction of the Yang-Baxter equation at the
one-loop order. This counterterm also restores the group factorisation property 
of the S-matrix which then exhibits a quantum-deformed supersymmetry. Motivated
by an analogy with the S-matrix of the  reduced \adss{2} model and assuming 
quantum integrability as well as the quantum-deformed supersymmetry we then
propose an expression for the exact quantum S-matrix for the perturbative
excitations of this theory.

In section \ref{sumofsym} the symmetries of the reduced \adss{2} and \adss{3}
theories are reviewed and their origin in the Pohlmeyer reduction of the
corresponding superstring theories  is emphasized. By analogy, we then suggest
that the physical symmetry of the reduced \adss{5} theory may be a quantum
deformation of the $[\mf{psu}(2|2)]^2 \ltimes \mbb{R}^2$ superalgebra.

In section \ref{seclag55} we consider the S-matrix of the reduced \adss{5}
theory. We demonstrate that the group-factorisation property is manifestly 
preserved at both the tree and one-loop level, indicating that no local
counterterms are  required here. At the same time, the S-matrix does not
satisfy the YBE, and this cannot be repaired  by adding  local counterterms. 
Motivated by the discussion of symmetries in  section \ref{sumofsym} we
investigate the similarity between the factorised one-loop S-matrix and the 
quantum-deformed $\mf{psu}(2|2) \ltimes \mbb{R}^2$  symmetric R-matrix of 
\ci{bk} (which by construction satisfies the YBE). We extend the trigonometric
relativistic classical limit of this  R-matrix \ci{bcl} to all orders in the
$1/k$  expansion  and conjecture that it should represent the quantum-deformed
S-matrix of the reduced \adss{5} theory.

Some concluding remarks are made in section \ref{conc}. Appendix \ref{ssgbits}
is an extension of the review of the $\mc{N}=2$ supersymmetric sine-Gordon
S-matrix in section \ref{seclag22}. In appendix \ref{ads3s3pohl} we investigate
the symmetries of the reduced \adss{3} theory. In appendix \ref{olcfd} we
discuss the origin of the one-loop counterterm  required in section \ref{pr33}
and demonstrate how it  can be derived from a functional determinant. Appendices
\ref{33expansion} and \ref{55expansion} give a detailed form of the factorised
S-matrices of the reduced \adss{3} and  \adss{5} theories. In appendix
\ref{qdefl} we extend the classical relativistic trigonometric limit \ci{bcl} of
the quantum-deformed $\mf{psu}(2|2) \ltimes \mbb{R}^2$ R-matrix of \ci{bk}
to all orders. Finally, in appendix \ref{bosonic} we present an updated
discussion of the  S-matrix  of the bosonic $G/H =SO(N+1)/SO(N)$ generalized
sine-Gordon theories studied earlier in \ci{ht2}.


\renewcommand{\theequation}{2.\arabic{equation}}
\setcounter{equation}{0}
\section{Perturbative S-matrix to one-loop order\la{intro}}

In this section we find the one-loop S-matrix for the perturbative excitations
of the Pohlmeyer reduction of the  superstring theory on \adss{n} for $n=2,3,5$
(the tree-level term in this S-matrix  was  found in \ci{ht1}). The reduced
theories are fermionic extensions of a generalised sine-Gordon model (gauged WZW
theory with an integrable potential) \ci{gt1,gt2} whose Lagrangian is given in
\eqref{gwzw}.

\subsection{General setup}

We parametrise the group field in \rf{gwzw} as 
\be\la{paramg}
g=\exp (X+\xi)\,,\hs{20pt}
\trm{where}\hs{10pt}X \in \mf{g}\ominus \mf{h}\,,\hs{10pt}
\xi\in\mf{h}\,.
\ee
Following \ci{ht1}, we fix the $A_+=0$ gauge in \eqref{gwzw} and integrate over
$A_-$. The resulting constraint equation allows us to perturbatively solve for
$\xi$ \eqref{paramg}, leaving the required $2(n-1)+2(n-1)$ massive degrees of
freedom.                              

With the help of integration by parts and field redefinitions that amount to use
of the linearised equations of motion we get  the following local quartic
Lagrangian \ci{ht1}
\be\la{spl}
\begin{split}
{L}
=
    \fr{k}{4\pi}\;\;&\trm{STr}\Big(\hs{3pt}
	\frac{1}{2}\dpl X\dm X-\frac{\mu^2}{2}X^2
	+\psl T\dpl\psl+\psr T\dm\psr+\mu\psl\psr
\\& \quad\hs{5pt} + \fr{1}{12}\com{X}{\dpl X}\com{X}{\dm X} 
		  +\frac{\mu^2}{24}\com{X}{\com{X}{T}}^2
\\& \quad\hs{5pt} -\fr{1}{4}\com{\psl T}{\psl}\com{X}{\dpl X}
		  -\fr{1}{4}\com{\psr}{T\psr}\com{X}{\dm X}		 
		  -\frac{\mu}{2}\com{X}{\psr}\com{X}{\psl}
\\& \quad\hs{5pt} + \fr{1}{2}\com{\psl T}{\psl}\com{\psr}{T\psr}
		  +\ldots\Big)\,.
\end{split}
\ee
This quartic Lagrangian will be sufficient to compute the one-loop 2-particle
S-matrix (see below). Considering the Pohlmeyer reduction of superstring theory
on \adss{5}, where the supergroup $\hat{F}= PSU(2,2|4)$ and $G=Sp(2,2) \x
Sp(4)$, one can expand \eqref{spl} in components \ci{ht1} (rescaling the fields
by $\sqrt{\fr{4\pi}{k}}$,\ $ L \to \Lag_5$)
{\allowdisplaybreaks
\bea
\Lag_{5} & = & \fr{1}{2}\dpl Y_m \dm Y_m 
\no          - \fr{\mu^2}{2} Y_m Y_m
             + \fr{1}{2}\dpl Z_m \dm Z_m
             - \fr{\mu^2}{2} Z_m Z_m
\\       & & + \fr{i}{2}\zel_m \dpl \zel_m
\no          + \fr{i}{2}\zer_m \dm \zer_m
             + i\mu \zer_m \zel_m
       + \fr{i}{2}\chl_m \dpl \chl_m
         + \fr{i}{2}\chr_m \dm \chr_m
             + i\mu \chr_m \chl_m
\\       & & + \fr{\pi}{2k}\Big[
\no          - \fr{2}{3}Y_m Y_m \dpl Y_n \dm Y_n
             + \fr{2}{3}Y_m \dpl Y_m Y_n \dm Y_n
             - \fr{\mu^2}{3}Y_m Y_m Y_n Y_n
\\       & & \hs{26pt}
\no          + \fr{2}{3}Z_m Z_m \dpl Z_n \dm Z_n
             - \fr{2}{3}Z_m \dpl Z_m Z_n \dm Z_n
             + \fr{\mu^2}{3}Z_m Z_m Z_n Z_n
\\       & & \hs{26pt}
\la{lag15}   + \fr{i}{2}(\g_{mnpq}+\e_{mnpq})
               (\zel_m \zel_n Y_p \dpl Y_q
               +\zer_m \zer_n Y_p \dm Y_q
\\       & & \hs{126pt}
\no            -\chl_m \chl_n Z_p \dpl Z_q
               -\chr_m \chr_n Z_p \dm Z_q)
\\       & & \hs{26pt}
\no          + \fr{i}{2}(\g_{mnpq}-\e_{mnpq})
               (\chl_m \chl_n Y_p \dpl Y_q
               +\chr_m \chr_n Y_p \dm Y_q
\\       & & \hs{126pt}
\no            -\zel_m \zel_n Z_p \dpl Z_q
               -\zer_m \zer_n Z_p \dm Z_q)
\\       & & \hs{26pt}
\no          +i\mu(\zer_m \zel_m+\chr_m \chl_m)
                  (Y_n Y_n - Z_n Z_n)
\\       & & \hs{26pt}
\no          -2i\mu (\e_{mnpq}+\de_{mn}\de_{pq}
                     -\de_{mp}\de_{nq}-\de_{mq}\de_{np})
                    (\zer_m \chl_n Y_p Z_q
                     -\chr_m \zel_n Z_p Y_q)
\\       & & \hs{26pt}
\no          + \e_{mnpq}(\zer_m \zer_n \zel_p \zel_q
                          -\chr_m \chr_n \chl_p \chl_q)\Big]
              +\ord{k^{-2}}\,.
\eea
}
$m,n,p,q$ are $SO(4)$ vector indices.\,\foot{Note that for notational
convenience we are using $SO(N)$ indices, see \ci{ht1} for conventions. The
group $G$ here is $Sp(2,2) \x Sp(4)$ and thus the index of the fundamental
representation $\nu$ in \rf{gwzw} is  equal to $\fr{1}{2}$. For a discussion of
the normalisation of $k$ see footnote \ref{norm} below.} We have also defined
the $SO(N)$ tensor $\g_{mnpq}$ as 
\be\la{gmnpq}
\g_{mnpq}=\de_{mp}\de_{nq}-\de_{mq}\de_{np}\,.
\ee
The fields $Y_m$ and $Z_m$ (which are components of $X$ in \rf{spl}) are
bosonic, while the fields $\zer_m$, $\zel_m$, $\chr_m$ and $\chl_m$ (originating
from  $ \psl,\psr$)  are 2-d Majorana-Weyl fermions. This Lagrangian describes
$8+8$ massive degrees of freedom. 

Up to a scaling ambiguity in $k$, the analogous Lagrangians for the reduced
\adss{2} and \adss{3} theories are given by restricting the indices to $m,n,p,q$
to take the values $1$ and $1,2$ respectively.\,\foot{\la{norm}For the reduced
\adss{3} theory we should also rescale $k \ra \fr{k}{2}$. The reason being that
in this case $G=SU(1,1) \x SU(2)$ and the dual Coxeter number of $SU(2)$,\
$c_{SU(2)} =2$ is twice that of $SO(3)$,\ $c_{SO(3)} =1$. For the reduced
\adss{5} theory we have $G=Sp(2,2) \x Sp(4)$. The dual Coxeter number of
$Sp(4)$,\  $c_{Sp(4)} =3$ is equal to that of $SO(5)$, \  $c_{SO(5)} =3$.
Therefore the reduced \adss{3} theory should have $k \ra \fr{k}{2}$ compared to
the $G/H = SO(3)/SO(2)$ theory in \ci{ht2}, whereas the reduced \adss{5} model
should have the same normalisation of $k$ as the $G/H = SO(5)/SO(4)$ theory of
\ci{ht2}. For the reduced \adss{2} case the group $G$ is abelian and thus there
is no quantization of $k$, i.e.  it can be arbitrarily rescaled. For convenience
we will  assume  the same normalisation as  in the reduced \adss{3} theory,
i.e. we will rescale $k \ra \fr{k}{2}$ in \eqref{lag15}. This is also the same 
normalisation as when  one takes $\nu=\fr{1}{2}$ in \eqref{gwzw}.}
The expansion of these Lagrangians to quartic order agrees with those that arise
from the $A_+=0$ gauge treatment up to field redefinitions \ci{ht1}.

Below we compute the one-loop two-particle S-matrix arising from the Lagrangian
\eqref{lag15} and its \adss{2} and \adss{3} truncations. Following \ci{ht1} we
use Feynman diagrams and standard perturbative quantum field theory. From the 
quadratic part of the action one can construct the asymptotic states for which
the  spatial momentum and energy eigenvalues are related by the usual
relativistic dispersion relation
\be
E=\sqrt{p^2+\mu^2}\,.
\ee
In 2-d relativistic theories it is convenient to  consider the corresponding
rapidity $\vt$, related to the on-shell spatial momentum as
\be
p=\mu\sinh\vt\,.
\ee

We will label the on-shell momenta of the incoming states as $p_1$ and $p_2$,
with the corresponding rapidities $\vt_1$ and $\vt_2$. As we are considering
integrable theories, the outgoing states should have the same momenta as the
incoming states and as the theories are relativistic the S-matrix should only
depend on the difference of the rapidities
\be
\q=\vt_1 - \vt_2\,.
\ee 
In the reduced \adss{n} theories there are $16\,(\x (n-1)^2)$ two-particle
states given by 
\begin{align}
&\ket{Y_m(p_1)Y_n(p_2)}\,,&
&\ket{Z_m(p_1)Z_n(p_2)}\,,&
&\ket{\z_m(p_1)\z_n(p_2)}\,,&
&\ket{\chi_m(p_1)\chi_n(p_2)}\,,&\no
\\
&\ket{Y_m(p_1)\z_n(p_2)}\,,&
&\ket{\z_m(p_1)Y_n(p_2)}\,,&
&\ket{Z_m(p_1)\chi_n(p_2)}\,,&
&\ket{\chi_m(p_1)Z_n(p_2)}\,,&\no
\\
&\ket{Y_m(p_1)\chi_n(p_2)}\,,&
&\ket{\chi_m(p_1)Y_n(p_2)}\,,&
&\ket{Z_m(p_1)\z_n(p_2)}\,,&
&\ket{\z_m(p_1)Z_n(p_2)}\,,&\la{scatstates}
\\
&\ket{Y_m(p_1)Z_n(p_2)}\,,&
&\ket{Z_m(p_1)Y_n(p_2)}\,,&
&\ket{\z_m(p_1)\chi_n(p_2)}\,,&
&\ket{\chi_m(p_1)\z_n(p_2)}\,.&\no
\end{align}
Naively we have $256$ amplitudes in the two-particle S-matrix; however, from the
Lagrangian \eqref{lag15} we see that at the tree level there are selection rules
such that the four rows of two-particle states in \eqref{scatstates} only
scatter amongst themselves. This is  a consequence of symmetries that are not
manifest in the Lagrangian \eqref{lag15}, as such these selection rules apply
beyond the tree level. We therefore have $4^2 \x 4=64$ non-zero scattering
processes. The symmetries are  discussed in greater detail for each of the
particular theories in later sections.

We will list  $40$ of these amplitudes for each of the \adss{n} reduced
theories, from which the remaining 24 ones can be easily derived. For example, 
to compute $B_{mnpq}(\q)$ in 
\be
\Sc\ket{\ze_m(p_1)Y_n(p_2)} = B_{mnpq}(\q)\ket{Y_p(p_1)\ze_q(p_2)} +\ldots\ , 
\ee
where $\Sc$ is the 2-particle S-matrix operator and the dots stand for other
possible terms, we may use the fact that we know $A_{mnpq}(\q)$ in
\begin{align}
\!\!\!\!\Sc\ket{Y_m(p_1)\ze_n(p_2)} 
= A_{mnpq}(\q)\ket{\ze_p(p_1)Y_q(p_2)} +\ldots&\hs{2pt}
\Ra\;\;\Sc\ket{\ze_m(p_2)Y_n(p_1)} 
= A_{nmqp}(\q)\ket{Y_p(p_2)\ze_q(p_1)}\la{der}
+\ldots\\&\hs{2pt}\Ra\;\;\Sc\ket{\ze_m(p_1)Y_n(p_2)} 
= A^*_{nmqp}(-\q)\ket{Y_p(p_1)\ze_q(p_2)}+\ldots \ , \no
\end{align}
implying that
\be
B_{mnpq}(\q)=A^*_{nmqp}(-\q)\,.
\ee
Clearly, if two fermions are passing through each other in \eqref{der} we pick
up a factor of $-1$.

As the reduced \adss{n} theories are classically integrable (there exists a Lax
connection \ci{gt1}) we expect them to be quantum-integrable so that the
two-particle S-matrix operator should satisfy the Yang-Baxter equation  
\be\la{Y}
\Sc_{12}(\q_{12})\Sc_{13}(\q_{13})\Sc_{23}(\q_{23})
= \Sc_{23}(\q_{23})\Sc_{13}(\q_{13})\Sc_{12}(\q_{12})
\ee
Here the triple operator products should be understood as acting  on a three
particle state with rapidities $\vt_1,\,\vt_2,\,\vt_3$. The subscripts on $\Sc$
label which particles in the state it is acting on, while the quantities
$\q_{ij}$ denote the rapidity differences,
\be
\q_{ij}=\vt_i -\vt_j\,.
\ee
We therefore have the usual relativistic relation
\be
\q_{13}=\q_{12}+\q_{23}\,.
\ee
For the reduced \adss{n} theories where we  are scattering fermions and bosons 
\eqref{Y} will have some fermionic grading as well. 

If we consider the S-matrix to a particular order in the weak coupling
($\fr{1}{k}\ll 1$) expansion, say $\ord{k^{-n-1}}$, then the corresponding order
of the Yang-Baxter equation is $\ord{k^{-n}}$: the contribution of the
$\ord{k^{-n}}$ part of the S-matrix to the $\ord{k^{-n}}$ part of the
Yang-Baxter equation \eqref{Y} vanishes trivially.

\subsection{One-loop results}

Below we will present the results of the direct one-loop computation of the
two-particle S-matrix for the reduced \adss{n} ($n=2,3,5$) theories. We use
standard Feynman diagram techniques starting with the Lagrangian \eqref{lag15}
or its various truncations.

The reduced \adss{5} theory is UV-finite \ci{rt}. Also, it was shown in \ci{gt1}
that the reduced \adss{2} theory is equivalent to the $\mc{N}=2$ supersymmetric
sine-Gordon theory and thus is UV-finite as well. Semiclassical computations as
in \ci{hiti} provide a check that the reduced \adss{3} theory is also UV-finite,
at least to the two-loop order. Thus in contrast to the purely bosonic theories 
there is no renormalisation of the mass parameter $\mu$.

To compute the one-loop S-matrix one need only consider the Feynman diagrams of
the form in Fig.1. This is because the finite part of the tadpole diagrams in
Fig.2 arising from the sextic terms in the Lagrangian \eqref{lag15} will vanish
in 2-d. 
\begin{figure}[t]
\begin{center}
\includegraphics[width=6cm]{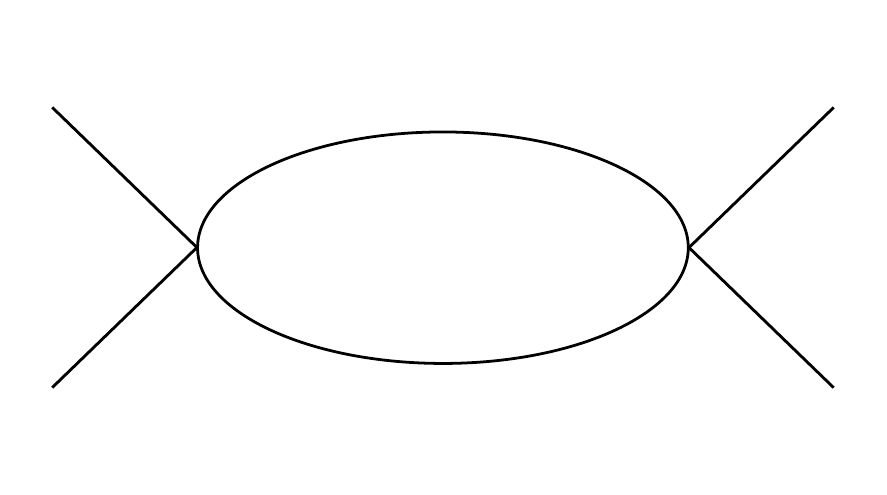}
\end{center}
\caption{Bubble diagram\la{fig1}}
\end{figure}
\begin{figure}[t]
\begin{center}
\includegraphics[width=6cm]{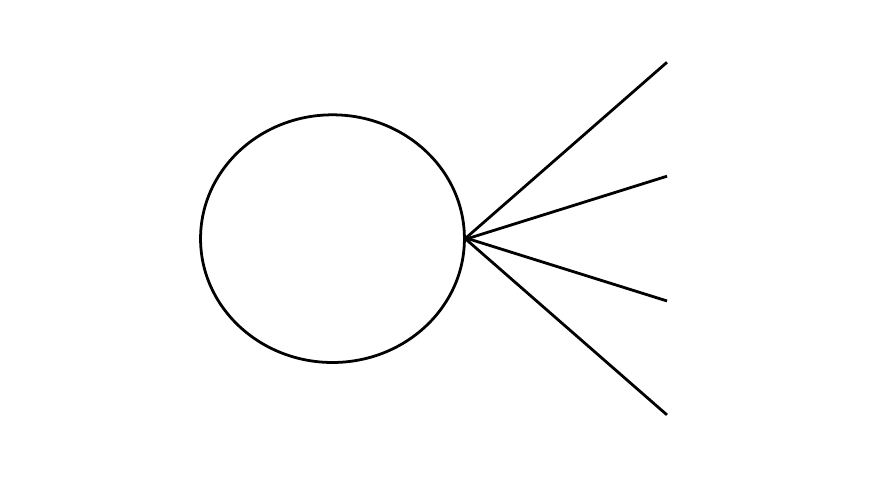}
\end{center}
\caption{Tadpole diagram\la{fig2}}
\end{figure}
Due to the form of the fermion-boson and fermion-fermion interactions the 
(gauged WZW model based) theories we consider here are naturally defined 
only in 2 dimensions so that dimensional regularisation is not suitable.
Instead we assume a direct momentum cut-off and ignore divergent terms. They
should cancel against the contributions of tadpole diagrams coming from the
sextic tadpoles, as the theories are UV-finite.

Below we will list the expressions for the one-loop S-matrices in each of the
three  reduced \adss{n} theories. It will be useful to extract the factor 
\be\la{qex}
\po(\q,k\,;c)=1+ \ c \ \fr{\pi\cosech \q}{2k^2} 
              \Big(i\big[ 2 + (i\pi-2\q)\coth\q\big]  -  \pi\cosech\q\Big)\ 
\ee
from the one-loop S-matrix (with different values of the constant $c$ depending
on the theory). This will set the $Y_m Z_n \ra Y_m Z_n$ and $\z_m \ch_n \ra
\z_m \ch_n$ amplitudes equal to one, at least to one-loop order.

\subsubsection{Reduced \adss{2} theory\la{secsmat22}}

Let us start with the Lagrangian \eqref{lag15} and restrict the indices
$m,n,p,q$ to only take a single value and rescale $k \ra \fr{k}{2}$. The
resulting Lagrangian is then that of the reduced \adss{2} theory \ci{gt1,ht1}
expanded to quartic order
\bea
\Lag_{2} & = & \fr{1}{2}\dpl Y \dm Y
\no         - \fr{\mu^2}{2} Y^2
             + \fr{1}{2}\dpl Z \dm Z
             - \fr{\mu^2}{2} Z^2
\\       & & + \fr{i}{2}\zel \dpl \zel
\la{lag12}        + \fr{i}{2}\zer \dm \zer
             + i\mu \zer\zel
             + \fr{i}{2}\chl \dpl \chl
             + \fr{i}{2}\chr \dm \chr
             + i\mu \chr\chl
\\       & & + \fr{\pi}{k}\Big[
 \no  \;- \fr{\mu^2}{3}Y^4
               + \fr{\mu^2}{3}Z^4
               + i\mu(\zer \zel+\chr \chl)(Y^2 - Z^2)
               + \,2i\mu(\zer \chl - \chr \zel) Y Z \Big]
             +\ord{k^{-2}}\,.
\eea
The resulting one-loop S-matrix is found to have the following structure:

{\allowdisplaybreaks
\nin\tbf{Boson-Boson}
\begin{align*}
\Sc\ket{Y(p_1)Y(p_2)}
= & f_1(\q,k)\ket{Y(p_1)Y(p_2)}
    +f_5(\q,k)\ket{Z(p_1)Z(p_2)}
\\& +f_6(\q,k)\ket{\z(p_1)\z(p_2)}
    +f_6(\q,k)\ket{\ch(p_1)\ch(p_2)}
\\\Sc\ket{Z(p_1)Z(p_2)}
= & f_1(\q,-k)\ket{Z(p_1)Z(p_2)}
    +f_5(\q,-k)\ket{Y(p_1)Y(p_2)}
\\& +f_6(\q,-k)\ket{\ch(p_1)\ch(p_2)}
    +f_6(\q,-k)\ket{\z(p_1)\z(p_2)}
\\\Sc\ket{Y(p_1)Z(p_2)}
= & f_3(\q,k)\ket{Y(p_1)Z(p_2)}
    +f_5(i\pi-\q,k)\ket{Z(p_1)Y(p_2)}
\\& +f_7(\q,k)\ket{\z(p_1)\ch(p_2)}
    -f_7(\q,k)\ket{\ch(p_1)\z(p_2)}
\end{align*}
\tbf{Boson-Fermion}
\begin{align*}
\Sc\ket{Y(p_1)\z(p_2)}
= & f_4(\q,k)\ket{Y(p_1)\z(p_2)}
    +if_6(i\pi-\q,k)\ket{\z(p_1)Y(p_2)}
\\& +f_8(\q,k)\ket{Z(p_1)\ch(p_2)}
    +if_7(i\pi-\q,k)\ket{\ch(p_1)Z(p_2)}
\\\Sc\ket{Y(p_1)\ch(p_2)}
= & f_4(\q,k)\ket{Y(p_1)\ch(p_2)}
    +if_6(i\pi-\q,k)\ket{\ch(p_1)Y(p_2)}
\\& -f_8(\q,k)\ket{Z(p_1)\z(p_2)}
    -if_7(i\pi-\q,k)\ket{\z(p_1)Z(p_2)}
\\\Sc\ket{Z(p_1)\z(p_2)}
= & f_4(\q,-k)\ket{Z(p_1)\z(p_2)}
    +if_6(i\pi-\q,-k)\ket{\z(p_1)Z(p_2)}
\\& -f_8(\q,-k)\ket{Y(p_1)\ch(p_2)}
    -if_7(i\pi-\q,-k)\ket{\ch(p_1)Y(p_2)}
\\\Sc\ket{Z(p_1)\ch(p_2)}
= & f_4(\q,-k)\ket{Z(p_1)\ch(p_2)}
    +if_6(i\pi-\q,-k)\ket{\ch(p_1)Z(p_2)}
\\& +f_8(\q,-k)\ket{Y(p_1)\z(p_2)}
    +if_7(i\pi-\q,-k)\ket{\z(p_1)Y(p_2)}
\end{align*}
\tbf{Fermion-Fermion}
\begin{align*}
\Sc\ket{\z(p_1)\z(p_2)}
= & \ f_2(\q,k)\ket{\z(p_1)\z(p_2)}
    -f_5(\q,k)\ket{\ch(p_1)\ch(p_2)}
\\& + f_6(\q,k)\ket{Y(p_1)Y(p_2)}
    +f_6(\q,-k)\ket{Z(p_1)Z(p_2)}
\\\Sc\ket{\ch(p_1)\ch(p_2)}
= & f_2(\q,-k)\ket{\ch(p_1)\ch(p_2)}
    -f_5(\q,-k)\ket{\z(p_1)\z(p_2)}
\\& +f_6(\q,-k)\ket{Z(p_1)Z(p_2)}   
    +f_6(\q,k)\ket{Y(p_1)Y(p_2)}
\\\Sc\ket{\z(p_1)\ch(p_2)}
= & f_3(\q,k)\ket{\z(p_1)\ch(p_2)}
    -f_5(i\pi-\q,k)\ket{\ch(p_1)\z(p_2)}
\\& +f_7(\q,k)\ket{Y(p_1)Z(p_2)}
    +f_7(\q,k)\ket{Z(p_1)Y(p_2)}
\end{align*}
\tbf{Functions}
\begin{align*}
\hat{f}_1(\q,k)=& 1-\fr{2i\pi}{k}\cosech \q
             -\fr{\pi\cosech \q}{k^2} 
               +\ord{k^{-3}}\,,\hs{15pt}
&\hat{f}_2(\q,k)=& 1+\fr{\pi\cosech^2 \q}{k^2} 
                +\ord{k^{-3}}\,,
\\\hat{f}_3(\q,k)=& 1+\ord{k^{-3}}\,,\hs{15pt}
&\hat{f}_4(\q,k)=& 1-\fr{i\pi}{k}\cosech \q
                   +\ord{k^{-3}}\,,
\\\hat{f}_5(\q,k)=&\fr{\pi^2}{4k^2}\sech^2\fr{\q}{2}
+\ord{k^{-3}}\,,\hs{15pt}
&\hat{f}_6(\q,k)=&\fr{i\pi}{2k}\sech\fr{\q}{2}
             +\fr{\pi^2}{2k^2}\cosech\q
                              \sech\fr{\q}{2}+\ord{k^{-3}}\,,
\\\hat{f}_7(\q,k)=&\fr{i\pi}{2k}\cosech\fr{\q}{2}+\ord{k^{-3}}\,,\hs{15pt}
&\hat{f}_8(\q,k)=&-\fr{\pi^2}{2k^2}\cosech\q+\ord{k^{-3}}\,.
\end{align*}}
The 8 functions $\hat{f}_i$ are related to $f_i$ entering $\Sc$ by the phase
factor $\po(\q,k\,;1)$ defined in \eqref{qex} 
\be \la{pp22}
{f}_i (\q,k) = {\po(\q,k\,;1)}\, \hat f_i (\q,k)  \  .
\ee
Here the $1/k$ terms in $f_i$ represent tree-level contributions to the
2-particle S-matrix ($k$ is the overall coefficient in \rf{gwzw}, see also
\rf{lag15}) and the $1/k^2$ terms  -- the one-loop contributions. 

\subsubsection{Reduced \adss{3} theory\la{secollsg33}}

Here we start  again  with the Lagrangian \eqref{lag15}, rescaling $k \ra
\fr{k}{2}$ and restricting the indices $m,n,p,q$ to take the values $1$ and $2$
(so they become $SO(2)$ vector indices). The resulting Lagrangian is then that
of the reduced \adss{3} theory \ci{gt2,ht1} expanded to quartic order
{\allowdisplaybreaks
\bea
\Lag_{3} & = & \fr{1}{2}\dpl Y_m \dm Y_m 
\no         - \fr{\mu^2}{2} Y_m Y_m
             + \fr{1}{2}\dpl Z_m \dm Z_m
             - \fr{\mu^2}{2} Z_m Z_m
\\       & & + \fr{i}{2}\zel_m \dpl \zel_m
\no          + \fr{i}{2}\zer_m \dm \zer_m
             + i\mu \zer_m \zel_m
         + \fr{i}{2}\chl_m \dpl \chl_m
           + \fr{i}{2}\chr_m \dm \chr_m
             + i\mu \chr_m \chl_m
\\       & & + \fr{\pi}{k}\Big[
\no          - \fr{2}{3}Y_m Y_m \dpl Y_n \dm Y_n
             + \fr{2}{3}Y_m \dpl Y_m Y_n \dm Y_n
             - \fr{\mu^2}{3}Y_m Y_m Y_n Y_n
\\       & & \hs{26pt}
\no          + \fr{2}{3}Z_m Z_m \dpl Z_n \dm Z_n
             - \fr{2}{3}Z_m \dpl Z_m Z_n \dm Z_n
             + \fr{\mu^2}{3}Z_m Z_m Z_n Z_n
\\       & & \hs{26pt}
\la{lag13}   + \fr{i}{2}\g_{mnpq}
               (\zel_m \zel_n Y_p \dpl Y_q
               +\zer_m \zer_n Y_p \dm Y_q
\\       & & \hs{126pt}
\no            -\chl_m \chl_n Z_p \dpl Z_q
               -\chr_m \chr_n Z_p \dm Z_q)
\\       & & \hs{26pt}
\no          + \fr{i}{2}\g_{mnpq}
               (\chl_m \chl_n Y_p \dpl Y_q
               +\chr_m \chr_n Y_p \dm Y_q
\\       & & \hs{126pt}
\no            -\zel_m \zel_n Z_p \dpl Z_q
               -\zer_m \zer_n Z_p \dm Z_q)
\\       & & \hs{26pt}
\no          +i\mu(\zer_m \zel_m+\chr_m \chl_m)
                  (Y_n Y_n - Z_n Z_n)
\\       & & \hs{26pt}
\no          -2i\mu (\de_{mn}\de_{pq}
                     -\de_{mp}\de_{nq}-\de_{mq}\de_{np})
                    (\zer_m \chl_n Y_p Z_q
                     -\chr_m \zel_n Z_p Y_q)\Big]
             +\ord{k^{-2}}
\eea}
The corresponding one-loop S-matrix is found to be: 

{\allowdisplaybreaks
\nin\tbf{Boson-Boson}
\begin{align*}
\Sc\ket{Y_m(p_1)Y_n(p_2)}
= & f_1{}_{mnpq}(\q,k)\ket{Y_p(p_1)Y_q(p_2)}
    +f_5{}_{mnpq}(\q,k)\ket{Z_p(p_1)Z_q(p_2)}
\\& +f_6{}_{mnpq}(\q,k)\ket{\z_p(p_1)\z_q(p_2)}
    +f_6{}_{mnpq}(\q,k)\ket{\ch_p(p_1)\ch_q(p_2)}
\\\Sc\ket{Z_m(p_1)Z_n(p_2)}
= &  f_1{}_{mnpq}(\q,-k)\ket{Z_p(p_1)Z_q(p_2)}
    +f_5{}_{mnpq}(\q,-k)\ket{Y_p(p_1)Y_q(p_2)}
\\& +f_6{}_{mnpq}(\q,-k)\ket{\ch_p(p_1)\ch_q(p_2)}
    +f_6{}_{mnpq}(\q,-k)\ket{\z_p(p_1)\z_q(p_2)}
\\\Sc\ket{Y_m(p_1)Z_n(p_2)}
= & f_3{}_{mnpq}(\q,k)\ket{Y_p(p_1)Z_q(p_2)}
    +f_5{}_{mqpn}(i\pi-\q,k)\ket{Z_p(p_1)Y_q(p_2)}
\\& +f_7{}_{mnpq}(\q,k)\ket{\z_p(p_1)\ch_q(p_2)}
    -f_7{}_{mnpq}(\q,k)\ket{\ch_p(p_1)\z_q(p_2)}
\end{align*}
\tbf{Boson-Fermion}
\begin{align*}
\Sc\ket{Y_m(p_1)\z_n(p_2)}
= & f_4{}_{mnpq}(\q,k)\ket{Y_p(p_1)\z_q(p_2)}
    +if_6{}_{mqpn}(i\pi-\q,k)\ket{\z_p(p_1)Y_q(p_2)}
\\& +f_8{}_{mnpq}(\q,k)\ket{Z_p(p_1)\ch_q(p_2)}
    +if_7{}_{mqpn}(i\pi-\q,k)\ket{\ch_p(p_1)Z_q(p_2)}
\\\Sc\ket{Y_m(p_1)\ch_n(p_2)}
= & f_4{}_{mnpq}(\q,k)\ket{Y_p(p_1)\ch_q(p_2)}
    +if_6{}_{mqpn}(i\pi-\q,k)\ket{\ch_p(p_1)Y_q(p_2)}
\\& -f_8{}_{mnpq}(\q,k)\ket{Z_p(p_1)\z_q(p_2)}
    -if_7{}_{mqpn}(i\pi-\q,k)\ket{\z_p(p_1)Z_q(p_2)}
\\\Sc\ket{Z_m(p_1)\z_n(p_2)}
= & f_4{}_{mnpq}(\q,-k)\ket{Z_p(p_1)\z_q(p_2)}
    +if_6{}_{mqpn}(i\pi-\q,-k)\ket{\z_p(p_1)Z_q(p_2)}
\\& -f_8{}_{mnpq}(\q,-k)\ket{Y_p(p_1)\ch_q(p_2)}
    -if_7{}_{mqpn}(i\pi-\q,-k)\ket{\ch_p(p_1)Y_q(p_2)}
\\\Sc\ket{Z_m(p_1)\ch_n(p_2)}
= & f_4{}_{mnpq}(\q,-k)\ket{Z_p(p_1)\ch_q(p_2)}
    +if_6{}_{mqpn}(i\pi-\q,-k)\ket{\ch_p(p_1)Z_q(p_2)}
\\& +f_8{}_{mnpq}(\q,-k)\ket{Y_p(p_1)\z_q(p_2)}
    +if_7{}_{mqpn}(i\pi-\q,-k)\ket{\z_p(p_1)Y_q(p_2)}
\end{align*}
\tbf{Fermion-Fermion}
\begin{align*}
\Sc\ket{\z_m(p_1)\z_n(p_2)}
= & f_2{}_{mnpq}(\q,k)\ket{\z_p(p_1)\z_q(p_2)}
    -f_{\td{5}}{}_{mnpq}(\q,k)\ket{\ch_p(p_1)\ch_q(p_2)}
\\ & +f_6{}_{mnpq}(\q,k)\ket{Y_p(p_1)Y_q(p_2)}
    +f_6{}_{mnpq}(\q,-k)\ket{Z_p(p_1)Z_q(p_2)}
\\\Sc\ket{\ch_m(p_1)\ch_n(p_2)}
= & f_2{}_{mnpq}(\q,-k)\ket{\ch_p(p_1)\ch_q(p_2)}
    -f_{\td{5}}{}_{mnpq}(\q,-k)\ket{\z_p(p_1)\z_q(p_2)}
\\ & +f_6{}_{mnpq}(\q,-k)\ket{Z_p(p_1)Z_q(p_2)}
     +f_6{}_{mnpq}(\q,k)\ket{Y_p(p_1)Y_q(p_2)}
\\\Sc\ket{\z_m(p_1)\ch_n(p_2)}
= & f_{\td{3}}{}_{mnpq}(\q,k)\ket{\z_p(p_1)\ch_q(p_2)}
    -f_{\td{5}}{}_{mqpn}(i\pi-\q,k)\ket{\ch_p(p_1)\z_q(p_2)}
\\ & +f_7{}_{mnpq}(\q,k)\ket{Y_p(p_1)Z_q(p_2)}
    +f_7{}_{mnpq}(\q,k)\ket{Z_p(p_1)Y_q(p_2)}
\end{align*}
\tbf{Functions}
\begin{align*}
\hat{f}_1{}_{mnpq}(\q,k)=&
\no \de_{mp}\de_{nq}\Big(1-\fr{2i\pi}{k}\cosech\q-\fr{2\pi^2}{k^2}\coth^2\q\Big)
\\&+\e_{mp}\e_{nq}\Big(\fr{2i\pi}{k}\coth\q -\fr{i\pi}{k^2}(i\pi-2\q)\cosech^2\q
\no                                   +\fr{3\pi^2}{k^2}\coth\q\cosech\q\Big)
\\&\underline{ -\fr{i\pi}{k^2}\Big(\e_{mn}\e_{pq} (\cosech\q-\coth\q)
\no                +\e_{mq}\e_{pn}(\cosech\q+\coth\q)\Big)}  +\ord{k^{-3}\ , }
\\
\hat{f}_2{}_{mnpq}(\q,k)=&
\no\de_{mp}\de_{nq}\Big(1+\fr{2\pi^2}{k^2}\cosech^2\q\Big)
\\&+\e_{mp}\e_{nq}\Big(-\fr{i\pi}{k^2}(i\pi-2\q)\cosech^2\q   
\no                      -\fr{\pi^2}{k^2}\coth\q\cosech\q \Big)
\\&\underline{+\fr{2i\pi}{k^2}\cosech\q\Big(\e_{mn}\e_{pq}+\e_{mq}\e_{pn}\Big)}
\no           +\ord{k^{-3}} \ , 
\\\hat{f}_3{}_{mnpq}(\q,k)=&\hat{f}_{\td{3}}{}_{mnpq}(\q,k)
=\de_{mp}\de_{nq} +\e_{mp}\e_{nq}\Big(-\fr{i\pi}{k^2}(i\pi-2\q)\cosech^2\q
\no                     +\fr{\pi^2}{k^2}\coth\q\cosech\q\Big)+\ord{k^{-3}} \ , 
\\\hat{f}_4{}_{mnpq}(\q,k)=&
\de_{mp}\de_{nq}\Big(1-\fr{i\pi}{k}\cosech\q -\fr{\pi^2}{2k^2}\Big)
\\&+\e_{mp}\e_{nq}\Big(\fr{i\pi}{k}\coth\q-\fr{i\pi}{k^2}(i\pi-2\q)\cosech^2\q
\no            +\fr{\pi^2}{k^2}\coth\q\cosech\q\Big)+\ord{k^{-3}}\ , 
\\\hat{f}_5{}_{mnpq}(\q,k)=&
\fr{\pi^2}{2k^2}\sech^2\fr{\q}{2}(\de_{mn}\de_{pq}+\e_{mn}\e_{pq})
\no             \underline{-\fr{i\pi}{k^2}\e_{mn}\e_{pq}
                               (\cosech\q-\coth\q)}+\ord{k^{-3}}\ , 
\\\hat{f}_{\td{5}}{}_{mnpq}(\q,k)=&
\fr{\pi^2}{2k^2}\sech^2\fr{\q}{2}(\de_{mn}\de_{pq}+\e_{mn}\e_{pq})
\no           \underline{-\fr{2i\pi}{k^2}\e_{mn}\e_{pq}
                                   \cosech\q}+\ord{k^{-3}}\ , 
\\\hat{f}_6{}_{mnpq}(\q,k)=& 
\no                     \Big(\fr{i\pi}{2k}\sech\fr{\q}{2}-
                       \fr{\pi^2}{2k^2}\sech\fr{\q}{2}\tanh\fr{\q}{2}\Big)
			(\de_{mn}\de_{pq}+\e_{mn}\e_{pq})
                       \underline{+\fr{i\pi}{k^2}\e_{mn}\e_{pq}
                       \sech\fr{\q}{2}}+\ord{k^{-3}}\ , 
\\\hat{f}_7{}_{mnpq}(\q,k)=& \fr{i\pi}{2k}\cosech\fr{\q}{2} 
\no          \big(-\de_{mn}\de_{pq}+\de_{mp}\de_{nq}+\de_{mq}\de_{np}\big)
              +\ord{k^{-3}}\ , 
\\  \hat{f}_8{}_{mnpq}(\q,k)=& \ord{k^{-3}}\ . 
\end{align*}}
As in the previous case \rf{pp22} the 10 tensor functions $(f_i)_{mnpq}$ are
related to $(\hat f_i)_{mnpq}$ by extracting the scalar phase factor
$\po(\q,k\,;2)$ \eqref{qex}
\be \la{pp33}
{f}_i (\q,k) = {\po(\q,k\,;2)}\, \hat{f}_i (\q,k) \ . 
\ee
Here the  $YYZZ$ and $\z\z\ch\ch$ amplitudes are not related in the same way as
they were in the \adss{2} case (the functions $f_5$ and $f_{\td{5}}$ are not
equal). 

This one-loop S-matrix does not satisfy the Yang-Baxter equation. Similarly to
the complex sine-Gordon case \ci{dvm1,ht2} one can find a local quartic
counterterm whose contribution cancels the underlined terms in the above 
S-matrix coefficients $\hat f_i$ and thus restores the validity of the
Yang-Baxter equation at the one-loop order. Adding this counterterm also
restores the equality between the coefficient functions $f_5$ and $f_{\td{5}}$. 
This will be discussed in detail in section \ref{pr33}.

\subsubsection{Reduced \adss{5} theory\la{secollsg55}}

The one-loop S-matrix computed starting with the Lagrangian \eqref{lag15} is

{\allowdisplaybreaks
\nin\tbf{Boson-Boson}
\begin{align*}
\Sc\ket{Y_m(p_1)Y_n(p_2)}
= & f_1{}_{mnpq}(\q,k)\ket{Y_p(p_1)Y_q(p_2)}
    +f_5{}_{mnpq}(\q,k)\ket{Z_p(p_1)Z_q(p_2)}
\\& +f_6^+{}_{mnpq}(\q,k)\ket{\z_p(p_1)\z_q(p_2)}
    +f_6^-{}_{mnpq}(\q,k)\ket{\ch_p(p_1)\ch_q(p_2)}
\\\Sc\ket{Z_m(p_1)Z_n(p_2)}
= & f_1{}_{mnpq}(\q,-k)\ket{Z_p(p_1)Z_q(p_2)}
   +f_5{}_{mnpq}(\q,-k)\ket{Y_p(p_1)Y_q(p_2)}
\\& +f_6^+{}_{mnpq}(\q,-k)\ket{\ch_p(p_1)\ch_q(p_2)}
     +f_6^-{}_{mnpq}(\q,-k)\ket{\z_p(p_1)\z_q(p_2)}
\\\Sc\ket{Y_m(p_1)Z_n(p_2)}
= & f_3{}_{mnpq}(\q,k)\ket{Y_p(p_1)Z_q(p_2)}
    +f_5{}_{mqpn}(i\pi-\q,k)\ket{Z_p(p_1)Y_q(p_2)}
\\& +f_7^+{}_{mnpq}(\q,k)\ket{\z_p(p_1)\ch_q(p_2)}
    -f_7{}^-_{mnpq}(\q,k)\ket{\ch_p(p_1)\z_q(p_2)}
\end{align*}
\tbf{Boson-Fermion}
\begin{align*}
\Sc\ket{Y_m(p_1)\z_n(p_2)}
= & f_4^+{}_{mnpq}(\q,k)\ket{Y_p(p_1)\z_q(p_2)}
    +if_6^+{}_{mqpn}(i\pi-\q,k)\ket{\z_p(p_1)Y_q(p_2)}
\\& +f_8^+{}_{mnpq}(\q,k)\ket{Z_p(p_1)\ch_q(p_2)}
    +if_7^-{}_{mqpn}(i\pi-\q,k)\ket{\ch_p(p_1)Z_q(p_2)}
\\\Sc\ket{Y_m(p_1)\ch_n(p_2)}
= & f_4^-{}_{mnpq}(\q,k)\ket{Y_p(p_1)\ch_q(p_2)}
    +if_6^-{}_{mqpn}(i\pi-\q,k)\ket{\ch_p(p_1)Y_q(p_2)}
\\& -f_8^-{}_{mnpq}(\q,k)\ket{Z_p(p_1)\z_q(p_2)}
    -if_7^+{}_{mqpn}(i\pi-\q,k)\ket{\z_p(p_1)Z_q(p_2)}
\\\Sc\ket{Z_m(p_1)\z_n(p_2)}
= & f_4^-{}_{mnpq}(\q,-k)\ket{Z_p(p_1)\z_q(p_2)}
    +if_6^-{}_{mqpn}(i\pi-\q,-k)\ket{\z_p(p_1)Z_q(p_2)}
\\& -f_8^-{}_{mnpq}(\q,-k)\ket{Y_p(p_1)\ch_q(p_2)}
    -if_7^+{}_{mqpn}(i\pi-\q,-k)\ket{\ch_p(p_1)Y_q(p_2)}
\\\Sc\ket{Z_m(p_1)\ch_n(p_2)}
= & f_4^+{}_{mnpq}(\q,-k)\ket{Z_p(p_1)\ch_q(p_2)}
    +if_6^+{}_{mqpn}(i\pi-\q,-k)\ket{\ch_p(p_1)Z_q(p_2)}
\\& +f_8^+{}_{mnpq}(\q,-k)\ket{Y_p(p_1)\z_q(p_2)}
    +if_7^-{}_{mqpn}(i\pi-\q,-k)\ket{\z_p(p_1)Y_q(p_2)}
\end{align*}
\tbf{Fermion-Fermion}
\begin{align*}
\Sc\ket{\z_m(p_1)\z_n(p_2)}
= & f_2{}_{mnpq}(\q,k)\ket{\z_p(p_1)\z_q(p_2)}
    -f_5{}_{mnpq}(\q,k)\ket{\ch_p(p_1)\ch_q(p_2)}
\\& +f_6^+{}_{mnpq}(\q,k)\ket{Y_p(p_1)Y_q(p_2)}
    +f_6^-{}_{mnpq}(\q,-k)\ket{Z_p(p_1)Z_q(p_2)}
\\\Sc\ket{\ch_m(p_1)\ch_n(p_2)}
= & f_2{}_{mnpq}(\q,-k)\ket{\ch_p(p_1)\ch_q(p_2)}
    -f_5{}_{mnpq}(\q,-k)\ket{\z_p(p_1)\z_q(p_2)}
\\& +f_6^+{}_{mnpq}(\q,-k)\ket{Z_p(p_1)Z_q(p_2)}
    +f_6^-{}_{mnpq}(\q,k)\ket{Y_p(p_1)Y_q(p_2)}
\\\Sc\ket{\z_m(p_1)\ch_n(p_2)}
= & f_3{}_{mnpq}(\q,k)\ket{\z_p(p_1)\ch_q(p_2)}
   -f_5{}_{mqpn}(i\pi-\q,k)\ket{\ch_p(p_1)\z_q(p_2)}
\\& +f_7^+{}_{mnpq}(\q,k)\ket{Y_p(p_1)Z_q(p_2)}
    +f_7^-{}_{mnpq}(\q,k)\ket{Z_p(p_1)Y_q(p_2)}
\end{align*}
\tbf{Functions}
\begin{align*}
\hat{f}_1{}_{mnpq}(\q,k)=&
    \de_{mp}\de_{nq}\big(1-\fr{i\pi}{k}\cosech\q 
       -\fr{\pi^2}{2k^2}\big)
\\&+\de_{mn}\de_{pq}\big(\fr{i\pi}{k}\coth\q
       +\fr{\pi}{k^2}(i(i\pi-\q)
       -\fr{\pi}{2}(\cosech\q-\coth\q)\cosech\q)\big)
\\&+\de_{mq}\de_{np}\big(-\fr{i\pi}{k}\coth\q
       +\fr{\pi}{k^2}(i\q
       -\fr{\pi}{2}(\cosech\q+\coth\q)\cosech\q)\big)+\ord{k^{-3}}\ , 
\\\hat{f}_2{}_{mnpq}(\q,k)=&
    \de_{mp}\de_{nq}\big(1 
      -\fr{\pi^2}{k^2}\big) -\fr{i\pi}{k}\e_{mnpq}
\coth\q
\\&+\de_{mn}\de_{pq}\big(\fr{\pi}{k^2}(-i\q
       +\fr{\pi}{2}(\cosech\q-\coth\q)\cosech\q)\big)
\\&+\de_{mq}\de_{np}\big(\fr{\pi}{k^2}(-i(i\pi-\q)
       +\fr{\pi}{2}(\cosech\q+\coth\q)\cosech\q)\big) +\ord{k^{-3}}\ , 
\\\hat{f}_3{}_{mnpq}(\q,k)=& 
    \de_{mp}\de_{nq}+\ord{k^{-3}}\ , 
\\\hat{f}_4^\pm{}_{mnpq}(\q,k)=& 
    \de_{mp}\de_{nq}\big(1-\fr{i\pi}{2k}\cosech\q
       -\fr{3\pi^2}{8k^2}\big)
\\&+(\de_{mn}\de_{pq}-\de_{mq}\de_{np}\mp\e_{mnpq})
       \big(\fr{i\pi}{2k}\coth\q
       +\fr{i\pi}{4k^2}(i\pi-2\q)\big)+\ord{k^{-3}}\ , 
\\\hat{f}_5{}_{mnpq}(\q,k)=& 
    \fr{\pi^2}{4k^2}\de_{mn}\de_{pq}\sech^2\fr{\q}{2}+\ord{k^{-3}}\ , 
\\\hat{f}_6^\pm{}_{mnpq}(\q,k)=&
    -\fr{\pi^2}{4k^2}\de_{mn}\de_{pq}\cosech\fr{\q}{2}
                           (1+\tanh^2\fr{\q}{2})
\\&+(\de_{mn}\de_{pq}+\de_{mp}\de_{nq}-\de_{mq}\de_{np}\pm\e_{mnpq})
    \big(\fr{i\pi}{4k}\sech\fr{\q}{2}
    +\fr{\pi^2}{8k^2}\cosech\fr{\q}{2}\big)+\ord{k^{-3}}\ , 
\\\hat{f}_7^\pm{}_{mnpq}(\q,k)=& \fr{i\pi}{4k}\cosech\fr{\q}{2}
   \big(\de_{mp}\de_{nq}+\de_{mq}\de_{np}
-\de_{mn}\de_{pq}\mp
\e_{mnpq}\big)+\ord{k^{-3}}\ , 
  \\\hat{f}_8^\pm{}_{mnpq}(\q,k)=& \fr{\pi^2}{4k^2}\cosech\q
  \big(\de_{mn}\de_{pq}+\de_{mq}\de_{np} -
\de_{mp}\de_{nq}\pm\e_{ mnpq}\big)+\ord{k^{-3}} \ , 
\end{align*}}
Here again we extracted the phase factor $\po(\q,k\,;1)$ defined in \eqref{qex},
i.e. all $f_i$ in the S-matrix are given in terms of the corresponding
$\hat{f}_i$ by 
\be \la{p55}
{f}_i (\q,k)  ={\po(\q,k\,;1)}\,  \hat{f}_i  (\q,k)  \ . 
\ee
The $1/k$ terms in $f_i$ are tree-level contributions found in \ci{ht1} while
$1/k^2$ terms are new one-loop contributions computed here.

For the reduced \adss{5} theory the $YYZZ$ and $\z\z\ch\ch$ amplitudes are 
related in the same way as they were for the reduced \adss{2} theory, i.e. 
in contrast to the \adss{3}  case, here $f_5 = \tilde{f}_5$. This equality could
be related to the group factorisation  property of the perturbative S-matrix
\ci{ht2} and may be  suggesting the presence of a hidden fermionic symmetry
relating  bosons and fermions (see also  below).


\renewcommand{\theequation}{3.\arabic{equation}}
\setcounter{equation}{0}
\section{S-matrix of the reduced \adss{2} theory\la{seclag22}}

In \ci{gt1} the reduced version of the \adss{2} superstring model based on the
$\fr{\hat F}{G}  =  \fr{PSU(1,1|2)}{SO(1,1) \x SO(2)}$ supercoset was shown to
be equivalent to the $\mc{N}=2$ supersymmetric sine-Gordon model, whose exact
S-matrix is known, \ci{sw,ahn,ku1,ku2}. In this section we review this S-matrix
and check that its perturbative expansion indeed  matches the one-loop result 
found  in  section \ref{secsmat22}. 

We also identify certain key features of this theory that will be useful in
analysing the reduced \adss{3} and \adss{5} theories. In particular, there is a
specific phase factor that also plays a r\^{o}le in the reduced \adss{5} theory.

\subsection{Symmetries\la{redsym22}}

Since in  the reduced \adss{2} theory $G=SO(1,1) \x SO(2)$, the gauge group $H$
is trivial and therefore the theory has no manifest bosonic symmetry (gauge or
global), other than the usual 2-d Poincar\'{e} symmetry. Parametrising the group
field $g$ in terms of an algebra-valued field \eqref{paramg} and expanding out
the Lagrangian \eqref{gwzw} in components we get 
\bea
L_{2}&=&\no \fr{k}{4 \pi} \Big[\dpl\phi\dm\phi+\dpl\vp\dm\vp
                + \fr{\mu^2}{2}\left(\cos 2\vp-\cosh 2\phi \right)
               +i\,\alpha\dm\alpha+i\,\delta\dm\delta
               +i\,\nu\dpl\nu+i\,\rho\dpl\rho
\\&&\la{lag32}
\hs{15pt}      -2i\mu\big(\cosh\phi\cos\vp\left(\nu\delta+\rho\alpha\right)
                +\sinh\phi\sin\vp(-\rho\delta+\nu\alpha)\big)\Big]\,.
\eea
Here $\phi,\,\vp$ are real bosonic  fields and $\alpha,\,\delta,\,\nu,\,\rho$
are real (hermitian) fermions. Expanding this Lagrangian to quartic order one
finds agreement with \eqref{lag12} up to  simple field and coupling constant 
redefinitions.

Furthermore, this Lagrangian is exactly that of $\mc{N}=2$ supersymmetric
sine-Gordon theory \ci{gt1}, i.e. this theory has $\mc{N}=2$ \ 2-d worldsheet
supersymmetry. The $\mc{N}=2$ supersymmetry algebra can be represented in the
following way\,\foot{Our notation for semi-direct sums and central extensions
is defined in footnote \ref{foot1}. Also, $ [\mf{psu}(1|1)]^2$  stands for
$\mf{psu}(1|1)\oplus \mf{psu}(1|1)$.} 
\be \label{symgroup22}
\so(1,1) \sds [\mf{psu}(1|1)]^2\ltimes \mbb{R}^2\,.
\ee
The superalgebra $\mf{psu}(1|1)$ has two anticommuting fermionic generators and
no bosonic generators. The $\mbb{R}^2$ central extensions correspond to the
light-cone combinations $ \mf{P}_\pm $ of the 2-momentum components, i.e. the
commutation relations can be written as 
\be\la{tak22}
\acom{\mf{Q}^i_R}{\mf{Q}^j_L}=0\,,\hs{20pt}
\acom{\mf{Q}^i_R}{\mf{Q}^j_R}=\de^{ij}\mf{P}_+\,,\hs{15pt}
\acom{\mf{Q}^i_L}{\mf{Q}^j_L}=\de^{ij}\mf{P}_-\,,\hs{15pt}
i,\,j=1,\,2\,.
\ee
The origin of this $\mc{N}=2$ 2-d  supersymmetry in the \adss{2} reduced theory
appears to be in the global target space supergroup used in the construction of
the \adss{2} superstring theory as a supercoset  GS sigma model. In particular, 
in the Pohlmeyer reduction \ci{gt1} the fermionic fields were redefined in such
a way that they became charged under the 2-d Lorentz symmetry of the reduced
theory (the original GS fermions were 2-d scalars).

We shall assume the scattering states of the theory to be eigenstates of the
momentum operator. As the direct-sum  superalgebra $[\mf{psu}(1|1)]^2$ commutes
with the momentum generators we expect the scattering states at fixed momenta to
transform in a bi-representation of this direct sum. Furthermore, as we are
dealing with an integrable theory we expect the S-matrix to factorise under the
corresponding  direct-product symmetry structure \ci{ku1,ku2}.

\subsection{Group factorisation of the S-matrix}

To confirm that the one-loop S-matrix of section \ref{secsmat22} agrees with the
perturbative expansion of the exact $\mathcal{N}=2$ supersymmetric sine-Gordon
S-matrix \ci{ku1,ku2} we relabel our states  as follows
\be\la{decc}\begin{split}
\ket{Z}=\ket{\Phi_{00}}\,,&\hs{20pt}\ket{\ch}=\ket{\Phi_{01}}\,,
\\\ket{Y}=\ket{\Phi_{11}}\,,&\hs{20pt}\ket{\z}=\ket{\Phi_{10}}\,.
\end{split}\ee
The $\mathcal{N}=2$ supersymmetry can be understood as two anticommuting (up to
central extensions) $\mathcal{N}=1$ supersymmetries that act on different
indices of $\ket{\Phi_{a\al}}$\ ($a,\,\al,\ldots=0,1$). We also take $0$ to be
a bosonic index and $1$ to be a fermionic index, so that the gradings are 
\be\la{22fg}
[0]=0\,,\hs{20pt}[1]=1\,.
\ee
The S-matrix can then be parametrised in the following way
\be\la{fact22}
\Sc\ket{\Phi_{a\al}\Phi_{b\bet}}=S_{ab,\al\bet}^{cd,\g\de}(\q,k)
                                          \ket{\Phi_{c\g}\Phi_{d\de}}\,.
\ee
As the reduced \adss{2} theory is integrable we expect (as discussed in section
\ref{redsym22}) the S-matrix to factorise under the direct-product symmetry 
group structure \eqref{symgroup22} as follows 
\be\la{factsss}
S_{a\al,b\bet}^{c\g,d\de}(\q,k)= (-1)^{[\al][b] +[\g][d]}
S_B(\q,k)S_{ab}^{cd}(\q,k)S_{\al\bet}^{\g\de}(\q,k)\,.
\ee
Here, following \ci{ku2}, an overall bosonic factor $S_B(\q,k)$ has been
extracted. The $\mc{N}=2$ supersymmetric sine-Gordon S-matrix can be understood
as a supersymmetrisation of the corresponding bosonic sine-Gordon S-matrix
\ci{ku2,ahn}. Therefore, we take this factor to be the same as the sine-Gordon
perturbative excitation S-matrix.\,\foot{The exact S-matrix for the sine-Gordon
perturbative excitation is given by 
\be\no
S_{sg}(\q,\De)= \fr{\sinh\q+i\sin\De}{\sinh\q-i\sin\De}\,,
\ee
where $\De$ is a function of the coupling $k$. In $\mc{N}=2$ supersymmetric
sine-Gordon this function is $\De=\fr{\pi}{k}$, see \ci{ku1,ku2} and below.} 
Its expansion to the one-loop order is 
\be\la{bosfact1}
S_B(\q,k)= 1 + \fr{2i\pi}{k}\cosech\q 
             - \fr{2\pi^2}{k^2}\cosech^2\q
             + \ord{k^{-3}}\,.
\ee
It is useful also to represent the factorised S-matrix \eqref{factsss} as acting
on a single field $\Phi_a$ as 
\be\la{smat22efact}
\Sc\ket{\Phi_a(p_1) \Phi_b(p_2)}
= S_{ab}^{cd}(\q,k)\ket{\Phi_c(p_1)\Phi_d(p_2)}\,,
\ee
where $\Phi_0$ is a  bosonic and $\Phi_1$ is a fermionic state (cf.
\eqref{22fg}). The one-loop S-matrix of section \ref{secsmat22} has the
factorised structure \eqref{fact22}. Taking into account the bosonic factor
\eqref{bosfact1}, the one-loop amplitudes of the factorised S-matrix are then
given by 
\be 
S_{ab}^{cd}(\q,k) =\ \pot(\q,k) \  \hat{S}_{ab}^{cd}(\q,k) \ , \la{ses}
\ee
where 
\be\la{smat22fact}\begin{split}
&\hat{S}_{00}^{00}(\q,k)=1+\fr{i\pi}{k}\cosech \q +\ord{k^{-3}}\,,
\hs{40pt}
\hat{S}_{11}^{11}(\q,k)=1-\fr{i\pi}{k}\cosech \q +\ord{k^{-3}}\,,
\\
&\hat{S}_{00}^{11}(\q,k)=-\fr{i\pi}{2k}\sech\fr{\q}{2} +\ord{k^{-3}}\,,
\hs{55pt}
\hat{S}_{11}^{00}(\q,k)=-\fr{i\pi}{2k}\sech\fr{\q}{2} +\ord{k^{-3}}\,,
\\
&\hat{S}_{01}^{01}(\q,k)=1 +\ord{k^{-3}}\,,
\hs{100pt}
\hat{S}_{01}^{10}(\q,k)=\fr{i\pi}{2k}\cosech\fr{\q}{2} +\ord{k^{-3}}\,.
\end{split}\ee
The overall factor that was extracted (see \eqref{factsss})
\be\la{pp22half}
\pot(\q,k)=1-\fr{i\pi}{k}\cosech\q+\fr{\pi\cosech \q}{4k^2} 
              \big(i(2 + (i\pi-2\q)\coth\q) -
                   3\pi\cosech\q\big)\,
\ee
satisfies the following equation
\be\la{eq1}
S_B(\q,k)\ \big[ \pot(\q,k)\big]^2\  = \po(\q,k\,;1) + \ord{k^{-3}}\,,
\ee
i.e. it matches the phase factor in section \ref{secsmat22}. 

Note that the choice of the phase factor \eqref{pp22half} implies that
$\hat{S}_{01}^{01}(\q,k)=1$ at the one-loop order. This structure continues to
hold to all orders, i.e. translating the factorised form \eqref{factsss} back to
the  original notation of section \ref{secsmat22} we conclude that the $YZ \ra
YZ$ and $\ze\chi \ra \ze\chi$ amplitudes are precisely equal to 
\be\la{eq2}
S_B(\q,k)\  \big[ S_{01}^{01}(\q,k)\big] ^2\,.
\ee
This is thus a natural choice for the phase factor and it will be useful also
in case of the reduced \adss{5} theory discussed below. 

It can be checked that the one-loop S-matrix \eqref{smat22efact} and
\eqref{smat22fact} satisfies the Yang-Baxter equation to one-loop order. In
terms of the tensor $S_{ab}^{cd}(\q,k)$ the YBE can be written as 
\be\begin{split}\la{rr22}
&\sum_{g,h,j=0}^1 \Big[  (-1)^{[h][g]+[d][j]+[e][f]}
S_{ab}^{hg}(\q_{12},k)S_{hc}^{dj}(\q_{13},k)
S_{gj}^{ef}(\q_{23},k)
\\&\hs{50pt}-(-1)^{[h][f]+[g][j]+[d][e]}
S_{bc}^{gj}(\q_{23},k)S_{aj}^{hf}(\q_{13},k)
S_{hg}^{de}(\q_{12},k)\Big] = 0\,.
\end{split}\ee
As an immediate consequence, the graded tensor product of two copies of the
S-matrix \eqref{factsss} will also satisfy the YBE. Therefore, the one-loop
S-matrix computed in section \ref{secsmat22} satisfies the YBE as expected.

\subsection{$\mc{N}=1$ supersymmetric sine-Gordon S-matrix\la{neq1ssgsmat}}

In the next two subsections we shall review the construction of the exact 
$\mc{N}=2$ supersymmetric sine-Gordon S-matrix \ci{ku2}. The first step is to
find  the $\mc{N}=1$ supersymmetric sine-Gordon S-matrix \ci{ahn,sw}. The 
$\mc{N}=1$ supersymmetric sine-Gordon Lagrangian describes one bosonic and one
fermionic degree of freedom. We will be interested in the S-matrix of  these
excitations\,\foot{Analogously to the bosonic sine-Gordon theory, here the
perturbative excitations of the fields in the Lagrangian  correspond not to 
the elementary excitations, which here are solitons, but to (a limit of) the 
bound states of the solitons. This is also the case in the  $\mc{N}=2$
supersymmetric sine-Gordon theory. A brief summary of other sectors of the
S-matrix is given in  appendix \ref{ssgbits}. For further details see
\ci{ahn,ku1}.} that can be denoted as 
\be\la{sttt}
\ket{\phi} \hs{5pt} \trm{and} \hs{5pt} \ket{\psi}\,.
\ee
The $\mc{N}=1$ supersymmetry  transforms $\phi \lra \psi$.

The S-matrix for this theory was first constructed in \ci{sw} where it was
diagonalised using the following change of basis of two-particle states
\bea
&&\ket{S}=\fr{1}{\sqrt{\cosh\fr{\q}{2}}}
\Big(\sinh\fr{\q}{4}\ket{\phi(p_1)\phi(p_2)} +
     \cosh\fr{\q}{4}\ket{\psi(p_1)\psi(p_2)} \Big)\,, \no \\ 
&& \ket{T}=\fr{1}{\sqrt{\cosh\fr{\q}{2}}}
\Big(\cosh\fr{\q}{4}\ket{\phi(p_1)\phi(p_2)}
     -\sinh\fr{\q}{4}\ket{\psi(p_1)\psi(p_2)} \Big)\,, \la{newbasis}\\
&&\ket{U}, \ket{V} =\fr{1}{\sqrt{2}}
\Big(\ket{\phi(p_1)\psi(p_2)}\mp \ket{\psi(p_1)\phi(p_2)}\Big)\,.\no
 \eea
The diagonalisation is a consequence of the S-matrix commuting with
supersymmetry. This further constrains the S-matrix by demanding that there are
only two independent amplitudes
\be\la{smatdiag}\begin{split}
  \Sc\ket{S}=\ S_{sg}(\q,\De)\ F_+(\q,\De)\ket{S}\,,&\hs{30pt}
  \Sc\ket{T}=\ S_{sg}(\q,\De)\ F_-(\q,\De)\ket{T}\,,
\\\Sc\ket{U}=\ S_{sg}(\q,\De)\ F_+(\q,\De)\ket{U}\,,&\hs{30pt}
  \Sc\ket{V}=\ S_{sg}(\q,\De)\ F_-(\q,\De)\ket{V}\,,
\end{split}\ee
where $\De$ is a function of the coupling $k$. The exact form of this function
depends on the particular theory. For example, in both the bosonic sine-Gordon
and the $\mc{N}=1$ supersymmetric sine-Gordon cases $k$ receives a finite shift.
In $\mathcal{N}=2$ supersymmetric sine-Gordon case there is no such shift. In
contrast to \ci{sw} we have extracted an overall factor $S_{sg}(\q,\De)$ which
is the S-matrix for the sine-Gordon perturbative excitation
\be\la{bosfact}
S_{sg}(\q,\De)=\fr{\sinh\q+i\sin\De}{\sinh\q-i\sin\De}\,.
\ee

The $\mc{N}=1$ supersymmetric sine-Gordon is an integrable theory and the
Yang-Baxter equation \eqref{rr22}  should be satisfied. This further constrains
the S-matrix by requiring that $F_\pm$ are related as \ci{sw}
\be\la{Fpm}
F_\pm(\q,\De)= 
\Big(1\mp \fr{i\sin\fr{\De}{2}}{\sinh\fr{\q}{2}}\Big)R(\q,\De)\,.
\ee
The common factor $R(\q,\De)$ can be further constrained by using the unitarity
and crossing relations \ci{sw}\,\foot{The unitarity and crossing constraints do
not have a unique solution. To choose the correct solution one should use
additional arguments related to the pole structure of the S-matrix \ci{sw}.
Alternatively, one can fix the ambiguity by matching the  S-matrix with the 
result of a perturbative field theory computation of scattering amplitudes.}
\be\begin{split}\la{r}
R(\q,\De)=& \fr{\sinh\q-i\sin\De}{\sinh\q+i\sin\De} \ Y(\q,\De)\ 
Y(i\pi-\q,\De)\,,
\\Y(\q,\De)=\prod_{l=1}^{\infty}&
\fr{\Gamma\big(\fr{\De}{2\pi}-\fr{i\q}{2\pi}+l\big)
    \Gamma\big(-\fr{\De}{2\pi}-\fr{i\q}{2\pi}+l-1\big)
    \Gamma\big(-\fr{i\q}{2\pi}+l-\fr{1}{2}\big)
    \Gamma\big(-\fr{i\q}{2\pi}+l+\fr{1}{2}\big)}
 {\Gamma\big(\fr{\De}{2\pi}-\fr{i\q}{2\pi}+l+\fr{1}{2}\big)
  \Gamma\big(-\fr{\De}{2\pi}-\fr{i\q}{2\pi}+l-\fr{1}{2}\big)
    \Gamma\big(-\fr{i\q}{2\pi}+l-1\big)
    \Gamma\big(-\fr{i\q}{2\pi}+l\big)}\,.
\end{split}\ee

Let us translate this exact S-matrix into the original basis of two-particle
states \eqref{sttt}, again extracting an overall bosonic factor
\be\la{smat22factalt}
\Sc\ket{\Phi_a(p_1)\Phi_b(p_2)}=\ S_{sg}(\q,\De)\
S_{\mc{N}_1}{}_{ab}^{cd}(\q,\De)
\ket{\Phi_c(p_1)\Phi_d(p_2)}\,.
\ee
Here $\Phi_0 = \phi$ and $\Phi_1 = \psi$, i.e. $0$ is a bosonic and $1$ is a
fermionic index and the components of $S_{\mc{N}_1}(\q,\De)$ are
\begin{align}
  \no S_{\mc{N}_1}{}_{00}^{00}(\q,\De)=
&\  R(\q,\De)\ (1+2i\sin\fr{\De}{2}\cosech\q)\ , 
&&& S_{\mc{N}_1}{}_{00}^{11}(\q,\De)=&
-i R(\q,\De)\  \sin\fr{\De}{2}\sech\fr{\q}{2}\ , 
\\S_{\mc{N}_1}{}_{11}^{00}(\q,\De)=&
  \  R(\q,\De)\ (1-2i\sin\fr{\De}{2}\cosech\q)\ , 
&&& S_{\mc{N}_1}{}_{11}^{11}(\q,\De)=&
   -i R(\q,\De)\  \sin\fr{\De}{2}\sech\fr{\q}{2}\ , 
\la{neq1susysmat}
\\\no S_{\mc{N}_1}{}_{01}^{01}(\q,\De)=&\ R(\q,\De)\ , 
&&& S_{\mc{N}_1}{}_{01}^{10}(\q,\De)=&
        i  R(\q,\De)\ \sin\fr{\De}{2}\cosech\fr{\q}{2}\ .
\end{align}
Motivated by \ci{ahn,ku2} we may think of $S_{\mc{N}_1}(\q,\De)$ as a minimal
$\mc{N}=1$ supersymmetric integrable S-matrix (denoted as $S_{RSG}^{(1,1)}$ in
\ci{ahn}). The S-matrix for the perturbative excitations of the $\mathcal{N}=1$
supersymmetric sine-Gordon model can then be thought of as the tensor product
of the S-matrix for the perturbative excitation of the sine-Gordon theory
with this supersymmetric S-matrix. This structure extends also to other sectors
of the theory \ci{ahn,ku2}, e.g., to  the soliton-soliton S-matrix. This is
discussed briefly in appendix \ref{ssgbits}.

It will be useful to write down  the expansion of $R(\q,\De)$ in \eqref{r} to
the one-loop order\,\foot{This can be derived  using  various polygamma
identities. Using this expression it is possible to check that the direct
result of the one-loop computation of the S-matrix for the perturbative
excitations of $\mc{N}=1$ supersymmetric sine-Gordon model  agrees with the
exact expression \ci{sw} given in \eqref{smatdiag}, \eqref{Fpm} and \eqref{r}.
The relation between $k$ and $\De$ for this theory is given by
\ci{ahn}
\be \no
k(\De)=\fr{\pi}{\De}+\fr{1}{2}\,, \hs{30pt} 
\De(k)=\fr{\pi}{k-\fr{1}{2}}\,.
\ee
This relation is the $\mathcal{N}=1$ supersymmetric sine-Gordon version of the
well-known bosonic sine-Gordon coupling constant renormalisation. The relation
to the couplings $\bet$ and $\g$ used in \ci{ahn} is as follows
$k=\fr{2\pi}{\bet^2}\,,\hs{2pt} \De=\fr{\g}{8}\,.$}
\be \la{rex}
R(\q,\De)=1-i\De \cosech \q 
           +\fr{\De^2\cosech\q}{4\pi} \Big(i \big[ 2+(i\pi-2\q)\coth\q\big] 
           -3\pi\cosech\q\Big) + \ord{\De^3} \,.
\ee

\subsection{$\mc{N}=2$ supersymmetric sine-Gordon S-matrix\la{neq2ssgsmat}}

If $\De(k)$ is given by\,\foot{Taking the relation to their couplings $\bet$ and
$\g$, $k=\fr{1}{\bet^2}\,,\hs{3pt}\De=\pi\g\,,$ this identification
\eqref{kde22} agrees with the relation between $\g$ and $\bet$ given in
\ci{ku1}, which was derived from central charge and quantum group arguments.}
\be\la{kde22}
k(\De)=\fr{\pi}{\De}\,,\hs{30pt}\De(k)=\fr{\pi}{k}\,,
\ee
i.e. $k$ is not shifted, then the expansion of \eqref{neq1susysmat} matches the
factorised result of the one-loop \adss{2} computation \eqref{smat22fact}.
Similarly, the expansions of $R(\q,\De)$ \eqref{rex} and the bosonic factor
\eqref{bosfact} match \eqref{pp22half} and \eqref{bosfact1} respectively. An
immediate consequence is  that the one-loop perturbative result of section
\ref{secsmat22} takes the form\,\foot{The symbol $\otimes_{_G}$ denotes the
graded tensor product, defined  with respect to  indices in \eqref{factsss}.}
\be\begin{split}\la{sss}
S_{sg}(\q,\fr{\pi}{k})\otimes
S_{\mc{N}_1}(\q&,\fr{\pi}{k})\otimes_{_G}
S_{\mc{N}_1}(\q,\fr{\pi}{k})\,.
\end{split}\ee
This agrees with the exact result for  the $\mc{N}=2$ supersymmetric sine-Gordon
S-matrix \ci{ku2}. As in  the $\mathcal{N}=1$ sine-Gordon  case, this can be
thought of as a supersymmetrisation of the bosonic sine-Gordon S-matrix. The
form of the S-matrix \eqref{sss} can be extended to other sectors, e.g., the
soliton-soliton S-matrix (see appendix \ref{ssgbits}).

The exact S-matrix \eqref{sss} is written as a tensor product of the three
S-matrices each satisfying the YBE  by construction. Therefore, it also
satisfies the Yang-Baxter equation.

\subsection{Phase factor}

In this subsection we identify a phase factor that will be useful in the
discussion of the S-matrix of the reduced \adss{5} theory. Motivated by the
factorised form of the S-matrix, \eqref{factsss}, \eqref{sss}, we consider 
\be\la{phase}
\Ph(\q,\De)=S_{sg}(\q,\De)\ \big[R(\q,\De)\big]^2\,.
\ee
$S_{sg}(\q,\De)$ and $R(\q,\De)$ are given in \eqref{bosfact} and \eqref{r}
respectively. As explained below \eqref{pp22half} this phase factor equals the
amplitudes for the $YZ \ra YZ$ and $\z\ch\ra\z\ch$ processes. This can be 
easily  seen from \eqref{eq2} and \eqref{smat22factalt}, \eqref{neq1susysmat} 
with 
\be
S_B(\q,k)=S_{sg}(\q,\fr{\pi}{k})\ ,  \hs{10pt}\trm{\ \ \ \ } \hs{10pt}
S_{01}^{01}(\q,k)=R(\q,\fr{\pi}{k})\,.
\ee
                      
For later  use let us record the  expansions of this phase factor and its square
root to the  one-loop ($\De^2$) order 
\be\la{phaseex}
\Ph(\q,\De)=1+\fr{\De^2}{2\pi} \cosech \q\big(i(2+(i\pi-2\q)
                               \coth \q)-\pi\cosech\q\big)+\ord{\De^3}\,.
\ee
\be\la{srphaseex}
\sqrt{\Ph(\q,\De)}=1+\fr{\De^2}{4\pi}\cosech \q\big(i(2+(i\pi-2\q)
                                     \coth \q)-\pi\cosech\q\big)+\ord{\De^3}\,.
\ee
As expected, \eqref{phaseex} matches the factor that was extracted from the
one-loop S-matrix in \eqref{pp22}. 

Finally, we present a few identities that are useful for checking the unitarity 
and crossing relations
\be\begin{split}\la{phaserel}
&R(\q,\De)R(-\q,\De)=\fr{\sinh^2\fr{\q}{2}}{\sinh^2\fr{\q}{2}+\sin^2\fr{\De}{2}}
\,, \hs{37pt}
\Ph(\q,\De)\Ph(-\q,\De)=\Big(\fr{\sinh^2\fr{\q}{2}}{\sinh^2\fr{\q}{2}
                                              +\sin^2\fr{\De}{2}}\Big)^2\,,
\\&R(\q,\De)=R(i\pi-\q,\De)\,,\hs{95pt}\Ph(\q,\De)=\Ph(i\pi-\q,\De)\,.
\end{split}\ee


\renewcommand{\theequation}{4.\arabic{equation}}
\setcounter{equation}{0}
\section{S-matrix of the reduced \adss{3} theory\la{pr33}}

In this section we shall investigate the S-matrix of the reduced \adss{3} theory
using the result of the one-loop computation of section \ref{secollsg33} as an
input. By a similar argument to that given in \ci{rt} for the reduced  \adss{5}
theory the S-matrix of this theory should also be UV finite.\,\foot{Indeed, 
using the expansion (including sextic terms) of the Lagrangian obtained by
integrating out $A_\pm$ in the axially gauged theory \ci{gt2}, we have checked 
the UV finiteness of  the one-loop $YY \ra YY$ scattering amplitude.}

The one-loop S-matrix found in  section \ref{secollsg33} does not satisfy the
Yang-Baxter equation. Similarly to the  complex sine-Gordon case we will show
that the integrability can be restored at the one-loop order  by the addition of
a local  counterterm \ci{dvm1}. As in \ci{dvm1,dh} we take integrability at the 
quantum level as our guiding principle and assume that such a counterterm  
should naturally appear in this theory. 

In \ci{ht2} the existence of the required counterterm in the complex sine-Gordon
case was understood as a consequence of starting with the gauged WZW formulation
\rf{gwzw}, gauge-fixing and integrating out unphysical fields. At present we do
not know how to trace the origin of the counterterm (given below) which is 
required in the reduced \adss{3} theory to a quantum contribution coming from 
path integral  based on the action  \eqref{gwzw}. Still, in appendix \ref{olcfd}
this  counterterm is shown to originate from a particular functional determinant
of an operator acting on algebra-valued fields of \eqref{gwzw}.

As we shall see below, the one-loop S-matrix  corrected to include the local
counterterm contribution group-factorises and also exhibits a novel
quantum-deformed supersymmetry. We propose that an exact S-matrix should be
fully constrained by demanding quantum-deformed supersymmetry, the Yang-Baxter
equation and  group factorisation along with the usual physical requirements of
unitarity and crossing.

\subsection{Bosonic symmetries\la{secsym33}}

The \adss{3} superstring sigma model is based on the supercoset 
\be \no 
\fr{\hat F}{G} = \fr{ PSU(1,1|2) \times  PSU(1,1|2)}{ SU(1,1) \x SU(2)}
\ee
and thus the corresponding  reduced  theory \ci{gt2} has  $G=SU(1,1) \x SU(2)$
and the gauge group $H=[SO(2)]^2$. One can also reformulate the reduced \adss{3}
theory such that it has $G =U(1,1) \x U(2)$ and the gauge group $H=[SO(2)]^4$,
see appendix \ref{ads3s3pohl}.\,\foot{The dimensions of both $G$ and $H$ have
been increased by two, thus there are no extra physical degrees of freedom in
the theory. The extra gauge degrees of freedom decouple from the rest of the
theory  and  therefore  can be ignored in the construction of the Lagrangian
\ci{gt2} and in the S-matrix computation.} The symmetry action of one of the
extra $SO(2)$s on the physical states is trivial\,\foot{The symmetry acts
non-trivially on the gauge field, allowing one to eliminate the corresponding
degree of freedom.}  but  the action of the other one  is non-trivial and thus
the non-trivial subgroup of $H$ is $[SO(2)]^3$. 

It is a feature of theories with an abelian gauge group $H$ that the Lagrangian
\eqref{gwzw} possesses both an  $H$-gauge symmetry and an additional global $H$ 
symmetry \cite{mist,fpgm,mirat}. The fields on which the global part of the
gauge symmetry has a linear action are field redefinitions of the fields on
which the global $H$ symmetry has a linear action. Therefore, the physical
symmetry of on-shell states is a single copy of $H$.

The Lagrangian \eqref{lag13} is written in the form that has manifest global
$SO(2)$ symmetry. To uncover  the full bosonic symmetry group we observe that
when $m,n,p,q$ are $SO(2)$ vector indices we have
\be
\g_{mnpq}=\e_{mn}\e_{pq}\,,
\ee
where $\e_{mn}$ is the usual antisymmetric $SO(2)$ tensor. We can then 
immediately see that all but the last line of \eqref{lag13} is invariant under
four separate $SO(2)$s, each of which only acts on one species of field,
($Y,Z,\z,\ch$). The last line is invariant when these four $SO(2)$s are
identified. There are also two additional $SO(2)$s defined as follows ($\Lambda
\in SO(2)$)
\be\begin{split}
&Y_m \ra \Lambda_{mn}Y_n\,, \hs{30pt}
Z_m  \ra Z_n\Lambda_{nm}\,,\hs{30pt}
\z_m \ra \z_m\,,\hs{30pt}\ch_m \ra \ch_n\,,
\\
&Y_m \ra Y_n\,, \hs{30pt}
Z_m  \ra Z_n\,,\hs{30pt}
\z_m \ra \z_m\Lambda_{mn}\,,
\hs{30pt}\ch_m \ra \ch_n\Lambda_{nm}\,,
\end{split}\ee
One can check that these are symmetries of \eqref{lag13} using the following
identity
\be
\de_{mn}(\Lambda^2)_{qp}-  \Lambda_{mp}\Lambda_{qn}-\Lambda_{qm}\Lambda_{np}
=
\de_{mn}\de_{pq}-\de_{mp}\de_{nq}-\de_{mq}\de_{np}\,.
\ee
The three $SO(2)$ symmetries and their action on the fields are thus given
by\,\foot{The notation is as follows: if the fields $Y_m,\,Z_m$ transform in the
$\2,-\2$ representations, then they transform as ($\Lambda \in SO(2)$) $Y \ra
\Lambda Y$, $Z \ra Z \Lambda$.}
\be \la{tabsym}\ba{cccc}
       & SO(2)_C & SO(2)_B & SO(2)_F
\\ Y   &   \2    &   \2    &   0
\\ Z   &   \2    &   -\2   &   0
\\\ze  &   \2    &   0    &   \2
\\\chi &   \2    &   0    &   -\2
\ea\ee
Let us digress and demonstrate that, as was claimed in section \ref{intro}, the
truncated Lagrangian \eqref{lag13} agrees (up to field redefinitions) with the
the Lagrangian obtained by fixing a gauge on $g$ and integrating out $A_\pm$ at
the classical level \ci{gt2} (as the gauge group $H$ is abelian here we choose
the axial gauging in \rf{gwzw}; the resulting Lagrangian can then be expanded
about the trivial vacuum) 
\bea
L _{3} &=&\no \fr{k}{4\pi} \Big[
  \dpl\phi\dm\phi + \tanh^2\phi\ \dpl v \dm  v 
+ \dpl\vp\dm\vp   + \tan^2\vp\ \dpl u\dm u
+ \fr{\mu^2}{2} \left(\cos 2\vp-\cosh 2\phi\right)
\\ && \no 
+ i\,\alpha\dm\alpha + i\,\beta\dm\beta
+ i\,\gamma\dm\gamma + i\,\delta\dm\delta
+ i\,\lambda\dpl\lambda + i\,\nu\dpl\nu
+ i\,\rho\dpl\rho + i\,\sigma\dpl\sigma 
\\ && \no
-i\tanh^2\phi\big[\dpl v (\lambda\nu-\rho\sigma)
                  -\dm v(\alpha\beta-\gamma\delta)\big]
+i\tan^2\vp\big[\dpl u (\lambda\nu-\rho\sigma)
                 -\dm u(\alpha\beta-\gamma\delta)\big] 
\\ && \la{lag33}
+ ({\sec^2\vp}-{\sech^2\phi}) (\alpha\beta-\gamma\delta)(\lambda\nu-\rho\sigma)
\\ && \no
-2i\mu\Big(\cosh\phi\cos\vp(\lambda\gamma+\nu\delta+\rho\alpha+\sigma\beta)
\\ && \no
\hs{30pt}+\sinh\phi\sin\vp\big[
\cos(v+u)(\rho\delta-\sigma\gamma +\lambda\beta-\nu\alpha)
-\sin(v+u)(\lambda\alpha+\nu\beta-\rho\gamma-\sigma\delta)\big]\Big)\Big]\,.
\eea
Here $\phi,\,\vp,\,v,\, u$ are real commuting fields and
$\alpha,\,\beta,\,\gamma,\,\delta,\,\lambda,\,\nu,\,\rho,\,\sigma$ are real
anticommuting fields. The $[SO(2)]^3$ symmetry of \eqref{lag13} (summarised in
\eqref{tabsym}) is the global part of the gauge group. When a gauge is fixed on
$g$ and $A_\pm$ are integrated out this symmetry is completely broken. In
theories with abelian gauge groups (e.g. the complex sine-Gordon model) there
is also a  global $H$ symmetry in addition to the  $H$ gauge symmetry. Here it
acts as follows
\be\begin{split}
u\ra u + c_1+c_2\,, &\hs{30pt} v\ra v + c_1-c_2\,,
\\\al+i\bet \ra e^{i(c_1+c_3)} (\al+i\bet)\,,&\hs{30pt}
  \rho + i\s \ra e^{i(c_1+c_3)} (\rho+i\s)\,,
\\\de+i\g \ra e^{i(c_1-c_3)} (\de+i\g)\,,&\hs{30pt}
  \nu + i\lambda \ra e^{i(c_1-c_3)} (\nu+i\lambda)\,,
\end{split}
\ee
where $c_1,\,c_2$ and $c_3$ are the three symmetry parameters. Expanding
\eqref{lag33} to quartic order in ``radial'' directions $\phi,\,\vp$ and using
the following field  redefinition 
\bea 
&& Y_1=  \phi\  \cos v \ , \ \ \ \ Y_2=  \phi\  \sin v \ , \ \ \ \ 
Z_1= \vp\  \cos u \ , \ \ \ \ Z_2= \vp\  \sin u \ ,  \no\\
&&( \zer_1,\ \zer_2,\ \zel_1,\ \zel_2, \ \chr_1,\ \chr_2,\ \chl_1,\ 
\chl_2) = (\alpha,\beta,\rho,\sigma,\delta, \gamma, \nu, \lambda)
\,,\la{redf}
\eea
we find the  agreement with \eqref{lag13} up to simple field and coupling
constant redefinitions as claimed.

\subsection{Quantum counterterms and the  Yang-Baxter equation\la{ybe33}}

The one-loop S-matrix of section \ref{secollsg33} does not satisfy the
Yang-Baxter equation \eqref{Y}. The YBE is related to  conservation of hidden
symmetry charges. As with  any global symmetry that is not manifestly  preserved
by a quantization procedure  one may try to maintain it  at the quantum level
by adding local  counterterms. 

This is what happens in a similar bosonic model -- the complex sine-Gordon
theory (whose quartic expansion is a truncation of \eqref{lag13}) --  where
there exists a local quantum counterterm that restores the satisfaction of the
YBE at the one-loop level \ci{dvm1}. Assuming YBE, the exact soliton S-matrix
for the complex sine-Gordon model was constructed in \ci{dh}. The correct
interpretation of the theory was conjectured to be based on the gauged WZW
theory (a special case of \rf{gwzw}) with the required quantum counterterm 
possibly appearing as a consequence of starting with this gauged WZW action in
the path integral. Indeed, in \ci{ht2} it was shown that the counterterm
required to reproduce the YBE-consistent S-matrix of \ci{dh} can indeed be
derived in this way. 

Here we find local counterterms that similarly restore the satisfaction of the
YBE at one-loop for the S-matrix of the reduced \adss{3} theory. In appendix
\ref{olcfd}  we suggest a functional determinant origin for these counterterms,
yet it appears that they cannot be naively interpreted as arising from a  
gauge-fixing procedure or integrating out unphysical fields in path integral
for \eqref{gwzw}. There may still be an alternative Lagrangian formulation of
this reduced theory that leads to the required counterterms. As explained in
appendix \ref{olcfd} such Lagrangian would involve unphysical fields in
fermionic subspaces of the superalgebra $\hat{\mf{f}}$.

To restore the YBE the underlined terms in the coefficients $\hat f_i$ of the 
one-loop S-matrix of section \ref{secollsg33}  need to be cancelled. These terms
are of the form that could arise from a set of local quartic counterterms. As
well as cancelling the underlined terms one can add the following arbitrary
correction to the coefficient functions $f_1,\,f_2,\,f_3,\, f_{\td{3}},\,f_4$
\be\la{addcont}\fr{i\pi}{k^2}
\Big[\de_{mp}\de_{nq}\ c_1(\q,k)+\e_{mp}\e_{nq}\ c_2(\q,k)\Big]
\ee
without affecting the YBE. 

By analogy with the reduced \adss{2} theory and the complex sine-Gordon model
\ci{ht2} we may propose some assumptions that possible counterterms should
satisfy: 
(i) the counterterms should be second order in derivatives and local; \ \ 
(ii) the function $c_1(\q,k)$ in \eqref{addcont} that may be interpreted as an
additional contribution to the phase factor should vanish. As we already have a
phase factor (the amplitude of  $Y_m Z_n \ra Y_m Z_n$ and $\z_m \chi_n \ra \z_m
\chi_n$ processes) that fits into a pattern with the reduced \adss{2} theory
it seems sensible to assume that it is not altered;\ \
(iii) the counterterms should factorise into two parts -- one transforming 
under 2-d Lorentz symmetry like $\dpl$ and the other like $\dm$, with each part
separately invariant under the $[SO(2)]^3$ global symmetry \eqref{tabsym}.

The last requirement is motivated by the origin of the complex sine-Gordon
counterterms found in \ci{ht2}. It implies that there should be no counterterms
involving four different ``species'' ($Y,\,Z,\,\z\;\trm{and}\;\chi$).
Consequently, there should be no counterterm-induced shift of the S-matrix 
proportional to the tree-level S-matrix (such a shift could be reinterpreted as
a  shift of  the coupling $k$). Also, $f_8$  coefficient should then remain
zero at one-loop. It turns out that the group factorisation of the S-matrix
discussed in section \ref{grpfact33} implies that $f_8$ should be identically
zero to all orders.

Let us consider the following counterterm which satisfies all of the above
requirements\,\foot{Note that this counterterm may be written as 
\be\no
\fr{\pi}{k^2}\mc{J}_+\mc{J}_-
\ee
where $\mc{J}_\pm$ is the conserved current associated to the manifest global
$SO(2)$ symmetry of the Lagrangian \eqref{lag13} (to quadratic order and up to 
a rescaling of fields).}
\be\la{counterterm}
\De \Lag_3 = \fr{\pi}{k^2}\e_{mn}\e_{pq}
(Y_m\dpl Y_n + Z_m\dpl Z_n - i\zer_m\zer_n - i\chr_m\chr_n)
(Y_p\dm Y_q + Z_p\dm Z_q - i\zel_p\zel_q - i\chl_p\chr_q)\,.
\ee

It produces the following corrections to the functions parametrising the
S-matrix:
{\allowdisplaybreaks
\begin{align}
\De f_1{}_{mnpq}(\q,k)=&\fr{i\pi}{k^2}\Big(\e_{mn}\e_{pq}
(\cosech\q-\coth\q)\no
+\e_{mq}\e_{pn}(\cosech\q+\coth\q)-2\e_{mp}\e_{nq}\coth\q\Big)\,,
\\\De f_2{}_{mnpq}(\q,k)=& - \fr{2i\pi}{k^2}\Big((\e_{mn}\no
\e_{pq }+\e_{mq} \e_{pn})\cosech\q +\e_{mp}\e_{nq}\coth\q\Big)\,,
\\\De f_3{}_{mnpq}(\q,k)=&\De \tilde{f}_3{}_{mnpq}(\q,k)
=- \fr{2i\pi}{k^2}\e_{mp}\e_{nq}\coth\q\,,\no
\\\De f_4{}_{mnpq}(\q,k)=&- \fr{2i\pi}{k^2}\e_{mp}\e_{nq}\coth\q\,,
\\\De f_5{}_{mnpq}(\q,k)=&
 \fr{i\pi}{k^2}\e_{mn}\e_{pq}(\cosech\q-\coth\q)\,,\hs{30pt}
\De{f}_{ \td 5} {}_{mnpq}(\q,k)=
     \fr{2i\pi}{k^2}\e_{mn}\e_{pq}\cosech\q \,,\no
\\\De f_6{}_{mnpq}(\q,k)=&-\fr{i\pi}{k^2}\e_{mn}\e_{pq}\sech\fr{\q}{2}
\,,\hs{40pt}
\De f_7{}_{mnpq}(\q,k)= 0\,,\hs{45pt}
\De f_8{}_{mnpq}(\q,k)= 0\,.\no
\end{align}}
The corrected functions $\hat f_i$ parametrising the one-loop S-matrix of
section \ref{secollsg33} are
{\allowdisplaybreaks
\begin{align}
\no\hat{f}_1{}_{mnpq}(\q,k)=&\de_{mp}\de_{nq}
\Big(1-\fr{2i\pi}{k} \cosech\q
            -\fr{2\pi^2}{k^2}\coth^2\q\Big)
\\\no&+\e_{mp}\e_{nq}\Big(\fr{2i\pi}{k}\coth\q
				- \fr{2i\pi}{k^2}\coth\q
               -\fr{i\pi}{k^2}(i\pi-2\q)\cosech^2\q
      +\fr{3\pi^2}{k^2}\coth\q\cosech\q\Big)
\\\no\hat{f}_2{}_{mnpq}(\q,k)=&\de_{mp}\de_{nq}
\Big(1      +\fr{2\pi^2}{k^2}\cosech^2\q\Big)
\\\no&+\e_{mp}\e_{nq}\Big(- \fr{2i\pi}{k^2}\coth\q
                     -\fr{i\pi}{k^2}(i\pi-2\q)\cosech^2\q
     -\fr{\pi^2}{k^2}\coth\q\cosech\q
          \Big)
\\\no\hat{f}_3{}_{mnpq}(\q,k)=&\hat{f}_{\td{3}}{}_{mnpq}(\q,k)=\de_{mp}
\de_ { nq }
+\e_{mp}\e_{nq}\Big(- \fr{2i\pi}{k^2}\coth\q
             -\fr{i\pi}{k^2}(i\pi-2\q)\cosech^2\q
      +\fr{\pi^2}{k^2}\coth\q\cosech\q\Big)
\\\hat{f}_4{}_{mnpq}(\q,k)=&\de_{mp}\de_{nq}
\Big(1-\fr{i\pi}{k}\cosech\q
      -\fr{\pi^2}{2k^2}\Big)\la{correctfunctions}
\\\no&+\e_{mp}\e_{nq}\Big(\fr{i\pi}{k}\coth\q
			- \fr{2i\pi}{k^2}\coth\q
             -\fr{i\pi}{k^2}(i\pi-2\q)\cosech^2\q
      +\fr{\pi^2}{k^2}\coth\q\cosech\q\Big)
\\\no\hat{f}_5{}_{mnpq}(\q,k)=&\hat{f}_{\td{5}}{}_{mnpq}(\q,k)=
  \fr{\pi^2}{2k^2}\sech^2\fr{\q}{2}\big( \de_{mn}\de_{pq}+
\e_{mn}\e_{pq}\big)
\\\no\hat{f}_6{}_{mnpq}(\q,k)=&
     \Big(\fr{i\pi}{2k}\sech\fr{\q}{2}-
      \fr{\pi^2}{2k^2}\sech\fr{\q}{2}\tanh\fr{\q}{2}\Big)
\big( \de_{mn}\de_{pq}+
\e_{mn}\e_{pq}\big)
\\\no\hat{f}_7{}_{mnpq}(\q,k)=& 
\fr{i\pi}{2k}\cosech\fr{\q}{2}\big(
-\de_{mn}\de_{pq}+\de_{mp}\de_{nq}+\de_{mq}\de_{np}\big)
\\\no\hat{f}_8{}_{mnpq}(\q,k)=& 0
\end{align}}
The addition of the counterterm has restored the relation between the $YYZZ$
and $\z\z\ch\ch$ amplitudes, i.e. now $f_5=f_{\td{5}}$. This may indicate that
as well as integrability, a fermionic symmetry relating the bosons and the 
fermions is restored when the required counterterm is added. Indeed, in section
\ref{supsy33} this S-matrix  will be  shown to commute with a quantum-deformed
supersymmetry. 

One may wonder if the above counterterm can be derived from the path integral 
for the corresponding gauged WZW-based  theory \rf{gwzw} as was the case for the
complex sine-Gordon  model \ci{ht2}. If we perform a similar analysis to
\ci{ht2} starting with the reduced \adss{3} theory we only get a bosonic
counterterm that produces part of the correction to $f_1$
\be
\De f_1{}_{mnpq}(\q,k)=-\fr{2i\pi}{k^2}
\Big(\de_{mn}\de_{pq}(\cosech\q+\coth\q)+
\de_{mq}\de_{np}(\cosech\q-\coth\q)
\Big)\,.
\ee
It is possible that there is an alternative way of formulating the Lagrangian
\eqref{gwzw} which treats bosons and fermions on a more equal footing. In this
case one may be  able to  obtain  the counterterm  \eqref{counterterm} as a
contribution of some functional determinant in the one-loop path integral  as in
\ci{ht2}. This is discussed further in appendix \ref{olcfd} (note that some of
the notation in appendix \ref{olcfd} is defined in section
\ref{sumofsym}).\,\foot{As discussed in section \ref{seclag22} the reduced
\adss{2} theory is equivalent to $\mc{N}=2$  supersymmetric sine-Gordon for
which the exact S-matrix has been derived \ci{ku2}. The perturbative computation
precisely matches this result, i.e.  there should be no  additional one-loop
corrections from local counterterms. For the reduced \adss{5} theory the
one-loop S-matrix group factorises. It seems likely that there should be no
one-loop counterterm  corrections there either, or at least they should respect
this group factorisation property. Any Lagrangian formulation of the reduced
\adss{3} theory that gives the required corrections at the one-loop order 
should then  produce  no corrections  when  applied to the cases  of the 
reduced \adss{2} and \adss{5} theories. The functional determinant 
contribution discussed in appendix \ref{olcfd} satisfies this property.}

\subsection{Group factorisation of the S-matrix\la{grpfact33}}

Having added the above counterterm it is possible to repackage the fields in
such a way that the resulting S-matrix factorises under some group structure.
Consider the following set of $SO(2)$ transformations which are symmetries of
the theory
\be\la{table} \ba{ccccc}
         & SO(2)_1 & SO(2)_2 & SO(2)_{\dot{1}} & SO(2)_{\dot{2}}
\\ Y_m   &   \2     &   0     &    \2       &    0
\\ Z_m   &   0     &   \2     &    0       &    \2
\\\ze_m  &   \2     &   0     &    0       &    \2
\\\chi_m &   0     &   \2     &    \2       &    0
\ea\ee
Any one of these $SO(2)$ transformations can be rewritten as a combination of
the other three,  agreeing with the symmetry analysis of section \ref{secsym33},
where global bosonic symmetry of the theory was shown to be $[SO(2)]^3$.

We relabel the fields in terms of their transformations under the four $SO(2)$s
\eqref{table}
\be\la{rearr}
Y_{a\adt}\,,\hs{20pt}
Z_{\al\agdt}\,,\hs{20pt}
\z_{a\agdt}\,,\hs{20pt}
\chi_{\al\adt}\,,
\ee
where the indices $a,\,\al,\,\adt,\,\agdt$ ($a=1,\,2$, $\al=3,\,4$) are vector
indices of $SO(2)_1,\, SO(2)_2 ,\, SO(2)_{\dot{1}} ,\, SO(2)_{\dot{2}}$
respectively, with the following fermionic grading
\be\la{fermgr33}
\bll a\brr =\bll \adt\brr =0\,,\hs{30pt} \bll \al\brr =\bll \agdt\brr =1\,.
\ee
Taking the fields to be real each has four degrees of freedom, whereas they
should only have two. To take care of this we impose the following constraints
\be\begin{split}
Y_{a\adt}=-\e_{ab}\e_{\adt\bdt}Y_{b\bdt}\,,\hs{20pt}&
Z_{\al\agdt}=-\e_{\al\bet}\e_{\agdt\bgdt}Z_{\bet\bgdt}\,,
\\\ze_{a\agdt}=-\e_{ab}\e_{\agdt\bgdt}\z_{b\bgdt}\,,\hs{20pt}&
\ch_{\al\adt}=-\e_{\al\bet}\e_{\adt\bdt}\ch_{\bet\bdt}\,.
\end{split}\ee
For example,
\be
Y_{1\dot{1}}=-Y_{2\dot{2}} \;\;(=\fr{1}{\sqrt{2}}Y_1)
\hs{10pt}\trm{and}\hs{10pt}
Y_{1\dot{2}}=Y_{2\dot{1}} \;\; (= \fr{1}{\sqrt{2}}Y_2)\,.
\ee
These constraints are necessary for the fields to respect all the symmetries
given in \eqref{table}.

The field rearrangement \eqref{rearr} allows us to consider the single field
\be
\Phi_{A \Adt}\,,\hs{30pt} A=(a,\al)\,\,\,\trm{and}\,\,\,\Adt=(\adt,\agdt)\,,
\ee
that encodes all four species of field $Y,\,Z,\,\ze$ and $\chi$ in the natural
way.

The counterterm-corrected one-loop S-matrix for the reduced \adss{3} theory
then factorises as follows
\be\la{33fact}
\Sc\ket{\Phi_{A\Adt}(p_1)\Phi_{B\Bdt}(p_2)}
=(-1)^{[\Adt][B]+[\Cdt][D]}S_{AB}^{CD}S_{\Adt\Bdt}^{\Cdt\Ddt}
\ket{\Phi_{C\Cdt}(p_1)\Phi_{D\Ddt}(p_2)}
\ee
\be\la{Sparam33}
S_{AB}^{CD} =
\left\{
\ba{l} L_1 \de_{ac}\de_{bd}
              + L_2 \e_{ac}\e_{bd}\,, 
\\            L_3 \de_{\al\g}\de_{\bet\de}
              + L_4 \e_{\al\g}\e_{\bet\de}\,, 
\\             L_5 \de_{ac}\de_{\bet\de}
              + L_6 \e_{ac}\e_{\bet\de}\,, 
\\             L_7 \de_{\al\g}\de_{bd}
              + L_8 \e_{\al\g}\e_{bd}\,, 
\\             L_{9} (\de_{ab} \de_{\g\de}
                     +\e_{ab} \e_{\g\de})\,, 
\\              L_{10} (\de_{\al\bet} \de_{cd}
                     +\e_{\al\bet} \e_{cd})\,,
\\             L_{11} (\de_{ad} \de_{\g\bet}
                      +\e_{ad} \e_{\g\bet})\,, 
\\                L_{12} (\de_{\al\de} \de_{cb}
                       +\e_{\al\de} \e_{cb})\,,
\ea \right.
\ee
where $L_i$ are functions of $\q$ and the coupling $k$.

It is useful to understand the factorised S-matrix \eqref{Sparam33} as acting
on a single field
\be\la{smat33fact}
\Sc\ket{\Phi_A(p_1) \Phi_B(p_2)}
= S_{AB}^{CD}(\q,k)\ket{\Phi_C(p_1)\Phi_D(p_2)}\,,\hs{20pt}
\Phi_A = (\phi_a,\psi_\al)\,,
\ee
where $\phi_a$ are bosonic and $\psi_\al$ are fermionic. This S-matrix should
satisfy the usual physical requirements of unitarity and crossing. Unitarity
implies
\be\la{unittens33}
(-1)^{[C] [D] + [ E] [ F] }\ S_{AB}^{CD}(\q,k)\
S_{DC}^{FE}(-\q,k)=
\de_{A}^E\de_B^F\,.
\ee
For crossing symmetry we need to introduce the crossed S-matrix denoted by
$\bar{S}_{AB}^{CD}(\q,k)$ which is identical to $S_{AB}^{CD}(\q,k)$ in 
\eqref{smat33fact} except with $(L_9,\,L_{10},\,L_{11},\,L_{12})$ replaced by
$i(L_9,\,L_{10},\,L_{11},\,L_{12})$. Crossing symmetry then implies
\be\la{crosstens33}
S_{AB}^{CD}(\q,k)=\sum_{E,F=1}^4 (-1)^{[A][B] +[ C][D]}\ C_B^E\
\bar{S}_{FA}^{EC}
(i\pi-\q)\ C^{-1}{}_F^D\,,
\ee
where $C_A^B$ is defined by 
\be\la{chco33}\begin{split}
&\mc{C}\ket{\Phi_A}=C_A^B\ket{\Phi_B}\,,
\\\mc{C}\ket{\phi_1}=-\ket{\phi_2}\,,\hs{20pt}
&\mc{C}\ket{\psi_3}=-\ket{\psi_4}\,,\hs{20pt}
\mc{C}\ket{\phi_2}=\ket{\phi_1}\,,\hs{20pt}
\mc{C}\ket{\psi_4}=\ket{\psi_3}\,,
\end{split}
\ee
and similarly for $C^{-1}{}_A^B$. Crossing symmetry requires the following
relations between the functions $L_i$,
\be\la{crossl}\begin{split}
L_1(i\pi-\q,k)=L_1(\q,k)\,,&\hs{20pt}
L_2(i\pi-\q,k)=-L_2(\q,k)\,,
\\L_5(i\pi-\q,k)= L_5(\q,k)\,,&\hs{20pt}
L_6(i\pi-\q,k)=-L_6(\q,k)\,,
\\L_9(i\pi-\q, k&)=i L_{11}(\q,k)\,,
\end{split}\ee
and similarly for $L_3,\,L_4,\,L_7,\,L_8,\,L_{10},\,L_{12}$.

For consistency (for example, between $\Sc\ket{Y_m(p_1)\z_n(p_2)}$ and
$\Sc\ket{\z_m(p_1)Y_n(p_2)}$, see appendix \ref{33expansion}) the functions
$L_i$ should also obey the conjugation relations
\be\la{lcong}\begin{split}
&L_i (\q,k)=L_i^*(-\q,k) \ , \ \ \ \ \ \ \ \ \ i= 1, 2, ..., 8\\
&L_9(\q,k)=-L_9^*(-\q,k)\,, \hs{20pt}L_{10}(\q,k)=-L_{10}^*(-\q,k)\,,
 \ \ \ \ \ \ \ \ L_{11}( \q, k)=L_{12}^*(-\q,k)\,.
\end{split}\ee
The Lagrangian \eqref{lag13} has also a $\mbb{Z}_2$ symmetry 
\be\la{z23}\begin{split}
Y \leftrightarrow Z\,,\hs{30pt}
\ze \leftrightarrow \chi\,,\hs{30pt}
k &\rightarrow -k\,,
\end{split}\ee
implying the following relations between the functions $L_i$
\begin{align}
&L_1(\q,k)=L_3(\q,-k)\,,
&\no
&L_2(\q,k)=L_4(\q,-k)\,,
\\
&L_5(\q,k)=L_7(\q,-k)\,,
&
&L_6(\q,k)=L_8(\q,-k)\,,
\\
&L_9(\q,k)=-L_{10}(\q,-k)\,,
&\no
&L_{11}(\q,k)=-L_{12}(\q,-k)\,.
\end{align}
In appendix \ref{33expansion} equation \eqref{Sparam33} is expanded and
rewritten in the original $SO(2)$ notation to enable comparison to the S-matrix
of section \ref{secollsg33} with the corrected functions
\eqref{correctfunctions}. Key features of the corrected one-loop S-matrix of
section \ref{ybe33} are the equality of the $YYZZ$ and $\z\z\ch\ch$ amplitudes
(i.e. $f_3=f_{\td{3}}$ and $f_5=f_{\td{5}}$) and the vanishing of the function
$f_8$. These along with other relationships between the parametrising functions
are required for the one-loop S-matrix to factorise as in \eqref{33fact},
\eqref{Sparam33}. The one-loop result gives the following $L_i$
\be 
L_i (\q,k) = \po(\q,k\,;1) \  \hat{L}_i(\q,k)  \ ,   \la{pfe}
\ee
where the phase factor $\po(\q,k\,;1)$ was defined in 
\eqref{qex}  and 
\begin{align}
&\hat{L}_1(\q,k)=\hat{L}_3(\q,-k)=1 -\fr{i\pi}{k}\cosech\q-\fr{\pi^2}{2k^2}
+\ord{k^{-3}}\no
\\
&\hat{L}_2(\q,k)=\hat{L}_4(\q,-k)=\fr{i\pi}{k}\coth\q-\fr{i\pi}{k^2}\coth\q
     -\fr{i\pi}{2k^2}(i\pi-2\q)(\cosech \q)^2
     +\fr{\pi^2}{2k^2}\coth\q\cosech\q
                     +\ord{k^{-3}}\no
\\
&\hat{L}_5(\q,k)=\hat{L}_7(\q,-k)=1
                     +\ord{k^{-3}}\la{l110}
\\
&\hat{L}_6(\q,k)=\hat{L}_8(\q,-k)=-\fr{i\pi}{k^2}\coth\q
     -\fr{i\pi}{2k^2}(i\pi-2\q) (\cosech \q)^2
     +\fr{\pi^2}{2k^2}\coth\q\cosech\q
                     +\ord{k^{-3}}\no
\\
&\hat{L}_9(\q,k)=-\hat{L}_{10}(\q,-k)=\fr{i\pi}{2k}\sech\fr{\q}{2}              
+\ord{k^{-3}}\,,\hs{20pt}\no
\hat{L}_{11}(\q,k)=-\hat{L}_{12}(\q,-k)=-\fr{i\pi}{2k}\cosech\fr{\q}{2}        
+\ord{k^{-3}}\no
\end{align}
The phase factor $\po(\q,k\,;1)$ is the one-loop expansion of the square root of
the factor that was identified in the one-loop S-matrix of the complete reduced
\adss{3} theory, \eqref{pp33}: the S-matrix has been factorised into two parts,
each of which we take with the same phase  factor.

The choice of the phase factor in \eqref{pfe} retains the structure
$\hat{L}_{5,7}=1+\ord{k^{-3}}$ to the one-loop order. From the expanded S-matrix
given in appendix \ref{33expansion} we see that it is not possible to choose a
phase factor such that the amplitudes for $Y_m Z_n \ra Y_m Z_n$ and $\z_m \chi_n
\ra \z_m \chi_n$ scattering processes and $\hat{L}_{5,7}$ both equal one to all
orders. This is different from the case of the reduced \adss{2} theory
\eqref{eq2} and suggests that these two theories are not part of the same
``family''.\,\foot{This is not such a surprise when looking at the supercosets
of the corresponding  superstring sigma models. For \adss{2} we have
$\hat{F}=PSU(1,1|2)$ whereas for \adss{3} we have $\hat{F}=PS([U(1,1|2)]^2)$,
i.e. a direct product. As discussed in \ci{gt2}, the Pohlmeyer reduction of
models with a direct product as the numerator group of the supercoset is
somewhat  different compared to the Pohlmeyer reduction of models with a single
group as the numerator group. The reduced \adss{5} theory for which
$\hat{F}=PSU(2,2|4)$ has a stronger relation to the reduced \adss{2} theory.} 

As well as satisfying the Yang-Baxter equation (which has the same form as in 
\rf{rr22}) to the  one-loop order, the perturbative S-matrix \eqref{Sparam33},
\eqref{smat33fact}, \eqref{l110}  also satisfies the unitarity, crossing and
conjugation relations in equations  \eqref{unittens33}, \eqref{crosstens33},
\eqref{crossl}, \eqref{lcong}. 

\

For the purpose of discussing the quantum-deformed supersymmetry in the next
subsection it is useful to rewrite the S-matrix \eqref{smat33fact} in terms of
the complex fields\,\foot{The S-matrix has a manifest $U(1) \x U(1)$ symmetry.
The field $\phi$ is then charged under the first $U(1)$ and $\psi$ under the
second. Thus it would be more natural to write $\phi_{\upm \mathbf{0}} =
\phi_1\pm i\phi_2\,,\hs{3pt} \psi_{\mathbf{0}\upm}=\psi_3 \pm i \psi_4\,.$
This notation is cluttered and we will suppress the $0$ index. The global
$U(1)$ indices are in bold to distinguish them from the Lorentz light-cone
indices.}
\be
\phi_\up = \phi_1 +  i\phi_2\,,\hs{10pt}
\phi_\um = \phi_1 -  i\phi_2\,,\hs{10pt}
\psi_\up = \psi_3  + i\psi_4\,,\hs{10pt}
\psi_\um = \psi_3   -  i\psi_4\,.
\ee
The S-matrix \eqref{smat33fact} acting on these fields is
\be\la{smat33factu1u1}\begin{split}
\Sc\ket{\phi_\up\phi_\up} = & (L_1(\q,k)-L_2(\q,k))\ket{\phi_\up\phi_\up}
\\\Sc\ket{\phi_\up\phi_\um} = & (L_1(\q,k)+L_2(\q,k))\ket{\phi_\up\phi_\um}
                                   +2L_9(\q,k)\ket{\psi_\up\psi_\um}
\\\Sc\ket{\psi_\up\psi_\up} = & (L_3(\q,k)-L_4(\q,k))\ket{\psi_\up\psi_\up}
\\\Sc\ket{\psi_\up\psi_\um} = & (L_3(\q,k)+L_4(\q,k))\ket{\psi_\up\psi_\um}
                                     +2L_{10}(\q,k)\ket{\phi_\up\phi_\um}
\\\Sc\ket{\phi_\up\psi_\up} = & (L_5(\q,k)-L_6(\q,k))\ket{\phi_\up\psi_\up}
                                        +2L_{11}(\q,k)\ket{\psi_\up\phi_\up}
\\\Sc\ket{\phi_\up\psi_\um} = & (L_5(\q,k)+L_6(\q,k))\ket{\phi_\up\psi_\um}
\\\Sc\ket{\psi_\up\phi_\up} = & (L_7(\q,k)-L_8(\q,k))\ket{\psi_\up\phi_\up}
                                        +2L_{12}(\q,k)\ket{\phi_\up\psi_\up}
\\\Sc\ket{\psi_\up\phi_\um} = & (L_7(\q,k)+L_8(\q,k))\ket{\psi_\up\phi_\um}
\end{split}\ee
This S-matrix clearly respects the $U(1) \x U(1)$ bosonic symmetry under which
$\phi_\up$ has charges $(1,0)$ and $\psi_\up$ has charges $(0,1)$.

\subsection{Quantum-deformed supersymmetry\la{supsy33}}

In this section the invariance of the factorised one-loop S-matrix
\eqref{Sparam33}, \eqref{smat33fact}, \eqref{l110} under a quantum-deformed
supersymmetry is demonstrated.

The reduced \adss{2} theory of section \ref{seclag22}  has a $\mc{N}=2$
worldsheet supersymmetry \ci{gt1} and one may expect to find a similar 2-d 
supersymmetry in larger models \ci{gt1}. This is suggested also by the
integrability of the model implying the existence of conserved fermionic charges
\ci{schm}. Very recently the existence of a (non-local)  on-shell  supersymmetry
in the theory \rf{gwzw} was pointed out in \ci{iv,hms} and the off-shell
generalisation demonstrated also in \ci{hms}.

Here we take an alternative approach: the idea is to find supersymmetry as a
symmetry of the S-matrix for on-shell states. The supersymmetry we shall find
below appears to be quantum-deformed and thus it is not immediately clear how
it should act on the off-shell fields present in the Lagrangian.

The classical supersymmetry algebra we shall consider below (denoted as $\SA$)
is the maximal sub-superalgebra of $\mf{psu}(2|2)\ltimes \mbb{R}^2$ such that
the bosonic subalgebra is $[\mf{so}(2)]^2 \oplus \mbb{R}^2$. This is motivated
by the fact that the reduced \adss{3} theory is  a truncation of the reduced
\adss{5} theory, for which there is a similarity with the quantum-deformed
$\mf{psu}(2|2)\ltimes \mbb{R}^3$ R-matrix \ci{bk,bcl}. We also take into account
that the global bosonic symmetry of \eqref{smat33factu1u1} is $[\mf{so}(2)]^2$.

The generators of the this classical supersymmetry algebra are:
two $SO(2)$ generators, denoted $\mf{R}$ and $\mf{L}$;
two positive chirality supercharges, $\mf{Q}_{\upm\ump}$;\,\foot{Note that here
the labels $\up$ and $\um$ do not denote the chirality of the supercharges, but
rather the charges under the $SO(2) \x SO(2)$ bosonic subalgebra.}
two negative chirality supercharges, $\mf{S}_{\upm\ump}$;
two central extension generators $\ppmn$ (which are related to the light-cone 
components of the 2-d momenta, cf. \rf{tak22}).\,\foot{The bar in $\ppmn$
indicates that these generators may have different $k$-dependent normalization
compared to those in \rf{tak22}.}
The corresponding commutation relations are given by
\begin{align}
&\com{\mf{R}}{\mf{R}}
=0\,,
&\no
&\com{\mf{L}}{\mf{L}}
=0\,,
\\&\com{\mf{R}}{\mf{Q}_{\upm\ump}}
=\pm i \mf{Q}_{\upm\ump}\,,
&\no
&\com{\mf{L}}{\mf{Q}_{\upm\ump}}
=\mp i \mf{Q}_{\upm\ump}\,,
\\&\com{\mf{R}}{\mf{S}_{\upm\ump}}
=\pm i \mf{S}_{\upm\ump}\,,
&\no
&\com{\mf{L}}{\mf{S}_{\upm\ump}}
=\mp i \mf{S}_{\upm\ump}\,,
\\&\acom{\mf{S}_{\upm\ump}}{\mf{Q}_{\upm\ump}}
=0\,,
&\la{commm3}
&\acom{\mf{S}_{\upm\ump}}{\mf{Q}_{\ump\upm}}
=\pm \fr{i}{2}(\mf{R}+\mf{L})=\pm\mf{A}\,,
\\&\acom{\mf{Q}_{\upm\ump}}{\mf{Q}_{\upm\ump}}
=0\,,
&\no
&\acom{\mf{Q}_{\upm\ump}}{\mf{Q}_{\ump\upm}}=-\ppn\,,
\\&\acom{\mf{S}_{\upm\ump}}{\mf{S}_{\upm\ump}}
=0\,,
&\no
&\acom{\mf{S}_{\upm\ump}}{\mf{S}_{\ump\upm}}=\pmn\,.
\end{align}
The linear combination 
\be
\mf{A}\equiv \fr{i}{2}(\mf{R}+\mf{L})
\ee
commutes with all other generators. This superalgebra may be represented as 
\be
\SA= \JA \ltimes \mf{so}(2) \ltimes \mbb{R}^2\,,
\ee
where $\mf{so}(2)$ corresponds to $\mf{A}$, \  $\mbb{R}^2$ to $\ppmn$ and the 
superalgebra $\JA$ has a bosonic subalgebra $\mf{so}(2)$, generated by
\be
\mf{B}\equiv\mf{R}-\mf{L}\,.
\ee
The commutation relations for the superalgebra $\JA$ are
\be\begin{split}\la{commm34}
\com{\mf{B}}{&\mf{B}}
=0\,,
\\\com{\mf{B}}{\mf{Q}_{\upm\ump}}
=\pm i \mf{Q}_{\upm\ump}&\,,
\hs{12pt}\com{\mf{B}}{\mf{S}_{\upm\ump}}
=\pm i \mf{S}_{\upm\ump}\,,
\\\acom{\mf{S}_{\upm\ump}}{\mf{Q}_{\upm\ump}}
=0\,,
&\hs{20pt}
\acom{\mf{S}_{\upm\ump}}{\mf{Q}_{\ump\upm}}
=0\,,
\\\acom{\mf{Q}_{\upm\ump}}{\mf{Q}_{\upm\ump}}
=0\,,
&\hs{20pt}
\acom{\mf{Q}_{\upm\ump}}{\mf{Q}_{\ump\upm}}=0\,,
\\\acom{\mf{S}_{\upm\ump}}{\mf{S}_{\upm\ump}}
=0\,,
&\hs{20pt}
\acom{\mf{S}_{\upm\ump}}{\mf{S}_{\ump\upm}}=0\,.
\end{split}\ee
This superalgebra (which apparently does not have a standard name) is a
semi-direct sum of $\mf{so}(2)$ with two copies of the
$\mf{psu}(1|1)$\,\foot{The appearance of this algebra may be expected given the
origin of the reduced theory. In the reduced \adss{2} model we start with the
supercoset $\fr{PSU(1,1|2)}{SO(1,1) \x SO(2)}$ and end up with the symmetry
algebra $[\mf{psu}(1|1)]^2\ltimes \mbb{R}^2$ \eqref{symgroup22} in the reduced
theory. The reduced \adss{3} theory arises from the Pohlmeyer reduction of the 
2-d sigma model with the target space $\fr{PS([U(1,1|2)]^2)}{U(1,1) \x U(2)}$
(we use notation defined in appendix \ref{ads3s3pohl}). In the \adss{3} case we
have two copies of $U(1,1|2)$ and thus we may expect part of the symmetry to be
given by $[PSU(1|1)]^4$. This is discussed in detail in section
\ref{symmetriess}.}
\be
\JA=\mf{so}(2)\sds\bll \mf{psu}(1|1)\brr ^2 \,.
\ee
We expect $\mf{A}$ to be a central extension because it and its corresponding
symmetry acting on the other half of the factorised S-matrix are actually the
same symmetry (which can be seen from \eqref{table}). Similarly to the
$\mbb{R}^2$ central extensions  $\ppmn$ we do not have two copies of this
$\mf{so}(2)$ central extension when we consider the symmetry of the full
S-matrix \eqref{33fact}
\be
\left[\JA\right]^2 \ltimes \mf{so}(2) \ltimes \mbb{R}^2\,.
\ee
This is also in agreement with the fact that the global bosonic symmetry is
$[SO(2)]^3$.

Given that the bosonic subalgebra  of $\SA$ defined by \rf{commm34} is abelian, 
it should not be altered by a quantum deformation (the S-matrix satisfies the
Yang-Baxter equation while respecting the classical $SO(2) \x SO(2)$ symmetry). 
As a result the {\it quantum deformation} of $\SA$ we are interested in is
rather simple. To construct it we  replace the anticommutation relation for
$\mf{S}_{\upm\ump}$ and $\mf{Q}_{\ump\upm}$ with 
\be\la{qddef}
\acom{\mf{S}_{\upm\ump}}{\mf{Q}_{\ump\upm}}= \pm \bll \mf{A}\brr _q\,,
\ee
where $q$ is a quantum deformation parameter and we use the standard notation
\be
\bll x\brr _q\equiv \fr{q^x -q^{-x}}{q-q^{-1}}\,.
\ee
The generators then have the following action on the one-particle states
\begin{align}
&\mf{R}\ket{\phi_\upm}=\pm i\ket{\phi_\upm}\,,
&\no
&\mf{R}\ket{\psi_\upm}=0\,,
\\
&\mf{L}\ket{\phi_\upm}=0\,,
&\no
&\mf{L}\ket{\psi_\upm}=\pm i\ket{\psi_\upm}\,,
\\
&\mf{Q}_{\upm\ump}\ket{\phi_\upm}=0\,,
&\no
&\mf{Q}_{\upm\ump}\ket{\psi_\upm}=d(\vt,k)\ket{\phi_\upm}\,,
\\
&\mf{Q}_{\upm\ump}\ket{\phi_\ump}=c(\vt,k)\ket{\psi_\ump}\,, 
&
&\mf{Q}_{\upm\ump}\ket{\psi_\ump}=0\,,\label{abcd}
\\
&\mf{S}_{\upm\ump}\ket{\phi_\upm}=0\,,
&\no
&\mf{S}_{\upm\ump}\ket{\psi_\upm}=b(\vt,k)\ket{\phi_\upm}\,,
\\
&\mf{S}_{\upm\ump}\ket{\phi_\ump}=a(\vt,k)\ket{\psi_\ump}\,,
&\no
&\mf{S}_{\upm\ump}\ket{\psi_\ump}=0\,,
\end{align}
\be\la{pppp}
\ppmn \ket{\Phi}= \ppmne(\vt,k)
\ket{\Phi}\,.
\ee
For the closure of the quantum-deformed supersymmetry algebra we require
\be\begin{split}\la{clos}
ab=\pmne\,,\hs{20pt}cd=-\ppne\,,\hs{20pt}&
ad=\big[\fr{1}{2}\big]_q\,,\hs{20pt}bc=-\big[\fr{1}{2}\big]_q\,.
\end{split}\ee
To consider the action of the quantum-deformed supersymmetry on the two-particle
states a  coproduct $\De$ is required (see, e.g., \ci{tor}). This coproduct
should respect the quantum-deformed (anti-)commutation relations
\eqref{qddef}\,\foot{As was noted  above, the  linear combination $\mf{A}=
\fr{i}{2} ( \mf{R}+\mf{L})$ commutes with all the other generators including
the fermionic ones.} 
\be\begin{split}\la{33coprod}
\De(\mf{R})= \mf{R} \otimes \mbb{I} + \mbb{I}\otimes \mf{R}\,,
\hs{20pt}&\De(\mf{L})= \mf{L} \otimes \mbb{I} + \mbb{I}\otimes \mf{L}\,,
\\
\De(\mf{Q}_{\upm\ump})= \mf{Q}_{\upm\ump} \otimes & q^{-\mf{A}} +
\mbb{I}\otimes \mf{Q}_{\upm\ump}\,,
\\
\De(\mf{S}_{\upm\ump})= \mf{S}_{\upm\ump} \otimes & \mbb{I} +
q^{\mf{A}}\otimes \mf{S}_{\upm\ump}\,,
\\
\De(\ppmn)= \ppmn \otimes & \mbb{I} + \mbb{I}\otimes \ppmn\,.
\end{split}
\ee
This coproduct co-commutes with the S-matrix\,\foot{The opposite coproduct,
$\De_{\trm{op}}(\mf{J})$, is obtained by acting with the graded permutation
operator for the tensor product on the original coproduct $\De(\mf{J})$.}
\be\la{ops}
\De_{op}(\mf{J})\ \Sc\ =\  \Sc\ \De(\mf{J})
\ee
for an appropriate choice of $a,b,c,d$ in \rf{abcd} and $q$. Assuming that the
quantum deformation parameter is related to the coupling $k$ by (see also 
below) 
\be\la{quq}
q=\exp(-\fr{2\pi   i }{k})\,,
\ee 
we find 
\be\la{bbbb}
a(\vt,k)=\sqrt{\fr{1}{2}\sec\fr{\pi}{k}}\ 
           e^{-\fr{\vt}{2}-\fr{i\pi}{2k}}\,, \ \ \ \ \ 
b(\vt,k) = a^*(\vt,k) \ , \ \ \ \ 
c(\vt,k) = -e^{\vt}  \ a(\vt,k) \ , \ \ \ \ 
d(\vt,k) = e^{\vt}  \ a^*(\vt,k) \ . \ \ \ \ 
\ee
Using \eqref{clos} this implies that the eigenvalues of the central charges 
$\ppmn$ are 
\be\la{pep}
\ppmne(\vt,k) = \fr{1}{2}\sec\fr{\pi}{k}\ e^{\pm\vt}\,,
\ee
i.e. $\ppmn$  can  indeed  be identified with the lightcone momentum generators
up to normalisation. This suggests that the algebra \eqref{commm3} is an 
$\mc{N}=4$ (i.e. $(4,4)$) 2-d supersymmetry with a non-trivial global bosonic
R-symmetry  subalgebra. The supercharges (whose anticommutator is proportional
to the 2-d momentum operator) are charged under both the Lorentz group, and the 
global bosonic symmetry group. It is the existence of this global bosonic
R-symmetry that allows for a quantum deformation of the supersymmetry algebra. 

Expanding \rf{qddef} at large $k$ we have  
\be \la{rrrr}
\bll \mf{A}\brr _q =   \mf{A}  + \fr{2 \pi^2}{3 k^2} ( \mf{A} - \mf{A}^3) + 
\ord{k^{-3}} \ , \ee
so that  the supersymmetry algebra remains standard at the leading tree-level
order. This may  be related to a recent suggestion about the existence of an 
on-shell supersymmetry as part of integrable hierarchy in this classical theory
in \ci{schm} (see also \ci{iv,hms}). Note, however, that for large $k$ the
non-trivial coproduct in \rf{33coprod} differs from the standard one already by 
$1/k$ terms, for example, 
\be \la{uuu}
\De(\mf{Q}_{\upm\ump})= \mf{Q}_{\upm\ump} \otimes  \mbb{I}  +
\mbb{I}\otimes \mf{Q}_{\upm\ump} +
\fr{2 \pi i}{ k}    \mf{Q}_{\upm\ump} \otimes   {\mf{A}}
 + \ord{k^{-2}} \ . 
\ee
The $1/k$ terms are required for the tree-level  S-matrix (given by the $1/k$
terms in $f_i$ in \rf{correctfunctions}) to be invariant under the 
undeformed supersymmetry algebra \rf{commm3}. In this sense the supersymmetry of
the S-matrix is deformed already at the tree level. This is different to the
\adss{2} case discussed in section \ref{seclag22}.

While the reason for a quantum deformation of the supersymmetry of the S-matrix
of the reduced \adss{3} theory is not completely clear the above simple
construction appears to be consistent and suggests that a similar
quantum-deformed supersymmetry  may  also  be present in the reduced  \adss{5}
theory. Let us note  that only the bosonic symmetries of the algebra $\SA$ are
obvious  symmetries of the Lagrangian \eqref{lag13}. As these bosonic
symmetries are abelian, they act on the S-matrix with the standard coproduct.
The \adss{5} case, for which the bosonic symmetries are non-abelian, should be 
more non-trivial as the coproduct of the bosonic symmetry generators will also 
be quantum-deformed (see below).

\subsection{Exact S-matrix conjecture}

Assuming the quantum-deformed supersymmetry discussed in the previous 
subsection \ref{supsy33} exists to all orders in the $1/k$ expansion one can
conjecture an exact S-matrix for the perturbative excitations of the reduced
\adss{3} theory. Co-commutativity of the S-matrix with the quantum group
coproduct \eqref{33coprod} constrains the form of the S-matrix up to two
functions  $P_1(\q,k), \ P_1(\q,k)$. The most general functions $L_i$
parametrising a relativistic S-matrix \eqref{smat33fact}, \eqref{Sparam33},
which co-commutes with the quantum-deformed supersymmetry of section
\ref{supsy33} are given by  
{\allowdisplaybreaks
\begin{align}
\no&L_{1,3}(\q,k)=\fr{1}{2}\Big[P_1(\q,k)\cosh\Big(\fr{\q}{2}
\pm\fr{i\pi}{k}\Big)\sech\fr{\q}{2} 
+ P_2(\q,k)\sinh\Big(\fr{\q}{2}\mp\fr{i\pi}{k}\Big)\cosech\fr{\q}{2} \Big]\ , 
\\
\no&L_{2,4}(\q,k)=\fr{1}{2}\Big[P_1(\q,k)\cosh\Big(\fr{\q}{2}
\pm\fr{i\pi}{k}\Big)\sech\fr{\q}{2} 
- P_2(\q,k)\sinh\Big(\fr{\q}{2}\mp\fr{i\pi}{k}\Big)\cosech\fr{\q}{2} \Big]\ , 
\\
&L_{5,7}(\q,k)=\fr{1}{2}\Big[P_1(\q,k)+P_2(\q,k)\Big]\,,\la{l111}
\hs{18pt}L_{6,8}(\q,k)=\fr{1}{2}\Big[P_1(\q,k)-P_2(\q,k)\Big], 
\\
\no&L_{9,10}(\q,k)=\fr{i}{2} P_1(\q, k)  \sin\fr{\pi}{k}
\sech\fr{\q}{2}\,,
\hs{14pt}L_{11,12}(\q,k)=-\fr{i}{2}P_2(\q,k) \sin\fr{\pi}{k} 
\cosech\fr{\q}{2}
\end{align}}
It can be checked that this S-matrix satisfies the Yang-Baxter equation, which
has the same form as in \eqref{rr22}. The two phase factors  $P_1, P_2$ will 
be fixed by using  the conditions of crossing, unitarity and consistency with
the perturbative one-loop result we obtained above. 

To  match the one-loop  S-matrix \eqref{l110} we require
\bea\no
P_1(\q,k)\!\!\!&= \!\!\!&1 - \fr{i\pi}{2k^2}(\sech \fr{\q}{2})^2 \ (\q+\sinh\q)
+\ord{\fr{1}{k^4}}\,,
\\\la{pertp1p2}P_2(\q,k)\!\!\!&= \!\!\!&
1 + \fr{i\pi}{2k^2}( \cosech \fr{\q}{2})^2 \
\big[(i\pi-\q)+\sinh\q\big]+\ord{\fr{1}{k^4}}\,.\eea
Crossing symmetry \eqref{crosstens33} and unitarity \eqref{unittens33} imply the
following constraints on the phase factors
\bea\la{crossp1p2}
&&P_1(i\pi-\q,k)=P_2(\q,k)\ , \\  
\la{unitp1p2}
&& P_1(\q,k)P_1(-\q,k)=1\,,\hs{30pt}P_2(\q,k)P_2(-\q,k)=
\fr{\sinh^2\fr{\q}{2}}{\sinh^2\fr{\q}{2}+\sin^2\fr{\pi}{k}}\,.
\eea
The $\mbb{Z}_2$ symmetry \eqref{z23}  implies 
\be
P_{1,2}(\q,k)=P_{1,2}(\q,-k)\,. \la{zaz}
\ee
As expected, the perturbative expressions \eqref{pertp1p2} satisfy these
relations.

To solve \eqref{unitp1p2}, \eqref{crossp1p2} we use the ansatz \ci{ansatz}
\be
P_2(\q,k)=p_2(\q,k)\prod_{l=1}^{\infty}
\fr{\rho(\q+2i\pi l,k)}{\rho(-\q+2i\pi( l+1),k)}\,,
\ee
where $\rho(\q,k)$ is an arbitrary function and we assume that  $p_2(\q,k)$
satisfies the following relations 
\be\la{p2rel}
p_2(\q,k)\ p_2(-\q,k)=1\,,\hs{20pt}p_2(i\pi-\q,k)\ p_2(i\pi +\q,k)=1\,.
\ee
The crossing relation \eqref{crossp1p2} implies
\be
P_1(\q,k)=p_2(i\pi-\q,k)\prod_{l=1}^{\infty}
\fr{\rho(\q+2i\pi (l+\fr{1}{2}),k)}{\rho(-\q+2i\pi(l+\fr{1}{2}),k)}\,.
\ee
The second relation in \eqref{p2rel} implies that the first equation in
\eqref{unitp1p2} is satisfied by construction. The second equation in
\eqref{unitp1p2} implies
\be
\rho(\q+2i\pi,k)\ \rho(-\q+2i\pi,k)=\fr{\sinh^2\fr{\q}{2}}{\sinh^2\fr{\q}{2}
+\sin^2\fr{\pi}{k}}=\fr{\sinh^2\fr{\q}{2}}
{\sinh\big(\fr{\q}{2}+\fr{i\pi}{k} \big)
\sinh\big(\fr{\q}{2}-\fr{i\pi}{k} \big)}\,.
\ee
Using the gamma function reflection formula this equation is solved
by\,\foot{There are alternative solutions \ci{ansatz}; the obvious ones are
gotten by considering $k \ra -k$ and $\q \ra -\q$. As we require that
$P_1(\q,k)=P_1(\q,-k)$ and $P_2(\q,k) = P_2(\q,-k)$ we can ignore the
$k \ra -k$ solution. We also disregard the $\q \ra - \q$ solution as its
expansion does not match the perturbative result \eqref{pertp1p2}.
For a similar reason we ignore the solution 
\be 
\no\rho(\q,k)=\pm\fr{\sinh\fr{\q}{2}}{\sinh\big(\fr{\q}{2}\pm\fr{i\pi}{k}
\big)}
\,. 
\ee}
\be
\rho(\q,k)=\fr{\Gamma(-\fr{i\q}{2\pi}-\fr{1}{k}-1)\Gamma(-\fr{i\q}{2\pi}
+\fr{1}{k})}{\Gamma(-\fr{i\q}{2\pi}-1)\Gamma(-\fr{i\q}{2\pi})}\,.
\ee
To fix $p_2(\q,k)$ we use the relation \rf{zaz}, implying
\be\la{p2rel21}
p_2(\q,k)=\fr{\sinh\big(\fr{\q}{2}-\fr{i\pi}{k}\big)}
             {\sinh\big(\fr{\q}{2}+\fr{i\pi}{k}\big)}\ p_2(\q,-k)\,.
\ee
The minimal choice for the function $p_2(\q,k)$ that satisfies \eqref{p2rel}
and \eqref{p2rel21} is
\be
p_2(\q,k)=\sqrt{\fr{\sinh\big(\fr{\q}{2}-\fr{i\pi}{k}\big)}
                    {\sinh\big(\fr{\q}{2}+\fr{i\pi}{k}\big)}}\,.
\ee
We end up with the following solution for the two functions $P_1,P_2$: 
\be\la{P1P2}\begin{split}
P_1(\q,k) = & \sqrt{\fr{\cosh\big(\fr{\q}{2}+\fr{i\pi}{k}\big)}
                    {\cosh\big(\fr{\q}{2}-\fr{i\pi}{k}\big)}}
\prod_{l=1}^{\infty}
\fr{\Gamma(\fr{i\q}{2\pi}-\fr{1}{k}+l-\fr{1}{2})\Gamma(\fr{i\q}{2\pi}
+\fr{1}{k}+l+\fr{1}{2})}{\Gamma(-\fr{i\q}{2\pi}-\fr{1}{k}+l-\fr{1}{2}
)\Gamma(-\fr
{ i\q } {2\pi }+\fr{1}{k}+l+\fr{1}{2})}
\fr{\Gamma(-\fr{i\q}{2\pi}+l-\fr{1}{2})\Gamma(-\fr{i\q}{
2\pi}+l+\fr{1}{2})}{\Gamma(\fr{i\q}{2\pi}+l-\fr{1}{2})\Gamma(\fr{i\q}{
2\pi}+l+\fr{1}{2})}\,,
\\P_2(\q,k) = &\sqrt{\fr{\sinh\big(\fr{\q}{2}-\fr{i\pi}{k}\big)}
                    {\sinh\big(\fr{\q}{2}+\fr{i\pi}{k}\big)}}
\prod_{l=1}^{\infty}
\fr{\Gamma(-\fr{i\q}{2\pi}-\fr{1}{k}+l-1)\Gamma(-\fr{i\q}{2\pi}
+\fr{1}{k}+l)}{\Gamma(\fr{i\q}{2\pi}-\fr{1}{k}+l)\Gamma(\fr{i\q}{
2\pi }+\fr{1}{k}+l+1)}
\fr{\Gamma(\fr{i\q}{2\pi}+l)\Gamma(\fr{i\q}{
2\pi}+l+1)}{\Gamma(-\fr{i\q}{2\pi}+l-1)\Gamma(-\fr{i\q}{
2\pi}+l)}\,.
\end{split}\ee
It can be checked directly that \eqref{P1P2} matches the perturbative
expansions \eqref{pertp1p2}. We therefore conjecture that the exact S-matrix
for the perturbative excitations of the reduced \adss{3} theory is given by
\eqref{33fact}, \eqref{Sparam33}, \eqref{l111} with phase factors \eqref{P1P2}.

Note that translating the factorised form \eqref{33fact} back to the original
notation of section \ref{secollsg33} neither the exact $Y_m Z_n \ra Y_m Z_n$
and $\ze_m \chi_n \ra \ze_m \chi_n$ amplitudes nor the square of $L_{5,7}$ 
are equal to \eqref{phase}. This indicates  that the reduced \adss{3} theory is
not exactly in the same  class of models as the reduced \adss{2} and \adss{5} 
theories.

Still, the product of the two phase factors \eqref{P1P2} is equal to the square
root of \eqref{phase} with $\De = \fr{2\pi}{k}$. Also, the factors in front
of the products of gamma functions in \eqref{P1P2} are square roots of the
amplitudes in the  complex sine-Gordon S-matrix, see \ci{dh}. This suggests, by
analogy with the reduced \adss{2} theory which may be interpreted as an
$\mc{N}=2$ supersymmetric dressing of the bosonic sine-Gordon theory, that the
reduced \adss{3} theory may  be  interpreted  as a quantum-deformed $\mc{N}=4$
supersymmetric dressing of the complex sine-Gordon model.


\renewcommand{\theequation}{5.\arabic{equation}}
\setcounter{equation}{0}
\section{Symmetries of the S-matrices\la{sumofsym}}

In this section we shall discuss and compare various global symmetries of the 
reduced \adss{2}, \adss{3} and \adss{5} theories with a motivation to understand
the  expected symmetry of the S-matrix of the \adss{5} case. We shall see that
the latter may be related to the quantum-deformed $\mf{psu}(2|2)\ltimes
\mbb{R}^3$ R-matrix of \ci{bk,bcl}.

The reduced \adss{2} and \adss{3} theories are invariant under a wide range of
different types of symmetries -- 2-d Poincar\'{e}, global bosonic symmetries,
gauge symmetries, classical supersymmetries and quantum-deformed symmetries.
The  2-d Poincar\'{e}  algebra\,\foot{$\sds$ denotes the semi-direct sum defined
in footnote \ref{foot1}.} 
\be
\mf{iso}(1,1) = \mf{so}(1,1) \sds \mbb{R}^2\,.
\ee
contains one Lorentz boost, space translation and time translation under which
all the reduced \adss{n} theories are invariant. In the reduced \adss{2} and
\adss{3} theories the symmetry that acts on the two-particle states and
co-commutes with the S-matrix in each case is based on a superalgebra of the
form
\be\la{dsd}
\mf{c}\ltimes \mbb{R}^2\,.
\ee
Here $\mbb{R}^2$ corresponds to the lightcone momenta and $\mf{c}$ contains 
the fermionic generators (charged under the Lorentz group) whose anticommutator
is proportional to the momenta. The fermionic generators are also charged under
the bosonic subalgebra $\mf{b}$ of $\mf{c}$. The algebra \eqref{dsd} thus has
the same structure as a 2-d supersymmetry algebra with a bosonic R-symmetry 
algebra given by $\mf{b}$.

This symmetry appears to originate from the global target space supersymmetry in
the associated superstring theory -- the algebra $\mf{c}$ is a particular
sub-superalgebra of the latter symmetry. Under the reduction procedure the
fermionic target space supersymmetry generators become charged under the
Lorentz group  and behave like  generators of 2-d supersymmetry in the reduced
theory.

In the case of the  reduced \adss{2} theory the bosonic subalgebra $\mf{b}$ is
absent and all the fermionic generators of $\mf{c}$ anticommute up to the
central extension generators. Thus a quantum deformation of the corresponding
algebra  of a kind discussed  in  section \ref{supsy33} is trivial here. In both
the reduced \adss{2} and \adss{3} theories we can write the physical symmetry of
the corresponding S-matrix as\,\foot{Here $U$ denotes the universal enveloping
algebra. The subscript $q$ then stands for the quantum deformation of this
algebra, (which has no effect in the \adss{2} case). That is $U_q(\mf{c})$ is
the quantum group.}
\be\la{fermsym}
U_q\big(\mf{so}(1,1) \sds (\mf{c}\ltimes \mbb{R}^2)\big)\,.
\ee

\subsection{Algebraic structure of Pohlmeyer reduction\la{prgr}}

In order to understand the origin of the superalgebra $\mf{c}$ it is useful to
review the algebraic structure of Pohlmeyer reduction \ci{gt1}. The reduced
\adss{n} theories  are Pohlmeyer reductions of the GS superstring sigma model
based on a supercoset space ${\hat{F}}/{G}\,$ where $\hat{F}$ is a supergroup
and $G$ is some bosonic subgroup. The superalgebra of $\hat{F}$ is required to
have a $\mbb{Z}_4$ decomposition
\be
\hat{\mf{f}}
=\hat{\mf{f}}_0 + \hat{\mf{f}}_1 +
 \hat{\mf{f}}_2 + \hat{\mf{f}}_3\,,\hs{30pt}
\com{\hat{\mf{f}}_i}{\hat{\mf{f}}_j}\subset
\hat{\mf{f}}_{i+j\trm{\,mod\,} 4}\,,
\ee
where even/odd subscripts denote bosonic/fermionic subspaces. The algebra
$\mf{g}$ of the group $G$ is identified as
\be
\mf{g}=\hat{\mf{f}}_0\,.
\ee
The Pohlmeyer reduction procedure involves solving the Virasoro constraints 
using a constant element
\be
T \in \mf{a} \subset \hat{\mf{f}}_2\,,
\ee
where $\mf{a}$ is the maximal abelian subalgebra of $\hat{\mf{f}}_2$.\,\foot{In
the reduced \adss{n} theories $\mf{a}$ is always 2-dimensional. In the
non-degenerate case $T$ is non-zero in both the $AdS_n$ and $S^n$ parts of the
algebra.} This element $T$ induces a further $\mbb{Z}_2$ decomposition on the
algebra
\begin{align}\no
 \hat{\mf{f}}=\hat{\mf{f}}^\perp +  \hat{\mf{f}}^\parallel\,,\hs{20pt}
 \trm{where}&\hs{20pt}
 \hat{\mf{f}}^\perp = \{ T , \{ T, \hat{\mf{f}} \} \}
 \,,\hs{10pt}
 \hat{\mf{f}}^\parallel = \com{T}{\com{T}{\hat{\mf{f}}}}\,,
 \\&\Ra\hs{10pt}\la{commmt}
 \com{\mf{f}^\perp}{T}=0\,,\hs{40pt}  \{ \mf{f}^{\parallel},{T} \}
 =0\,.
 \end{align}
The reduced theory is then identified with a fermionic extension of a gauged
WZW model with an integrable potential \rf{gwzw}. The gauged WZW model is based
on the coset space ${G}/{H}$, where the algebra of $H$ is given by 
\be
\mf{h}= \hat{\mf{f}}_0^\perp\,.
\ee 
Let us also consider a particular sub-superalgebra of $\hat{\mf{f}}$
\be\la{hhht}
\hat{\mf{f}}^\perp =  \hat{\mf{h}} \ltimes T\,,  \ \ \ \ \ \   \ \ \ \ \ \ 
\hat{\mf{h}} = \mf{h} \oplus \hat{\mf{f}}_1^\perp \oplus \hat{\mf{f}}_3^\perp 
\ , \ee 
where  
\be 
\com{\hat{\mf{f}}_{1}^\perp}{\hat{\mf{f}}_{1}^\perp}\,,\;
\com{\hat{\mf{f}}_{3}^\perp}{\hat{\mf{f}}_{3}^\perp}\subset
T \,,\hs{30pt}
\com{\hat{\mf{f}}_{1}^\perp}{\hat{\mf{f}}_{3}^\perp}\,,\;
\com{\hat{\mf{f}}_{3}^\perp}{\hat{\mf{f}}_{1}^\perp}
\subset 
\mf{h}\,.
\ee
Here $T$ behaves like a central extension as it is
abelian and commutes with all other generators, \eqref{commmt}. The algebra 
$\hat{\mf{h}}$\,\foot{This algebra can be found via an
Wigner-$\dot{\trm{I}}$n$\ddot{\trm{o}}$n$\ddot{\trm{u}}$ contraction, i.e. by 
rescaling the central extension generators by a constant and sending it to
zero.} will play a r\^{o}le in the  discussion of the quantum-deformed
supersymmetry of the reduced theories.

\subsection{Reduced \adss{2} theory}

As explained in section \ref{redsym22}, the reduced \adss{2} theory has manifest
$\mc{N}=2$\ 2-d supersymmetry with the superalgebra that can be written as 
\be\la{ads2symex}
\mf{so}(1,1) \sds ([\mf{psu}(1|1)]^2\ltimes \mbb{R}^2)\,.
\ee
This agrees with the form of \eqref{fermsym} with no quantum deformation. As the
$\mc{N}=2$ supersymmetry is manifest in the action, one should not indeed  
expect any quantum deformation in this case. As discussed earlier, ignoring the
Lorentz part of the algebra, up to central extensions all the generators of the
algebra commute/anticommute. The quantum deformation of the type considered in
the reduced \adss{3} theory would thus have no effect on \eqref{ads2symex}.

The \adss{2} superstring model can be written as a GS sigma-model with the 
target space 
\be
\fr{PSU(1,1|2)}{SO(1,1) \x SO(2)}\,.
\ee
Here $\hat{F}=PSU(1,1|2)$ is a global symmetry and  $G =SO(1,1) \x SO(2)$ is a
gauge symmetry. In the group $H$ of the Pohlmeyer reduced theory here is 
trivial. Then the subspaces $\hat{f}_1^\perp$ and  $\hat{f}_3^\perp$ are both 
superalgebras equivalent to $\mf{psu}(1|1)$. Therefore, here $\hat{\mf{h}}$ in
\rf{hhht} is 
\be\la{hhat22}
\hat{\mf{h}}=[\mf{psu}(1|1)]^2\,,
\ee
i.e. is the same algebra as $\mf{c}$ in \eqref{fermsym}, \eqref{ads2symex}.

\subsection{Reduced \adss{3} theory}

As discussed in section \ref{secsym33} and appendix \ref{ads3s3pohl} the
manifest bosonic symmetry  of the Lagrangian of the reduced \adss{3} theory has
the following algebra 
\be\la{ads33symmb2}
\mf{iso}(1,1) \oplus [\mf{u}(1)]^3 \oplus [\mf{u}^{(g)}(1)]^3\,,
\ee
where the superscript $(g)$ denotes a gauge symmetry. The fields on which the
global part of the gauge symmetry has a linear action are field redefinitions of
the fields on which the global $H$ symmetry has a linear action. Therefore, the
physical symmetry acting on on-shell states is
\be\la{ads33symmb}
\mf{iso}(1,1) \oplus [\mf{u}(1)]^3\,.
\ee
In section \ref{supsy33} an on-shell quantum-deformed supersymmetry was shown to
co-commute with the one-loop S-matrix. This quantum supersymmetry extends the
physical symmetry of the theory to 
\be\la{ads3symex}
U_q\Big(\mf{so}(1,1) \sds
\big(\big[\mf{u}(1)\sds[\mf{psu}(1|1)]^2\big]^2
\ltimes \mf{u}(1) \ltimes \mbb{R}^2\big)\Big)\,.
\ee
Due to the abelian nature of the bosonic subgroup (up to central extensions), 
only the action of the supersymmetry generators on two-particle states is
quantum deformed.

The \adss{3}  superstring theory can be written as a 2-d sigma-model with the 
target space\,\foot{Here we use a somewhat  non-standard  form of the supercoset
leading to an equivalent Lagrangian. As explained in appendix \ref{ads3s3pohl}
the symmetry analysis of this theory is more systematic if we consider the coset
\eqref{cosetalt}. For a discussion of  definitions of various projections of
central elements see appendix \ref{ads3s3pohl}.}
\be\la{cosetalt}
\fr{PS[U(1,1|2) \x  U(1,1|2)]}{U(1,1) \x U(2)}\,.
\ee
This theory has a global $\hat{F}=PS([U(1,1|2)]^2)$ symmetry and $G = U(1,1) \x
U(2)$ gauge symmetry. Then in the reduced theory $H = [U(1)]^4$. As explained 
in appendix \ref{ads3s3pohl} one of these $U(1)$s acts trivially on all the
fields and can thus be ignored leaving us with $H=[U(1)]^3$.

Once this extra $\mf{u}(1)$ is projected out we find that $\hat{\mf{h}}$ in
\rf{hhht} is 
\be
\hat{\mf{h}}=\big[\mf{u}(1)\sds [\mf{psu}(1|1)]^2\big]^2
\ltimes \mf{u}(1)\,.
\ee
We see that again $\hat{\mf{h}}$ is the same algebra as $\mf{c}$ in
\eqref{fermsym},\eqref{ads3symex}.

\subsection{Reduced \adss{5} theory}

Let us now try to use an analogy with the lower-dimensional cases to understand
which symmetries should appear  in the \adss{5} case. The \adss{5} superstring
theory is based on the supercoset   
\be\la{coset55}
\fr{PSU(2,2|4)}{Sp(2,2) \x Sp(4)}\,, 
\ee
i.e. here $\hat{F}=PSU(2,2|4)$ is the  global symmetry and  $G = Sp(2,2) \x
Sp(4)$ is the gauge symmetry. The gauge group of the reduced theory is
$H=[SU(2)]^4$. Then the algebra $\hat{\mf{h}}$  in \rf{hhht} is\,\foot{This
same sub-superalgebra arises also when considering the expansion of the 
superstring action around the BMN vacuum: the manifest symmetry of the
Lagrangian is broken to $[SU(2)]^4$ while the on-shell symmetry (under which
the superstring S-matrix is invariant) is precisely $\hat{\mf{h}}$.}
\be\la{hath55}
\hat{\mf{h}}=\big[\mf{psu}(2|2)\big]^2\,.
\ee
The manifest bosonic symmetry algebra of the reduced \adss{5} theory Lagrangian
is given by 
\be\la{ads55symmb2}
\mf{iso}(1,1) \oplus \big[\mf{su}^{(g)}(2)\big]^4\,,
\ee
where again the superscript $(g)$ denotes  gauge symmetry. There is no
additional global symmetry in contrast to the abelian $H$ case. 

The perturbative S-matrix found in section \ref{secollsg55} is constructed in 
such a way that it has manifest global symmetry $[SU(2)]^4$, which is the same
as the global part of the gauge group. This is true at tree-level \ci{ht1} and
also at the  one-loop level as will be shown in sections \ref{secsym55} and
\ref{secfac}. A manifest (i.e. acting with the standard coproduct) non-abelian
global symmetry of a relativistic, trigonometric S-matrix for the theories
\rf{gwzw} is already in conflict \ci{ht2} with the Yang-Baxter equation at the
tree level.

Motivated by the \adss{3} example, where the symmetry group $\hat{\mf{h}}$ was
quantum-deformed, one may conjecture that the same should happen also in the 
\adss{5} case, i.e. the S-matrix should be invariant under the corresponding 
quantum group 
\be\la{ads5symex}
U_q\Big(\mf{so}(1,1) \sds
\big(\big[\mf{psu}(2|2)\big]^2 \ltimes \mbb{R}^2\big)\Big)\,,
\ee
where we have replaced $\mf{c}$ in \eqref{fermsym} with $\hat{\mf{h}}$ from
\eqref{hath55}. 

An R-matrix invariant under $U_q(\mf{psu}(2|2) \ltimes \mbb{R}^3)$ has been
studied in \ci{bk,bcl}. It was observed that there is a particular classical
limit of the R-matrix \ci{bcl} that bears strong resemblance to the tree-level 
S-matrix found in \ci{ht1}. We will extend this limit to all orders in $1/k$ in 
section \ref{bkcomp}, finding a relativistic trigonometric R-matrix satisfying 
unitarity, crossing and the Yang-Baxter equation. In this limit the third
central extension vanishes. The resulting R-matrix has similarities to the
one-loop S-matrix of section \ref{secollsg55}. It is then natural to consider
this R-matrix as a candidate for the physical S-matrix of the perturbative
excitations of the reduced \adss{5} theory.

Unlike the reduced \adss{3} theory here the group $H=[SU(2)]^4$ is non-abelian
and therefore the quantum deformation will non-trivially affect its action on
the two-particle states (the action of $H$ on the one-loop perturbative S-matrix
was assumed to be given by the standard coproduct). Understanding the origin of 
this quantum deformation is an important open question. More generally, this
question applies also to similar bosonic models with a non-abelian gauge group
\ci{ht2,hm1,hm2,hm3} discussed in appendix \ref{bosonic}.


\renewcommand{\theequation}{6.\arabic{equation}}
\setcounter{equation}{0}
\section{S-matrix of the reduced \adss{5} theory\la{seclag55}} 

In this section we finally  consider the case of  prime interest -- the reduced 
\adss{5} theory --  with the aim of understanding the structure of the
corresponding quantum S-matrix.

We shall first demonstrate that the group factorisation property of the
perturbative S-matrix (see section \ref{secollsg55}), first  observed at
tree-level in \ci{ht1}, continues  to  one-loop level. The factorised S-matrix
can then  be compared with the quantum-deformed $\mf{psu}(2|2)\ltimes
\mbb{R}^3$ R-matrix of \ci{bk}. A particular classical limit of this R-matrix
was identified in \ci{bcl} whose form is very similar to that of the tree-level
S-matrix in  \ci{ht1}. We will extend this limit to all orders and show, in
particular, that this similarity continues also to the  one-loop order. It is
then natural to consider this R-matrix as a candidate for the physical S-matrix
of the perturbative excitations of the reduced \adss{5} theory. 

Other than the similarity with the tree-level S-matrix the main motivation for
considering this R-matrix is an analogy with the S-matrix in the \adss{3} case 
which is invariant under a quantum-deformed supersymmetry. The choice of the
symmetry algebra, $\mf{psu}(2|2)\ltimes \mbb{R}^2$, is explained in section
\ref{sumofsym}. Also, in bosonic theories similar to \rf{gwzw} with a
non-abelian  group $H$ the quantum deformation has been conjectured to be the
physical symmetry of the theory \ci{hm1,hm2,hm3}. 

The one-loop S-matrix computed in \ci{ht1} and in section \ref{secollsg55} is 
invariant under the maximal bosonic subalgebra of $[\mf{psu}(2|2)]^2$. This
subalgebra is non-abelian. However, an S-matrix invariant under the quantum
deformation is not invariant under this non-abelian symmetry with the standard
coproduct for the bosonic generators. Hence there are differences between the
perturbative  S-matrix originating from the  action \rf{gwzw} and the
quantum-deformed S-matrix seen already  at the tree level \ci{bcl,ht1}. This is 
not surprising as the quantum-deformed S-matrix satisfies the Yang-Baxter
equation while the perturbative $H$-invariant one-loop S-matrix does not. We
conclude this section by investigating a relation between these two S-matrices.

\subsection{Perturbative S-matrix at one-loop order\la{secsym55}}

In the reduced \adss{5} theory \ci{gt1} we have  $G=Sp(2,2) \x Sp(4)$ and the 
gauge group $H= [SU(2)]^4$. The Lagrangian \eqref{lag15} is written with a
manifest global $SO(4)$ symmetry. This symmetry is a subgroup of the global part
of the gauge group $H$. As the $A_+=0$ gauge fixing preserves the global part of
$H$ the Lagrangian is expected to be  invariant under the full $H=[SU(2)]^4$
global symmetry. This symmetry can be made manifest by using the field
redefinitions \ci{ht1}\,\foot{The 2-indices are raised and lowered with the
antisymmetric tensors $\epsilon^{ab}$, etc., i.e. $F^a = \epsilon^{ab}  F_b, \ 
F_b= \epsilon_{bc} F^c$. Dotted and undotted indices are assumed to be
completely independent. We use the convention that
$\epsilon^{12}=1,\,\epsilon_{12}=-1$, $\epsilon^{ab} \epsilon_{bc}=\delta^a_c$
and the rescaled  set of Pauli matrices 
$$\s^1=\bar{\s}^1=\fr{1}{\sqrt{2}}\bp1&0\\ 0&1\emp \ ,  
 \ \ \ \   \s^2=-\bar{\s}^2=\fr{1}{\sqrt{2}}\bp  0&1\\-1&0\emp \,, \ \ \ \  
           \s^3=-\bar{\s}^3=\fr{1}{\sqrt{2}}\bp 0&i\\i&0\emp \,, \ \ \ \ 
           \s^4=-\bar{\s}^4=\fr{1}{\sqrt{2}}\bp i&0\\0&-i\emp \,.
$$}
\be\begin{split}
Y_m = (\bar{\s}_m)^{\adt a}Y_{a \adt}\,,
&\hs{30pt}Y_{a\adt}=(\s_m)_{a\adt}Y_m\,,
\\Z_m = (\bar{\s}_m)^{\agdt \al}Z_{\al \agdt}\,,
&\hs{30pt}Z_{\al\agdt}=(\s_m)_{\al\agdt}Z_m\,,
\\\z_m = (\bar{\s}_m)^{\agdt a}\z_{a \agdt}\,,
&\hs{30pt}Y_{a\agdt}=(\s_m)_{a\agdt}Y_m\,,
\\\ch_m = (\bar{\s}_m)^{\adt \al}\ch_{\al \adt}\,,
&\hs{30pt}\ch_{\al\adt}=(\s_m)_{\al\adt}\ch_m\,,
\end{split}\ee
where $(Y_{a\adt})^*=Y^{\adt a}$, etc. The translation of \eqref{lag15} into the
manifestly $[SU(2)]^4$ invariant form is \ci{ht1} 
{\allowdisplaybreaks
\bea
\Lag_5&=&   \fr{1}{2}     \dpl Y_{a\adt}    \dm Y^{\adt a}
\no         - \fr{\mu^2}{2} Y_{a\adt}         Y^{\adt a}
            + \fr{1}{2}     \dpl Z_{\al\agdt} \dm Z^{\agdt\al}
            - \fr{\mu^2}{2} Z_{\al\agdt}      Z^{\agdt\al}
\\ &&       + \fr{i}{2}     \zel_{a\agdt}    \dpl \zel^{\agdt a}
\no         + \fr{i}{2}     \zer_{a\agdt}    \dm \zer^{\agdt a}
            - i\mu          \zel_{a\agdt}    \zer^{\agdt a}
\\ &&       + \fr{i}{2}     \chl_{\al\adt}   \dpl \chl^{\adt \al}
\no            + \fr{i}{2}     \chr_{\al\adt}   \dm \chr^{\adt \al}
            - i\mu          \chl_{\al\adt}   \chr^{\adt\al}
\\ && +\fr{\pi}{2k}\bigg[
\no    - \fr{2}{3}   \big(Y_{a\adt}Y^{\adt a}\dpl Y_{b\bdt}\dm Y^{\bdt b}
                         -Y_{a\adt}\dpl Y^{\adt a} Y_{b\bdt}\dm Y^{\bdt b}
                      +\fr{\mu^2}{2}Y_{a\adt}Y^{\adt a}Y_{b\bdt}Y^{\bdt b}\big)
\\ && \hs{30pt}
  + \fr{2}{3} \big(Z_{\al\agdt}Z^{\agdt \al}\dpl Z_{\bet\bgdt}\dm Z^{\bgdt \bet}
\no        -Z_{\al\agdt}\dpl Z^{\agdt\al} Z_{\bet\bgdt}\dm Z^{\bgdt \bet}
     +\fr{\mu^2}{2}Z_{\al\agdt}Z^{\agdt \al}Z_{\bet\bgdt}Z^{\bgdt \bet}\big) 
\\ && \hs{30pt} 
  + i \big(\zel_{a\agdt}\zel^{\agdt b}Y^{\bdt a}\dpl Y_{b\bdt}
\la{lag35}                          
         + \zer_{a\agdt}\zer^{\agdt b}Y^{\bdt a}\dm Y_{b\bdt}
         + \mu\,\zer_{a\agdt}\zel^{\agdt a}Y_{b\bdt}Y^{\bdt b}\big)
\\ && \hs{30pt} 
  - i \big(\zel_{a\agdt}\zel^{\bgdt a}Z^{\agdt \bet}\dpl Z_{\bet \bgdt}
\no      + \zer_{a\agdt}\zer^{\bgdt a}Z^{\agdt \bet}\dm Z_{\bet \bgdt}
   + \mu\,\zer_{a\agdt}\zel^{\agdt a}Z_{\bet\bgdt}Z^{\bgdt \bet}\big)
\\ && \hs{30pt} 
   + i  \big(\chl_{\al\adt}\chl^{\bdt \al}Y^{\adt b}\dpl Y_{b\bdt}
\no        + \chr_{\al\adt}\chr^{\bdt \al}Y^{\adt b}\dm Y_{b\bdt}
     + \mu\,\chr_{\al\adt}\chl^{\adt \al}Y_{b\bdt}Y^{\bdt b}\big)
\\ && \hs{30pt} 
  - i \big(\chl_{\al\adt}\chl^{\adt \bet}Z^{\bgdt \al}\dpl Z_{\bet\bgdt}
\no   +\chr_{\al\adt}\chr^{\adt\bet}Z^{\bgdt \al}\dm Z_{\bet\bgdt}
   +\mu\,\chr_{\al\adt}\chl^{\adt \al}Z_{\bet\bgdt}Z^{\bgdt \bet}\big)
\\ && \hs{30pt} 
  + 4i\mu  \big(\zer_{a\agdt}\chl_{\bet\bdt}Y^{\bdt a}Z^{\agdt \bet} 
\no           -\chr_{\al\adt}\zel_{b\bgdt}Y^{\adt b}Z^{\bgdt \al}\big)
\\&&\hs{30pt} 
   +2 \big(\zel_{a\agdt}\zel_{b\bgdt}\zer^{\agdt b}\zer^{\bgdt a}            
   -\chl_{\al\adt}\chl_{\bet\bgdt}\chr^{\adt\bet}\chr^{\bdt\al}\big)\bigg] 
\no  + \ord{k^{-2}}\,.
\eea}

\subsubsection{Group factorisation of the S-matrix\la{secfac}}

Consider the following factorised S-matrix \ci{ht1,kmrz}
\be\la{55fact}
\Sc\ket{\Phi_{A\Adt}(p_1)\Phi_{B\Bdt}(p_2)}
=(-1)^{[\Adt][B]+[\Cdt][D]}\ S_{AB}^{CD}\ S_{\Adt\Bdt}^{\Cdt\Ddt}\ 
\ket{\Phi_{C\Cdt}(p_1)\Phi_{D\Ddt}(p_2)} \ , 
\ee
\be\la{Sparam}
S_{AB}^{CD} =
\left\{
\ba{l} K_1 \de_a^c\de_b^d
              + K_2 \de_a^d\de_b^c\,, 
\\             K_3 \de_\al^\g\de_\bet^\de  
              + K_4 \de_\al^\de\de_\bet^\g\,,   
\\             K_5 \e_{ab}\e^{\g\de}\,, 
            \hs{9pt} K_6 \e_{\al\bet}\e^{cd}\,, 
\\             K_7 \de_a^d \de_\bet^\g\,, 
           \hs{15pt} K_8 \de_\al^\de \de_b^c\,, 
\\             K_9 \de_a^c \de_\bet^\de\,, 
            \hs{15pt} K_{10} \de_\al^\g \de_b^d\,,
\ea \right.
\ee
where $K_i$ are functions of $\q$ and the coupling $k$. Here $\Phi_{A\Adt}$
(with $A=(a,\al)$ and $\Adt=(\adt,\agdt)$, where $a,\al,\adt,\agdt$ are indices
of fundamental representations of the four $SU(2)$ groups comprising the global 
$[SU(2)]^4$ symmetry) encodes the fields  $Y_{a\adt}\,, \  Z_{\al\agdt}\,,\
\z_{a\agdt}\,, \  \chi_{\al\adt}\,.$ As for the reduced \adss{3} theory
\eqref{fermgr33} we assume the fermionic grading
\be
\bll a\brr =\bll \adt\brr =0\,,\hs{30pt}
\bll \al\brr =\bll \agdt\brr =1\,.
\ee
As in the reduced \adss{3} theory it is useful to consider ``half'' of this
factorised S-matrix \eqref{Sparam} as acting on a single field 
\be\la{smat55fact}
\Sc\ket{\Phi_A(p_1) \Phi_B(p_2)}
= S_{AB}^{CD}(\q,k)\ket{\Phi_C(p_1)\Phi_D(p_2)}\,,\hs{20pt}
\Phi_A = (\phi_a,\psi_\al)\,,
\ee
where $\phi_a$ are bosonic and $\psi_\al$ are fermionic. This S-matrix should
satisfy the usual physical requirements of unitarity and crossing. The unitarity
and crossing relations are the same as in \eqref{unittens33},
\eqref{crosstens33} and \eqref{chco33}. The crossed S-matrix
$\bar{S}_{AB}^{CD}(\q,k)$ is identical to $S_{AB}^{CD}(\q,k)$ in
\eqref{smat55fact} except with $(K_5,\,K_6,\,K_7,\,K_8)$ replaced by
$i(K_5,\,K_6,\,K_7,\,K_8)$. Crossing symmetry requires the following relations
between the functions $K_i$, 
\be\la{crossk}\begin{split}
&K_1(i\pi-\q,k)=K_1(\q,k)+K_2(\q,k)\,,\hs{20pt}
K_2(i\pi-\q,k)=-K_2(\q,k)\,,
\\&K_5(i\pi-\q,k)= i K_7(\q,k)\,,\hs{65pt}
K_7(i\pi-\q,k)=- i K_5(\q,k)\,,
\\&\hs{90pt}K_9(i\pi-\q,k)=K_9(\q,k)\,,
\end{split}\ee
and similarly for $K_3,\,K_4,\,K_6,\,K_8,\,K_{10}$.

For consistency (for example between $\Sc\ket{Y_m(p_1)\z_n(p_2)}$ and
$\Sc\ket{\z_m(p_1)Y_n(p_2)}$, see appendix \ref{55expansion}) the functions
$K_i$ should also obey the following  conjugation relations
\be\la{conggg}\begin{split}
K_1(\q,k)=K_1^*(-\q,k)\,,&\hs{20pt}K_3(\q,k)=K_3^*(-\q,k)\,,
\\K_2(\q,k)=K_2^*(-\q,k)\,,&\hs{20pt}K_4(\q,k)=K_4^*(-\q,k)\,,
\\K_5(\q,k)=-K_5^*(-\q,k)\,,&\hs{20pt}K_6(\q,k)=-K_6^*(-\q,k)\,,
\\K_7(\q,k)=K_8^*(-\q,k)\,,&\hs{20pt}K_9(\q,k)=K_{10}^*(-\q,k)\,.
\end{split}\ee
The Lagrangian \eqref{lag15}  has also  the $\mbb{Z}_2$ symmetry \eqref{z23}
implying the following relations  
\be\begin{split}
&K_1(\q,k)=K_3(\q,-k)\,,
\hs{28pt}
K_2(\q,k)=K_4(\q,-k)\,,
\\
&K_5(\q,k)=-K_6(\q,-k)\,,
\hs{20pt}
K_7(\q,k)=-K_8(\q,-k)\,,
\\
&\hs{65pt}K_9(\q,k)=K_{10}(\q,-k)\,.
\end{split}\ee
In appendix \ref{55expansion} equation \eqref{55fact} is expanded with
arbitrary $K_i$ and rewritten in the original $SO(4)$ notation to enable
comparison to the S-matrix of section \ref{secollsg55}. Along with other
relations between the parametrising functions, the relation between $YYZZ$ and
$\z\z\ch\ch$ amplitudes ($f_3$ and $f_5$ functions)  means the one-loop S-matrix
of section \ref{secollsg55} factorises as in \eqref{55fact}, \eqref{Sparam} 
with $K_i$ given by 
\be \la{faac}
K_i = \ \po(\q,k\,;{\textstyle\fr{1}{2}}) \  \hat{K}_i \ , \ee
\be\la{k110}\begin{split}
\hat{K}_1(\q,k)=&\hat{K}_3(\q,-k)=1 + \fr{i\pi}{2k}\tanh\fr{\q}{2}
                     -\fr{5\pi^2}{8k^2}-\fr{i\pi\q}{2k^2}
                     +\ord{k^{-3}}
\\
\hat{K}_2(\q,k)=&\hat{K}_4(\q,-k)=-\fr{i\pi}{k}\coth\q
+\fr{\pi^2}{2k^2}+\fr{i\pi\q}{k^2}
                     +\ord{k^{-3}}
\\
\hat{K}_5(\q,k)=&-\hat{K}_6(\q,-k)=-\fr{i\pi}{2k}\sech\fr{\q}{2}
                     +\ord{k^{-3}}
\\
\hat{K}_7(\q,k)=&-\hat{K}_8(\q,-k)=-\fr{i\pi}{2k}\cosech\fr{\q}{2}
                     +\ord{k^{-3}}
\\
\hat{K}_9(\q,k)=&\hat{K}_{10}(\q,-k)=1                  
+\ord{k^{-3}}\ .
\end{split}\ee
Here the phase
factor $\po(\q,k\,;\fr{1}{2})$  was defined in \eqref{qex} and represents the
square root of the factor in \eqref{p55} (where we considered the full S-matrix
rather than ``half'' of the factorised S-matrix).

The choice of the phase factor \eqref{p55} ensures that
$\hat{K}_{9,10}=1+\ord{k^{-3}}$ to one-loop order. This choice is convenient
for comparing to the quantum-deformed S-matrix. From the expanded S-matrix in
appendix \ref{55expansion} we see that  like in  the reduced \adss{2} case
(but unlike the reduced \adss{3} case) a phase factor can be extracted such
that both the amplitudes of  $Y_m Z_n \ra Y_m Z_n$ and $\z_m \ch_n \ra \z_m
\ch_n$ scattering processes and $K_{9,10}$ are all equal to 1.

The \adss{2} and \adss{5} superstring sigma models are of the same type being 
based on the supergroup 
\be
PSU(n,n|2n)\,, \ \ \ \ \ \ \   n=1, \ 2 \ .
\ee
It is thus natural to expect  that the S-matrices of the corresponding reduced
theories follow the same pattern, in particular, their phase factors are
similar. The phase factors that we extracted in these cases  in sections
\ref{secsmat22} and \ref{secollsg55} (given by the amplitude for the $Y_m Z_n
\ra Y_m Z_n$ and $\z_m \ch_n \ra \z_m \ch_n$ processes) were equal. We thus
conjecture that the phase factor of the reduced \adss{5} theory should be given
by the phase factor  \eqref{phase}  of the reduced \adss{2} theory, again with
$\De=\fr{\pi}{k}$  as in \rf{kde22}. Further justification for this choice is
presented in section \ref{bkcomp}, where we find  that this is precisely the
phase factor that follows from solving the conditions of unitarity and crossing
for the quantum-deformed S-matrix.

The one-loop perturbative S-matrix \eqref{Sparam}, \eqref{smat55fact},
\eqref{k110} satisfies the expected   unitarity, crossing and conjugation
relations, \eqref{unittens33}, \eqref{crosstens33}, \eqref{crossk},
\eqref{conggg} with the crossed S-matrix given above \eqref{crossk}.
Substituting the one-loop S-matrix \eqref{Sparam} and \eqref{k110} into the 
YBE, which has the same form as \eqref{rr22}, one finds that the result is
non-zero. This happened also for the bosonic models with a non-abelian symmetry 
$H$ \ci{ht2} where the r\^{o}le of the Yang-Baxter equation requires further
study. In the next  subsection we  will consider the closely related
quantum-deformed S-matrix \ci{bk,bcl} which by construction satisfies the
Yang-Baxter equation \eqref{rr22}.

Let us make two additional comments. In the purely bosonic theories, the
coupling $k$ (level of WZW theory) is generally shifted by a constant in
certain exact quantum relations. This shift is absent though in 2-d
supersymmetric models. There appears to be no shift of $k$ also in the 
reduced \adss{2} and \adss{3} theories discussed in sections \ref{seclag22}
and \ref{pr33}. The same  should be true also in the reduced \adss{5} theory.

In the reduced \adss{3} theory a quantum counterterm was required to restore
integrability, i.e. the satisfaction of the YBE  at one-loop order, see section
\ref{ybe33}. In the reduced \adss{5} theory counterterms are not  required  to
restore the group factorisation of the one-loop S-matrix\,\foot{This is not the
case for the bosonic $G/H=SO(5)/SO(4)$ theory discussed in \ci{gt2}. There the
quantum counterterms were required to restore group factorisation under $\so(4)
\cong \su(2)\oplus \su(2)$ at one-loop order.} and the similarity with the
quantum-deformed S-matrix of section \ref{bkcomp} suggests that indeed no
additional local counterterms should be present here.\,\foot{In appendix
\ref{olcfd} a particular functional determinant based on the group structure of
the reduced theories (section \ref{prgr}) is proposed. The contribution of this
functional determinant to the one-loop S-matrix vanishes in the reduced \adss{2}
and \adss{5} theories while giving the required correction \eqref{counterterm}
in the reduced \adss{3} theory.}

\subsection{Quantum-deformed $\mf{psu}(2|2) \ltimes \mbb{R}^3$\ symmetric
S-matrix\la{bkcomp}}

In \ci{bk} the fundamental R-matrix associated to the quantum deformation of the
universal enveloping algebra ($U_q$) of the centrally extended superalgebra
$\mf{psu}(2|2) \ltimes \mbb{R}^3$ was constructed. In appendix \ref{qdefl} we 
generalise to all orders the trigonometric relativistic classical limit
identified in \ci{bcl} that exhibited similarities to the tree-level S-matrix of
the reduced \adss{5} theory \ci{ht1}. The trigonometric relativistic limit 
corresponds to
(i) taking the global symmetry parameter $g$ of \ci{bk} to infinity and 
(ii) identifying the quantum deformation parameter $q$ with the coupling $k$ as
in \rf{quq}\,\foot{Such a parametrisation of the quantum deformation parameter
is familiar from quantum group structures in theories based on (deformations
of)  WZW  model (see, e.g., \ci{qd,hm1,hm2}).}
\be\la{qexp}
q=\exp\Big(-\fr{i\pi}{k}\Big)\,.
\ee
In this limit one of the central extensions vanishes leaving us with the
symmetry group \eqref{ads5symex} expected from the arguments of section
\ref{sumofsym} .

The quantum-deformed S-matrix in this limit takes the form 
{\allowdisplaybreaks
\begin{align}
\no&\mathcal{S}\ket{\phi_1\phi_1}=\big(J_1+J_2\big)\ket{\phi_1\phi_1}
\\\no&\mathcal{S}\ket{\phi_1\phi_2}=J_1\sec\fr{\pi}{k}\ket{\phi_1\phi_2}  
                +\big(J_2-iJ_1\tan\fr{\pi}{k}\big)\ket{\phi_2\phi_1}
                                -J_5\sec\fr{\pi}{k}\ket{\psi_3\psi_4}
+J_5(1+i\tan\fr{\pi}{k})\ket{\psi_4\psi_3}
\\\no&\mathcal{S}\ket{\phi_2\phi_1}=J_1\sec\fr{\pi}{k}\ket{\phi_2\phi_1}
+\big(J_2+iJ_1\tan\fr{\pi}{k}\big)\ket{\phi_1\phi_2}
                                -J_5\sec\fr{\pi}{k}\ket{\psi_4\psi_3}
+J_5(1-i\tan\fr{\pi}{k})\ket{\psi_3\psi_4}
\\\no&\mathcal{S}\ket{\phi_2\phi_2}=\big(J_1+J_2\big)\ket{\phi_2\phi_2}
\\&\mathcal{S}\ket{\psi_3\psi_3}=\big(J_3+J_4\big)\ket{\psi_3\psi_3}\la{rbk}
\\\no&\mathcal{S}\ket{\psi_3\psi_4}=J_3\sec\fr{\pi}{k}\ket{\psi_3\psi_4}
+\big(J_4-iJ_3\tan\fr{\pi}{k}\big)\ket{\psi_4\psi_3}
                                -J_6\sec\fr{\pi}{k}\ket{\phi_1\phi_2}
+J_6(1+i\tan\fr{\pi}{k})\ket{\phi_2\phi_1}
\\\no&\mathcal{S}\ket{\psi_4\psi_3}=J_3\sec\fr{\pi}{k}\ket{\psi_4\psi_3}
+\big(J_4+iJ_3\tan\fr{\pi}{k}\big)\ket{\psi_3\psi_4}
                                -J_6\sec\fr{\pi}{k}\ket{\phi_2\phi_1}
+J_6(1-i\tan\fr{\pi}{k})\ket{\phi_1\phi_2}
\\\no&\mathcal{S}\ket{\psi_4\psi_4}=\big(J_3+J_4\big)\ket{\psi_4\psi_4}
\\\no&\mathcal{S}\ket{\phi_a\psi_\bet}=J_7\;\de_a^d\de_\bet^\g\ket{
\psi_\g\phi_d}
+J_9\;\de_a^c\de_\bet^\de\ket{\phi_c\psi_\de}
\\\no&\mathcal{S}\ket{\psi_\al\phi_b}=J_8\;\de_\al^\de
                                           \de_b^c\ket{\phi_c\psi_\de}
                              +J_{10}\;\de_\al^\g\de_b^d\ket{\psi_\g\phi_d}\ ,
\end{align}
}
where 
{\allowdisplaybreaks
\begin{align}
\no
&J_{1,3}(\q,k)=P_0(\q,k)\cos\fr{\pi}{k}\sech\fr{\q}{2}
                             \cosh\Big(\fr{\q}{2}\pm\fr{i\pi}{2k}\Big)
\\\no&
J_{2,4}(\q,k)=\mp iP_0(\q,k)\Big[1-\cos\fr{\pi}{k}
+\cosh\q+\cosh\Big(\q\pm\fr{i\pi}{k}\Big)\Big]\sin\fr{\pi}{2k}\cosech\q
\\&\la{jfuncex}
J_{5,6}(\q,k)=-iP_0(\q,k)\cos\fr{\pi}{k}\sin\fr{\pi}{2k}
\sech\fr {\q}{2}
\\\no&
J_{7,8}(\q,k)=-iP_0(\q,k)\sin\fr{\pi}{2k}\cosech\fr{\q}{2}
\\\no&
J_{9,10}(\q,k)=P_0(\q,k)
\end{align}
}
The functions $J_i$  do not parametrise the quantum-deformed S-matrix in the
same way as the functions $K_i$ in  \eqref{k110} parametrise the perturbative
S-matrix given in  \eqref{Sparam}: there is an additional dependence on
$q=\exp(-\fr{i\pi}{k})$.  Consequently, the  $[SU(2)]^2$  global symmetry  is 
broken to  $[U(1)]^2$. The quantum-deformed  S-matrix satisfies the Yang-Baxter 
equation. 

Extracting  the phase factor $P_0(\q,k)$, 
\be  
J_i(\q,k) = P_0(\q,k) \ \hat{J}_i(\q,k)  \ , 
\ee
the $1/k$  expansion of the functions $\hat{J}_i(\q,k)$  is given, to
``one-loop'' order,  by
\be\la{j}\begin{split}
\hat{J}_1(\q,k)=&\hat{J}_3(\q,-k)=1 + \fr{i\pi}{2k}\tanh\fr{\q}{2}
                     -\fr{5\pi^2}{8k^2}
+\ord{k^{-3}}
\\
\hat{J}_2(\q,k)=&\hat{J}_4(\q,-k)=-\fr{i\pi}{k}\coth\q
+\fr{\pi^2}{4k^2}
                     +\ord{k^{-3}}
\\
\hat{J}_5(\q,k)=&-\hat{J}_6(\q,-k)=-\fr{i\pi}{2k}\sech\fr{\q}{2}
                     +\ord{k^{-3}}
\\
\hat{J}_7(\q,k)=&-\hat{J}_8(\q,-k)=-\fr{i\pi}{2k}\cosech\fr{\q}{2}
                     +\ord{k^{-3}}
\\
\hat{J}_9(\q,k)=&\hat{J}_{10}(\q,-k)=1                  
+\ord{k^{-3}}\ . 
\end{split}\ee
There is a strong similarity with \eqref{k110}. However, $K_{1,2,3,4}$ contain
some extra $\q$-dependent terms. The functions $J_i$ in \eqref{jfuncex} do not
satisfy the classical crossing symmetry relations obeyed  by $K_i$
\eqref{crossk}, rather they satisfy a set quantum-deformed relations
\eqref{crossj} given below.

\subsubsection{Phase factor}

To facilitate comparison with the one-loop S-matrix of sections \ref{secollsg55}
and \ref{secfac}, the phase factor $P_0$  has been chosen such that
$\hat{J}_9=\hat{J}_{10}=1$. This phase factor is fixed by the physical
requirements of unitarity, crossing and matrix unitarity \ci{bk} (see appendix
\ref{qdefl}). We give these relations in terms of the tensor function
$\Sq_{AB}^{CD}(\q,k)$ defined by 
\be
\mc{S}\ket{\Phi_A(p_1)\Phi_B(p_2)}=\Sq_{AB}^{CD}(\q,k)
\ket{\Phi_C(p_1) \Phi_D(p_2)}\,.
\ee
Unitarity implies (see  \eqref{unit})
\be\la{unittens}
(-1)^{[ C][ D] +[ E][ F]}\Sq_{AB}^{CD}(\q,k)\ 
\Sq_{DC}^{FE}(-\q,k)=
\de_{A}^E\de_B^F\,.
\ee
Substituting the quantum-deformed S-matrix \eqref{rbk}, \eqref{jfuncex} into
\eqref{unittens} gives 
\be\la{p0unit}
P_0(\q,k) \ P_0(-\q,k)=
\fr{\sinh^2\fr{\q}{2}}{\sinh^2\fr{\q}{2}+\sin^2\fr{\pi}{2k}}\,.
\ee
To formulate the crossing constraint let us introduce the crossed
quantum-deformed S-matrix $\bar{\Sq}_{AB}^{CD}(\q,k)$ which is identical to
$\Sq_{AB}^{CD}(\q,k)$ except with $(J_5,\,J_6,\,J_7,\,J_8)$ replaced by
$i(J_5,\,J_6,\,J_7,\,J_8)$. Then the crossing symmetry condition \eqref{crossss}
implies
\be\sum_{C,D,G,H=1}^4 \Sq_{AB}^{CD}(\q,k)\ C_C^G\ \bar{\Sq}_{HD}^{GF}(i\pi+\q,k)
\ C^{-1}{}_H^E=\de_A^E\de_B^F\,.
\ee 
In view of  the unitarity relation \eqref{unittens} this can be rewritten in the
usual form
\be 
\Sq_{AB}^{CD}(\q,k) = \sum_{E,F=1}^4 (-1)^{[ A][ B] + [ C] [ D] }\ C_B^E \
\bar{\Sq}_{FA}^{EC}
(i\pi-\q,k) \ C^{-1}{}_F^D\,.
\ee
The charge conjugation matrix $C_A^B$ is defined by 
\be\begin{split}
&\mc{C}\ket{\Phi_A}=C_A^B\ket{\Phi_B}\,,
\\\mc{C}\ket{\phi_1}=-q^{\fr{1}{2}}\ket{\phi_2}\,,\hs{20pt}
&\mc{C}\ket{\psi_3}=-q^{\fr{1}{2}}\ket{\psi_4}\,,\hs{20pt}
\mc{C}\ket{\phi_2}=q^{-\fr{1}{2}}\ket{\phi_1}\,,\hs{20pt}
\mc{C}\ket{\psi_4}=q^{-\fr{1}{2}}\ket{\psi_3}\,,
\end{split}
\ee
and similarly for $C^{-1}{}_A^B$. Substituting the quantum-deformed S-matrix
\eqref{rbk}, \eqref{jfuncex} into \eqref{unittens} gives the crossing relation
for the phase factor
\be
P_0(i\pi-\q,k)=P_0(\q,k)\,.
\ee
The crossing symmetry also requires the following relations between the
functions $J_i$,
\be\la{crossj}\begin{split}
&J_1(i\pi-\q,k)=\cos\fr{\pi}{k}\big[J_1(\q,k)+J_2(\q,k)\big]\,,
\hs{20pt}J_2(i\pi-\q,k)=-\cos\fr{\pi}{k}\big[  J_2(\q,k)
                -\tan^2\fr{\pi}{k}J_1(\q,k)\big]
\\&J_5(i\pi-\q,k)= i \cos\fr{\pi}{k}J_7(\q,k)\,,
\hs{69pt}J_7(i\pi-\q,k)=- i \sec\fr{\pi}{k}J_5(\q,k)
\\&\hs{170pt}J_9(i\pi-\q,k)=J_9(\q,k)\,,
\end{split}\ee
and similarly for $J_3,\,J_4,\,J_6,\,J_8,\,J_{10}$. The conjugation relations
\be\la{congj}\begin{split}
J_1(\q,k)=J_1^*(-\q,k)\,,&\hs{20pt}J_3(\q,k)=J_3^*(-\q,k)\,,
\\J_2(\q,k)=J_2^*(-\q,k)\,,&\hs{20pt}J_4(\q,k)=J_4^*(-\q,k)\,,
\\J_5(\q,k)=-J_5^*(-\q,k)\,,&\hs{20pt}J_6(\q,k)=-J_6^*(-\q,k)\,,
\\J_7(\q,k)=J_8^*(-\q,k)\,,&\hs{20pt}J_9(\q,k)=J_{10}^*(-\q,k)\,.
\end{split}\ee
still hold as they did for the functions $K_i$ \eqref{conggg} as long as
the phase factor satisfies 
\be
P_0(\q,k)=P_0^*(-\q,k)\,.
\ee

To summarize, the trigonometric relativistic quantum-deformed S-matrix
\eqref{rbk}, \eqref{jfuncex} is consistent with unitarity and the 
quantum-deformed crossing symmetry provided the phase factor $P_0(\q,k)$
satisfies  the following constraints 
\be\begin{split}\la{phaserel55}
P_0(\q,k)P_0(-\q,&k)=
\fr{\sinh^2\fr{\q}{2}}{\sinh^2\fr{\q}{2}+\sin^2\fr{\pi}{2k}}\,,
\\
P_0(i\pi-\q,k)=P_0(\q,k)\,,&\hs{50pt}
P_0(\q,k)=P_0^*(-\q,k)\,.
\end{split}\ee

In section \ref{secfac} $\Ph(\q,\fr{\pi}{k})$ in \eqref{phase} was conjectured
to be a candidate for the phase factor of the reduced \adss{5} theory based on
the one-loop computation and group-theory arguments. Here  we are considering
the factorised S-matrix so that the corresponding phase factor $P_0$ is the
square root of $\Ph$, i.e.
\be\begin{split}\la{phaseqd55}
P_0(\q,k)=& \sqrt{\Ph(\q,\fr{\pi}{k})} = \sqrt{\fr{\sinh\q
+i\sin\fr{\pi}{k}}{\sinh\q-i\sin\fr{\pi}{k}}}\ 
R(\q,\fr{\pi}{k})\,,\hs{20pt}
R(\q,\De)= \fr{\sinh\q-i\sin\De}{\sinh\q+i\sin\De}\ Y(\q,\De)\ 
Y(i\pi-\q,\De)\,,
\\Y(\q,\De)=&\prod_{l=1}^{\infty}
\fr{\Gamma\big(\fr{\De}{2\pi}-\fr{i\q}{2\pi}+l\big)
    \Gamma\big(-\fr{\De}{2\pi}-\fr{i\q}{2\pi}+l-1\big)
    \Gamma\big(-\fr{i\q}{2\pi}+l-\fr{1}{2}\big)
    \Gamma\big(-\fr{i\q}{2\pi}+l+\fr{1}{2}\big)}
 {\Gamma\big(\fr{\De}{2\pi}-\fr{i\q}{2\pi}+l+\fr{1}{2}\big)
  \Gamma\big(-\fr{\De}{2\pi}-\fr{i\q}{2\pi}+l-\fr{1}{2}\big)
    \Gamma\big(-\fr{i\q}{2\pi}+l-1\big)
    \Gamma\big(-\fr{i\q}{2\pi}+l\big)}\,, \ \ \ \De(k)=\fr{\pi}{k}\,.
\end{split}\ee
From the relations \eqref{phaserel} we see that this phase factor satisfies the
unitarity, crossing and conjugation relations \eqref{phaserel55}.

The small $\q$ expansion of the phase factor \eqref{phaseqd55} is
\be
P_0(\q,k) = -\fr{\q}{2\sin\fr{\pi}{2k}}\trm{sign}(\sin\fr{\pi}{k})+\ord{\q^2}\,,
\ee 
implying the initial condition for the quantum-deformed S-matrix (for
$k>0$)
\be
\Sq_{AB}^{CD}(0,k) = i(-1)^{[C][D]}\de_A^D \de_B^C
\;\trm{sign}(\sin\fr{\pi}{k})\,.
\ee
The quantum-deformed S-matrix \eqref{rbk}, \eqref{jfuncex} satisfies the 
graded Yang-Baxter equation \eqref{rr22} by construction \ci{bk}. We have also
checked this  explicitly.

The quantum-deformed S-matrix \eqref{rbk}, \eqref{jfuncex} along with the phase
factor \eqref{phaseqd55} is thus a candidate for the physical S-matrix for the
perturbative excitations of the reduced \adss{5} theory. There are many
additional properties of this S-matrix to investigate, for example, the pole
structure.

\subsubsection{Quantum-deformed symmetry\la{symmetriess}}

Let us  review the action of the quantum-deformed symmetry \ci{bk} on the
S-matrix \eqref{rbk}. The symmetry algebra is the quantum deformation of the
universal enveloping algebra, $U_q(\mf{psu}(2|2) \ltimes \mbb{R}^3)$. The
generators of $\mf{psu}(2|2) \ltimes \mbb{R}^3$ are:\,\foot{Compared to the
notation of \ci{bk} we have renamed $\mf{S} \lra \mf{Q}$ and $\mf{K},\mf{P} \ra
-\ppn,\pmn$.}
$\mf{R}_a^{\;\,b}$ -- generators of one bosonic $\su(2)$;
$\mf{L}_\al^{\;\,\bet}$ -- generators of the second bosonic $\su(2)$; 
$\mf{Q}_a^{\;\,\bet}$ -- one set of four fermionic generators mixing the two
$\su(2)$s;
$\mf{S}_\al^{\;\,b}$ -- the  second set of four fermionic generators;
$\mf{C},\,\ppmn$ -- the three central charges.
The non-trivial (anti-)commutation relations of these generators are
\be\begin{split}\la{commmm}
\com{\mf{R}_a^{\;\,b}}{\mf{R}_c^{\;\,d}}
=\de_c^b \mf{R}_a^{\;\,d}-\de_a^d \mf{R}_c^{\;\,d}\,,
&\hs{20pt}
\com{\mf{L}_\al^{\;\,\bet}}{\mf{L}_\g^{\;\,\de}}
=\de_\g^\bet \mf{L}_\al^{\;\,\de}-\de_\al^\de \mf{L}_\g^{\;\,\de}\,,
\\\com{\mf{R}_a^{\;\,b}}{\mf{Q}_c^{\;\,\de}}
=\de_c^b \mf{Q}_a^{\;\,\de} -\fr{1}{2}\de_a^b \mf{Q}_c^{\;\,\de}\ , 
&\hs{20pt}
\com{\mf{L}_\al^{\;\,\bet}}{\mf{Q}_c^{\;\,\de}}
=-\de_\al^\de \mf{Q}_c^{\;\,\bet} +\fr{1}{2}\de_\al^\bet \mf{Q}_c^{\;\,\de}\ , 
\\\com{\mf{R}_a^{\;\,b}}{\mf{S}_\g^{\;\,d}}
=-\de_a^d \mf{S}_\g^{\;\,b} +\fr{1}{2}\de_a^b \mf{S}_\g^{\;\,d}\ , 
&\hs{20pt}
\com{\mf{L}_\al^{\;\,\bet}}{\mf{S}_\g^{\;\,d}}
=\de_\g^\bet \mf{S}_\al^{\;\,d} -\fr{1}{2}\de_\al^\bet \mf{S}_\g^{\;\,d}\ 
\\\acom{\mf{S}_\al^{\;\,b}}{\mf{Q}_c^{\;\,\de}}
=\de_c^b& \mf{L}_\al^{\,\;\de}+\de_\al^\de \mf{R}_c^{\;\,b} +
\de_c^b\de_\al^\de\mf{C}\ , 
\\\acom{\mf{Q}_a^{\,\;\bet}}{\mf{Q}_c^{\;\,\de}}
=-\e_{ac}\e^{\bet\de}\ppn \ , 
&\hs{20pt}
\acom{ \mf{S}_\al^{\,\;b}}{ \mf{S}_\g^{\;\,d}}
=\e_{\al\g}\e^{bd}\pmn \ .
\end{split}\ee
The universal enveloping algebra, $U(\mf{psu}(2|2) \ltimes \mbb{R}^3)$ is
generated by polynomials of the Lie algebra generators. A minimal set of its
generators can be taken as follows\,\foot{Not all the generators of the Lie
algebra need to be kept explicitly as certain generators can be rewritten as
polynomials of other generators.}
\be\begin{array}{ccc}
\mf{H}_1 = \mf{R}_2^{\;\,2}-\mf{R}_1^{\;\,1}=2\mf{R}_2^{\;\,2}
\,,&\mf{E}_1 =\mf{R}_2^{\;\,1}\,,&\mf{F}_1=\mf{R}_1^{\;\,2}\,,
\\\mf{H}_2 = -\mf{C}-\fr{1}{2}\mf{H}_1-\fr{1}{2}\mf{H}_3
\,,&\mf{E}_2 =\mf{S}_4^{\;\,2}\,,&\mf{F}_2=\mf{Q}_2^{\;\,4}\,,
\\\mf{H}_3 = \mf{L}_4^{\;\,4}-\mf{L}_3^{\;\,3}=2\mf{L}_4^{\;\,4}
\,,&\mf{E}_3 =\mf{L}_4^{\;\,3}\,,&\mf{F}_3=\mf{L}_3^{\;\,4}\,.
\end{array}\ee
In this basis the symmetric Cartan matrix is
\be
A_{jk}=\left(\ba{ccc}2&-1&0\\-1&0&1\\0&1&-2\ea\right)
\ee
and the commutation relations with $\mf{H}_j$ are
\be
\com{\mf{H}_j}{\mf{H}_k}=0\,,\hs{20pt}
\com{\mf{H}_j}{\mf{E}_k}=A_{jk} \mf{E}_k\,,\hs{20pt}
\com{\mf{H}_j}{\mf{F}_k}=-A_{jk}\mf{F}_k\,.
\ee
The non-vanishing (anti-)commutators between $\mf{E}_j$ and $\mf{F}_k$ are
\be
\com{\mf{E}_1}{\mf{F}_1}=\mf{H}_1\,,\hs{20pt}
\acom{\mf{E}_2}{\mf{F}_2}=-\mf{H}_2\,,\hs{20pt}
\com{\mf{E}_3}{\mf{F}_3}=-\mf{H}_3\,.
\ee
The Serre relations, which we will not give here, are discussed in \ci{bk}.

In the quantum deformation, the commutation relations with $\mf{H}_j$ stay as
they are. The three non-vanishing (anti)-commutators between $\mf{E}_j$ and
$\mf{F}_k$ are deformed to
\be
\com{\mf{E}_1}{\mf{F}_1}=\bll \mf{H}_1\brr _q\,,\hs{20pt}
\acom{\mf{E}_2}{\mf{F}_2}=-\bll \mf{H}_2\brr _q\,,\hs{20pt}
\com{\mf{E}_3}{\mf{F}_3}=-\bll \mf{H}_3\brr _q\,.
\ee
Here again $\bll x\brr _q\equiv \fr{q^x -q^{-x}}{q-q^{-1}}$ and the quantum
deformation parameter $q$ is related to $k$ by \eqref{qexp}. The Serre relations
similarly become quantum-deformed, see \ci{bk}.

The quantum-deformed generators have the following action on the one-particle
states $\ket{\phi_a}$, $\ket{\psi_\al}$\,\foot{In appendix \ref{qdefl} the
notation $C_i$ is used in considering two particle states. It stands for the
function $C(\vt,k)$ evaluated with the rapidity $\vt_i$ of the first or the 
second particle.}
\small
\be\begin{array}{llll}\no
  \mf{H}_1\ket{\phi_1}=-\ket{\phi_1}\,,
& \mf{H}_1\ket{\phi_2}=\ket{\phi_2}\,,
& \mf{H}_1\ket{\psi_3}=0\,,
& \mf{H}_1\ket{\psi_4}=0\,,
\\\mf{H}_2\ket{\phi_1}=-(C-\fr{1}{2})\ket{\phi_1}\,,
& \mf{H}_2\ket{\phi_2}=-(C+\fr{1}{2})\ket{\phi_2}\,,
& \mf{H}_2\ket{\psi_3}=-(C-\fr{1}{2})\ket{\psi_3}\,,
& \mf{H}_2\ket{\psi_4}=-(C+\fr{1}{2})\ket{\psi_4}\,,
\\\mf{H}_3\ket{\phi_1}=0\,,
& \mf{H}_3\ket{\phi_2}=0\,,
& \mf{H}_3\ket{\psi_3}=-\ket{\psi_3}\,,
& \mf{H}_3\ket{\psi_4}=\ket{\psi_4}\,,
\\\mf{E}_1\ket{\phi_1}=q^{\fr{1}{2}}\ket{\phi_2}\,,
& \mf{E}_1\ket{\phi_2}=0\,,
& \mf{E}_1\ket{\psi_3}=0\,,
& \mf{E}_1\ket{\psi_4}=0\,,
\\\mf{E}_2\ket{\phi_1}=0\,,
& \mf{E}_2\ket{\phi_2}=a(\vt,k)\ket{\psi_4}\,,
& \mf{E}_2\ket{\psi_3}=b(\vt,k)\ket{\phi_1}\,,
& \mf{E}_2\ket{\psi_4}=0\,,
\\\mf{E}_3\ket{\phi_1}=0\,,
& \mf{E}_3\ket{\phi_2}=0\,,
& \mf{E}_3\ket{\psi_3}=0\,,
& \mf{E}_3\ket{\psi_4}=q^{-\fr{1}{2}}\ket{\psi_3}\,,
\\\mf{F}_1\ket{\phi_1}=0\,,
& \mf{F}_1\ket{\phi_2}=q^{-\fr{1}{2}}\ket{\phi_1}\,,
& \mf{F}_1\ket{\psi_3}=0\,,
& \mf{F}_1\ket{\psi_4}=0\,,
\\\mf{F}_2\ket{\phi_1}=c(\vt,k)\ket{\psi_3}\,,
& \mf{F}_2\ket{\phi_2}=0\,,
& \mf{F}_2\ket{\psi_3}=0\,,
& \mf{F}_2\ket{\psi_4}=d(\vt,k)\ket{\phi_1}\,,
\\\mf{F}_3\ket{\phi_1}=0\,,
& \mf{F}_3\ket{\phi_2}=0\,,
& \mf{F}_3\ket{\psi_3}=q^{\fr{1}{2}}\ket{\psi_4}\,,
& \mf{F}_3\ket{\psi_4}=0\,,
\end{array}\ee
\vs{-9pt}
\be\begin{array}{ll}
\mf{C}\ket{\Phi_A} =C(\vt,k)\ket{\Phi_A}\,,\  \ \ \ \  \  \ \ \ \  
&\ppmn\ket{\Phi_A} =\ppmne(\vt,k)\ket{\Phi_A}\,.
\end{array}\ee
\normalsize
The trigonometric relativistic limit of the R-matrix in \ci{bk} is found by
taking $g\ra \infty$. Taking this limit in the functions
$a,\,b,\,c,\,d,\,\ppmne$ and $C$ given in \ci{bk} leads to similar relations as
in\foot{The $g \ra \infty$ limit of $b,\,d,\,\ppmne$ in \ci{bk} is slightly
technical. The simplest way to compute them is to use $a$ and $c$ in the limit
$g\ra \infty$ and the relations coming from the closure of the supersymmetry
algebra on one-particle states. Alternatively, one can expand out the
expressions for $b,\,d,\,\ppmne$ as a power series in $g^{-1}$. For this one
needs the next-to-leading order corrections in $g^{-1}$ for $q^C,\, x^-$ and
$x^+$.} \rf{bbbb},\rf{pep}
\be\la{k}
a(\vt,k)=\sqrt{\fr{1}{2}\sec\fr{\pi}{2k}}\ 
           e^{-\fr{\vt}{2}-\fr{i\pi}{4k}}\,, \ \ \ \ \ 
b(\vt,k) = a^*(\vt,k) \ , \ \ \ \ 
c(\vt,k) = -e^{\vt}  \ a(\vt,k) \ , \ \ \ \ 
d(\vt,k) = e^{\vt}  \ a^*(\vt,k) \ , \ \ \ \ 
\ee
\be\ppmne(\vt,k)=\fr{1}{2}\sec\fr{\pi}{2k}\ 
            e^{\pm\vt}\,,
\hs{20pt} C(\vt ,k)=0\,. \la{kk}
\ee

The  vanishing of $C$ is consistent with \eqref{qcu} and confirms the claim
that the third central extension vanishes and \eqref{ads5symex} is left as the
symmetry algebra. The functions $a,b,c,d$ satisfy the four relations required
for the closure of the supersymmetry algebra,
\be
ad=\big[C+\fr{1}{2}\big]_q\,,\hs{20pt}
bc=\big[C-\fr{1}{2}\big]_q\,,\hs{20pt}
ab=\pmne\,,\hs{20pt}
cd=-\ppne\,.\ee
Since $\ppmne \sim e^{\pm\vt}$  we may again interpret $\ppmn$ as the lightcone
momentum symmetry generators up to a  normalisation. Once again, the resulting
symmetry algebra strongly resembles a 2-d supersymmetry algebra with a global
bosonic R-symmetry. The supersymmetry generators are charged under both the
Lorentz group (they anticommute to 2-d momentum) and the global bosonic symmetry
group. The existence of the global bosonic R-symmetry appears to quantum-deform
the supersymmetry. Unlike the reduced \adss{3} theory here the bosonic symmetry
is non-abelian and therefore is non-trivially altered by the quantum
deformation.

To define the action of these symmetries on the two particle states, the
coproduct is required. In \ci{bk} an additional braiding factor, $\mf{U}$, was 
introduced in the coproduct. In the $g\ra \infty$ limit the braiding factor
becomes trivial, see \eqref{qcu}. The usual quantum-deformed coproduct is
\begin{align}
&\De(\mf{H}_i)=\mf{H}_i\otimes \mbb{I} +  \mbb{I}\otimes \mf{H}_i\,,
&&\De(\mf{C})=\mf{C}\otimes \mbb{I} + \mbb{I} \otimes \mf{C}\,,\no
\\&\De(\mf{E}_i)=\mf{E}_i\otimes \mbb{I} + q^{-\mf{H}_i} \otimes \mf{E}_i\,,
&&\De(\pmn)=\pmn\otimes \mbb{I} + q^{2\mf{C}} \otimes \pmn\,,
\\&\De(\mf{F}_i)=\mf{F}_i\otimes q^{\mf{H}_i} + \mbb{I} \otimes \mf{F}_i\,,
&&\De(\ppn)=\ppn\otimes q^{-2\mf{C}} +\mbb{I} \otimes \ppn\,.\no
\end{align}
The quantum S-matrix \eqref{rbk} satisfies the co-commutativity relation with
the symmetry generators (which has  same form  as in \rf{ops})
\be
\De_{\trm{op}}(\mf{J})\ \mc{S} = \mc{S} \ \De(\mf{J})\,.
\ee

\subsection{Relating the perturbative S-matrix and the quantum-deformed  
S-matrix}

Let us now  discuss  a  possible  relation between the one-loop perturbative
S-matrix for the reduced \adss{5} theory in its factorised form given in 
section \ref{secfac}, and the quantum-deformed S-matrix of \ci{bk} in the
relativistic trigonometric limit \eqref{rbk}, \eqref{jfuncex}. 

Let us look at a particular sector of the quantum-deformed S-matrix,
\eqref{rbk}, \eqref{j} at leading (tree-level) order in the $1/k$
expansion\,\foot{Note that the phase factor $P_0=1+\ord{k^{-2}}$  and thus it
plays no r\^{o}le at the tree level.} 
\be \la{trunc}\ba{ccc}
  &\ba{cccc} \ket{\phi_1\phi_1}\hs{8pt}&\hs{8pt} \ket{\phi_1\phi_2}
\hs{18pt}&\hs{18pt} \ket{\phi_2\phi_1} \hs{8pt}&\hs{8pt}\ket{\phi_2\phi_2}
\ea &
\\  &  &
\\  S_q^{\trm{(tree)}} =  &    \left(\ba{cccc}
     1 & 0 &0 &0    
  \\ 0 & 1 + \fr{i\pi}{2k}\tanh\fr{\q}{2} &-\fr{i\pi}{k}\coth\q+\fr{i\pi}{k}&0
  \\ 0 &-\fr{i\pi}{k}\coth\q-\fr{i\pi}{k}  & 1 +\fr{i\pi}{2k}\tanh\fr{\q}{2}&0
  \\ 0 &0 &0 &1
\ea\right)
& \ba{c}\ket{\phi_1\phi_1} \\ \ket{\phi_1\phi_2}  \\ \ket{\phi_2\phi_1}\\
\ket{\phi_2\phi_2}  \ea
\ea
\ee
Compared to the tree-level terms in the perturbative S-matrix \eqref{k110}
here the off-diagonal entries  contain extra  $\pm \fr{i\pi}{k}$ terms. These
corrections obey the conjugation relations \eqref{conggg} and thus it appears
that one does not need to alter the standard commutation relations of creation
operators. It is unlikely that such  a correction  may  come from a standard
local quartic interaction  term (of the type that we considered in \ci{ht2} and
section \ref{ybe33}). 

Adding local counterterms and quantum deforming a symmetry appear to be two
disconnected possibilities. Counterterms may restore some properties of
integrable theories such as group factorisation. The Yang-Baxter equation, 
however, has a tree-level anomaly if the global symmetry $H$ is non-abelian and
that anomaly cannot be removed by adding a local counterterm. This anomaly can
be avoided  via a quantum deformation of the symmetry algebra, but how this
could happen in a Lagrangian formulation is not clear at the moment.

To relate the two S-matrices we may try an alternative approach -- to find a
non-unitary rotation of the deformed S-matrix  that restores the classical
$[SU(2)]^2$ symmetry  and maps it  into  the perturbative S-matrix. Below we
find a rotation which does this at tree-level. The rotated S-matrix does not
satisfy the unitarity relation, \eqref{unittens33} or crossing symmetry
\eqref{crosstens33} (the crossed S-matrix for the reduced \adss{5} theory is
given above \eqref{crossk}) implying this agreement does  not extend beyond the 
tree level. It is unclear if it is possible to generalise the rotation matrix in
such a way  that the  rotated S-matrix satisfies these physical requirements and
agrees with one-loop factorised S-matrix of section \ref{secfac}.

Whether the quantum-deformed S-matrix or its rotation is actually the physical
S-matrix of the perturbative excitations of this reduced theory is an open
question.

\subsubsection{Tree-level transformation}

To identify a  rotation matrix that restores the classical $[SU(2)]^2$ symmetry
when applied to  the quantum-deformed S-matrix  we consider the latter as a
matrix acting on the space of two-particle states.
{ \small 
\be\la{rotrr} 
\ba{c}
 \hs{-40pt}\ba{ccc}
  &\ba{cccc} \hs{1pt}\ket{\phi_1\phi_1} &  \hs{1pt}\ket{\phi_2\phi_2} &
             \hs{1pt}  \ket{\psi_3\psi_3} &\hs{1pt}  \ket{\psi_4\psi_4}\ea &
\\  &  &
\\  \Sq_1 =  & \left(\ba{cccc}
            J_1+J_2 &   0   &   0   &   0
          \\  0   & J_1+J_2 &   0   &   0
          \\  0   &   0   & J_3+J_4 &   0
          \\  0   &   0   &   0   & J_3+J_4 \ea\right)
& \ba{c}\ket{\phi_1\phi_2} \\ \ket{\phi_2\phi_1}  \\
                \ket{\psi_3\psi_4} \\ \ket{\psi_4\psi_3} \ea
\ea
\hs{15pt}
\ba{ccc}
  &\ba{cc} \ket{\phi_a\psi_\al} & \ket{\psi_\al\phi_a}\ea &
\\  &  &
\\  \Sq_3 =  &    \left(\ba{cc}
                J_9 & J_8
          \\    J_7 & J_{10} \ea\right)
& \ba{c}\ket{\phi_a\psi_\al}  \\ \ket{\psi_\al\phi_a}  \ea
\ea
\\
\\
\\
\\ \hs{-40pt}\ba{ccc}
  &\ba{cccc}\hs{18pt}\ket{\phi_1\phi_2} &\hs{64pt} \ket{\phi_2\phi_1}&
             \hs{64pt}\ket{\psi_3\psi_4} &\hs{64pt} \ket{\psi_4\psi_3}  \ea&
\\  &  &
\\  \Sq_2 =  &  \sec\fr{\pi}{k}  \left(\ba{cccc}
                J_1  &   (J_2+J_1) \cos \fr{\pi}{k} - e^{-\fr{i\pi}{k}}J_1
               &  -J_6  &    e^{-\fr{i\pi}{k}}J_6
          \\    (J_2+J_1) \cos \fr{\pi}{k} - e^{\fr{i\pi}{k}}J_1& J_1 
               &  e^{\fr{i\pi}{k}}J_6  &  -J_6
          \\    -J_5  &   e^{-\fr{i\pi}{k}}J_5   
               &  J_3 &  (J_4+J_3) \cos \fr{\pi}{k} - e^{-\fr{i\pi}{k}}J_3
          \\    e^{\fr{i\pi}{k}}J_5   &    -J_5   
               &   (J_4+J_3) \cos \fr{\pi}{k} - e^{\fr{i\pi}{k}}J_3   & J_3
\ea\right)
& \ba{c}\ket{\phi_1\phi_2} \\ \ket{\phi_2\phi_1} \\
                \ket{\psi_3\psi_4} \\ \ket{\psi_4\psi_3}  \ea 
\ea
\ea\ee
}
Rotating the quantum-deformed S-matrix with the following transformation
matrices
{\be
\ba{c}
U_1=\left(\ba{cccc}
  1&0&0&0
\\0&1&0&0
\\0&0&1&0
\\0&0&0&1\ea\right)\ , 
\hs{40pt}
U_3=\left(\ba{cccc}
  1&0
\\0&1
\ea\right)
\\
\\
\\
U_2=\sqrt{\sec\fr{\pi}{k}} 
\left(\ba{cccc}
  \cos\fr{\pi}{2k}&i\sin\fr{\pi}{2k}&0&0
\\-i\sin\fr{\pi}{2k}&\cos\fr{\pi}{2k}&0&0
\\0&0&\cos\fr{\pi}{2k}&i\sin\fr{\pi}{2k}
\\0&0&-i\sin\fr{\pi}{2k}&\cos\fr{\pi}{2k}\ea\right)
\ea
\ee}
gives
{\small  
\be\la{rotr} 
\ba{c}
 \hs{-10pt}\ba{ccc}
  &\ba{cccc} \hs{1pt}\ket{\phi_1\phi_1} &  \hs{1pt}\ket{\phi_2\phi_2} &
             \hs{1pt}  \ket{\psi_3\psi_3} &\hs{1pt}  \ket{\psi_4\psi_4}\ea &
\\  &  &
\\  U_1^\dagger \Sq_1 U_1 =  & \left(\ba{cccc}
            J_1+J_2 &   0   &   0   &   0
          \\  0   & J_1+J_2 &   0   &   0
          \\  0   &   0   & J_3+J_4 &   0
          \\  0   &   0   &   0   & J_3+J_4 \ea\right)
& \ba{c}\ket{\phi_1\phi_2} \\ \ket{\phi_2\phi_1}  \\
                \ket{\psi_3\psi_4} \\ \ket{\psi_4\psi_3} \ea
\ea
\hs{10pt}
\ba{ccc}
  &\ba{cc} \ket{\phi_a\psi_\al} & \ket{\psi_\al\phi_a}
\ea &
\\  &  &
\\  U_3^\dagger \Sq_3 U_3 =  &    \left(\ba{cc}
                J_9 & J_8
          \\    J_7 & J_{10} \ea\right)
& \ba{c}\ket{\phi_a\psi_\al}  \\ \ket{\psi_\al\phi_a}  \ea
\ea
\\
\\ \\
\ba{ccc}
  &\ba{cccc}\hs{15pt}\ket{\phi_1\phi_2} & \ket{\phi_2\phi_1}&
             \ket{\psi_3\psi_4} & \ket{\psi_4\psi_3}  \ea&
\\  &  &
\\   U_2^\dagger \Sq_2 U_2=  & 
\left(\ba{cccc}
                J_1  &   J_2
               &  -J_6  &    J_6
          \\    J_2 & J_1 
               &  J_6  &  -J_6
          \\    -J_5   &   J_5
               &  J_3 &  J_4
          \\    J_5  &    -J_5
               &   J_4  & J_3
\ea\right)
& \ba{c}\ket{\phi_1\phi_2} \\ \ket{\phi_2\phi_1} \\
                \ket{\psi_3\psi_4} \\ \ket{\psi_4\psi_3}  \ea
\ea \ea \ee
}
The rotation clearly transforms the quantum-deformed S-matrix such that it 
becomes  invariant under the classical symmetry group $[SU(2)]^2$.

The functions $J_i$ parametrise the rotated S-matrix \eqref{rotr} in the same
way the functions $K_i$ parametrised the one-loop perturbative S-matrix of
section \ref{secfac}, \eqref{smat55fact}, \eqref{Sparam}. Furthermore, the
leading terms  in the expansion of the functions $J_i$  and $K_i$ match  (see
\eqref{k110} and \eqref{j}), i.e. the rotated S-matrix \eqref{rotr} matches 
the perturbative S-matrix at the tree-level.

Note that because $U_2$ is not a unitary matrix there is no contradiction with
the fact that while the perturbative S-matrix does not satisfy the YBE at the 
tree-level, the quantum-deformed S-matrix does by construction.

\subsubsection{Beyond tree-level?}

Beyond tree-level the functions $J_i$ and $K_i$ disagree. This is expected as
the rotated S-matrix \eqref{rotr} does not satisfy the unitarity and crossing
relations \eqref{unittens33} and \eqref{crosstens33}. In particular, the
functions $J_i$ satisfy the crossing relations \eqref{crossj}, whereas the
crossing relations for $K_i$ are given by \eqref{crossk}.

It is natural to try to generalise the rotation matrices $U_k$ so that the
resulting rotated S-matrix still respects the classical group symmetry,
$[SU(2)]^2$, but also satisfies unitarity and crossing. Note that this will
introduce a rapidity dependence into the rotation matrices $U_k$. 

To get clues about how to do this one may try to find perturbative corrections
to the functions $J_i$ restoring unitarity and crossing symmetry order by order
in $\fr{1}{k}$. This procedure does not uniquely fix the corrections. The
simplest corrections that restore unitarity and crossing at one-loop are exactly
those that give the perturbative functions $K_i$ \eqref{k110}. At higher orders 
$J_5,\,J_6,\,J_7,\,J_8,\,J_9$ and $J_{10}$ all require corrections, implying
that both $U_1$ and $U_3$ will also be non-trivial.

To conclude, which transformation relates the perturbative $[SU(2)]^2$
invariant S-matrix to the quantum deformed S-matrix satisfying YBE and what is 
the origin of the quantum group symmetry remain central open questions. An
on-going  investigation of the integrable structure \ci{schm,hms} and the
solitons in similar theories \ci{hm1,hm2,hm3} may provide a deeper insight into
these issues.


\renewcommand{\theequation}{7.\arabic{equation}}
\setcounter{equation}{0}
\section{Concluding remarks\la{conc}}

In this paper we have studied the S-matrix for the perturbative 
(Lagrangian-field) excitations in a class of 2-d UV finite massive quantum 
field theories \rf{gwzw} which (at least at the classical level) may be 
interpreted as Pohlmeyer reductions of the \adss{n} superstring theories. 

We reviewed in detail the \adss{2} case where the reduced theory is equivalent
to the $\N=2$ supersymmetric sine-Gordon model and thus the exact S-matrix is
known (not only in the perturbative-excitation sector but also in the solitonic
sectors). Certain properties of this S-matrix (e.g., the structure of the phase 
factor) are shared by more complicated cases with $n > 2$ where also some new
features appear.

In the \adss{3} case (which may be viewed as a fermionic generalization of the 
sum of the complex sine-Gordon and complex sinh-Gordon models) we found that the
perturbative one-loop S-matrix derived directly from the Lagrangian formulation
requires a correction coming from a particular local counterterm in order to be
consistent with quantum integrability, i.e. to satisfy the Yang-Baxter equation.
It remains an interesting open question as to whether there is an alternative 
formulation to \rf{gwzw} of this theory (treating bosons and fermions in a more
symmetric way -- cf. fermionic gauge fields in supersymmetric gauged WZW theory
\ci{susygwzw,t1}) that automatically produces the required counterterm
\rf{counterterm} as happened in the complex sine-Gordon case \ci{ht2}. Such a
formulation may also make more manifest a (non-local) 2-d supersymmetry recently
observed at the level of the corresponding classical equations of motion
\ci{iv,hms} and also at the level of the Lagrangian \eqref{gwzw} in \ci{hms}.
In particular, non-localities like $(\dpl)^{-1}$ may be indicating that some
massless fields were solved for. 

Indeed, we discovered that at the  tree level the perturbative S-matrix is
invariant under a $(4,4)$ 2-d supersymmetry algebra originating from the
underlying supergroup structure, with the 2-d momentum operator playing the
r\^{o}le of a central extension. However, in contrast to the case of the
\adss{2} theory here this tree-level symmetry still requires a non-standard
coproduct. The supersymmetry algebra itself gets  quantum-deformed, starting at
the one-loop level -- the anticommutator  of the left and right  supercharges 
is modified by a non-linear term depending on the abelian global symmetry
generators. It would be important to understand the origin of this quantum
deformation, e.g., if it is somehow related to a non-locality of the classical
supersymmetry discussed in  \ci{iv,hms}. Under the assumption of
the quantum-deformed supersymmetry algebra we proposed the exact (all orders in
$1/k$) expression for the perturbative S-matrix of the reduced \adss{3} theory.
It remains to study its pole structure and to reconstruct the corresponding
solitonic sectors of the full physical S-matrix of this theory. 

Our prime  interest -- the reduced \adss{5} theory -- has an additional
non-trivial feature: the corresponding gauge group $H= [SU(2)]^4$ is non-abelian
and consequently the Lagrangian \rf{gwzw} does not have an additional global $H$
symmetry as in the abelian case. 

The perturbative S-matrix defined by the gauge-fixed Lagrangian has a manifest
$H=[SU(2)]^4$ global symmetry which is the global part of the gauge group. We
computed the one-loop contribution and have shown that the group factorisation
property  of the S-matrix discovered at the tree-level in \ci{ht1} continues 
to one-loop order. However, the Yang-Baxter equation which has a tree-level
anomaly \ci{ht1,ht2} continues to be violated. 

Motivated by the existence of the quantum-deformed supersymmetry in the \adss{3}
case we proposed that the factorised S-matrix of the \adss{5} theory may also 
have quantum-deformed symmetry while also satisfying the Yang-Baxter equation.  
We suggested that the corresponding ``quantum-deformed'' S-matrix  may be 
identified with a quantum-deformed R-matrix corresponding to the symmetries of
the \adss{5} case, i.e. with the fundamental R-matrix for the quantum
deformation of the universal enveloping algebra of $\mf{psu}(2|2)\ltimes
\mbb{R}^3$. This R-matrix was constructed in \ci{bk}, and in \ci{bcl} a
particular limit was identified in which its structure becomes similar to that
of the tree-level perturbative S-matrix \ci{ht1} of the reduced \adss{5} theory.
Here we extended this trigonometric relativistic limit to all orders and
demonstrated that the similarity between the resulting  R-matrix and the
perturbative S-matrix directly defined by the Lagrangian continues to be present
also at the subleading one-loop order. 

There are, however, significant differences between the quantum-deformed
S-matrix and the perturbative S-matrix. In particular, as the algebra of $H$ is
non-abelian, its action on the quantum-deformed S-matrix requires a non-trivial
coproduct, i.e. is also quantum-deformed. To try to relate the two S-matrices 
we constructed a non-unitary rotation that mapped one into the other at the 
tree level. Whether this rotation can be extended to higher orders in $1/k$ is
unclear, but if it is possible it will involve introducing a rapidity dependence
into the rotation matrix.

The need for this rotation may be related to some  conflict between the gauge
choice  $A_+=0$ and the conservation of  hidden  integrable charges: it is the
presence of the global non-abelian symmetry in the gauge-fixed action that leads
to a  tree-level anomaly in the YBE. This issue, as well the  reason for the 
quantum  deformation of the symmetry in the Lagrangian formulation, remain to
be clarified. 


\section*{Acknowledgements}
We are grateful to  G. Arutyunov, N. Beisert, T. Hollowood, Y. Iwashita, 
L. Miramontes, R. Roiban and A. Torrielli for many useful discussions. We would
also like to thank R. Roiban for helpful comments on an earlier draft and 
N. Beisert for pointing out the reference \ci{glzt}. Part of this work was
completed while A.A.T. was visiting the University of Western Australia in Perth
and he would like to thank S. Kuzenko for the hospitality and discussions. B.H.
is supported by EPSRC.


\appendices


\renewcommand{\theequation}{A.\arabic{equation}}
\setcounter{equation}{0}
\section{Complete S-matrices of the $\mc{N}=1$ and $\mc{N}=2$ supersymmetric  
sine-Gordon models\la{ssgbits}}

In this appendix we review the complete S-matrices of the $\mc{N}=1$ and
$\mc{N}=2$ supersymmetric sine-Gordon theories \ci{ahn,ku2}.

\subsection{$\mc{N}=1$ supersymmetric sine-Gordon\la{smatcom}}

The Lagrangian of $\mc{N}=1$ supersymmetric sine-Gordon model may be written as
(cf. \rf{lag32})
\bea
\Lag_{\mc{N}=1}^{(sSG)}&=&\la{lagn1susy}
\fr{k}{4 \pi}\Big(\dpl\vp\dm\vp+\fr{\mu^2}{2}\cos 2\vp
                  +i\,\delta\dm\delta+i\,\nu\dpl\nu
                  -2i\mu\,\nu\delta\,\cos\vp\Big)\,,
\eea
In section \ref{neq1ssgsmat} the S-matrix for the perturbative excitations was
reviewed. The complete S-matrix of this theory is much  larger due to the
existence of solitons and breathers. Schematically, it takes the following 
form \ci{ahn}
\be\begin{split}
\trm{soliton-soliton:}&\hs{30pt}S_{SG}(\q,\De)\otimes
S^{(2)}_{RSG}(\q,\De)\,,
\\\trm{soliton-breather:}&\hs{30pt}S^{(n)}_{SG}(\q,\De)\otimes
S^{(n)}_{RSG}(\q,\De)\,,
\\\trm{breather-breather:}&\hs{30pt}S^{(n,m)}_{SG}(\q,\De)\otimes
S^{(n,m)}_{RSG}(\q,\De)\,,
\end{split}\ee
with\,\foot{\la{sine}In the bosonic sine-Gordon theory one has
$\De=\fr{\pi}{k-1}\,.$}
\be \la{coup}
\De=\fr{\pi}{k-\fr{1}{2}}\,.
\ee
The S-matrix factorises into the  bosonic ($S_{SG}$) and supersymmetric
($S_{RSG}$) parts. The bosonic factor is always the S-matrix for the
corresponding excitations in the sine-Gordon model with the coupling
\eqref{coup}. The supersymmetric part is discussed in \ci{ahn}. The S-matrix for
perturbative excitations  we discuss in this paper  is  a special case of the 
breather-breather S-matrix with $n=m=1$, see \ci{ahn}. The bosonic $n=m=1$
factor is given by 
\be\la{ssg11}
S_{SG}^{(1,1)}(\q,\De)=S_{sg}(\q,\De)=\fr{\sinh\q+i\sin\De}
{\sinh\q-i\sin \De}\,.
\ee
Similarly, the supersymmetric $n=m=1$ S-matrix is given by \eqref{neq1susysmat}
\be\la{ide2}
S^{(1,1)}_{RSG}(\q,\De) = S_{\mc{N}_1}(\q,\De)\,.
\ee
Looking for poles and zeros on the physical strip amounts to investigating the
limit when $\q=i\De$. $S_{sg}(\q,\De)$ has a simple pole at $\q=i\De$. The $S$
and $U$ channels (defined in section \ref{neq1ssgsmat}) of
$S_{\mc{N}_1}(\q,\De)$ have a simple zero, while the $T$ and $V$ channels have
no pole or zero. Thus the $S$ and $U$ channels of the total perturbative
excitation S-matrix have no pole or zero, while the $T$ and $V$ channels have a
simple pole at $\q=i\De$ corresponding to the existence of a bound state.

\subsection{$\mathcal{N}=1$ supersymmetric sinh-Gordon\la{susysinh}}

A related theory is $\mathcal{N}=1$ supersymmetric sinh-Gordon. This theory is
formally related to $\mathcal{N}=1$ supersymmetric sine-Gordon \eqref{lagn1susy}
by the transformation
\be\begin{split}
\vp\ra i\phi\,,\hs{20pt} \nu &\ra i \rho\,,\hs{20pt}
\de \ra i \al\,,\hs{20pt}\trm{and}\hs{10pt}k \ra - k\,, 
\end{split}\ee
i.e. the $\mathcal{N}=1$ supersymmetric sinh-Gordon Lagrangian is (cf.
\rf{lag32})
\bea
\Lag_{\mc{N}=1}^{(sShG)}&=&\la{lagn1susyh}
k\big(\dpl\phi\dm\phi-\fr{\mu^2}{2}\cosh 2\phi
+i\,\al\dm\al+i\,\rho\dpl\rho
-2i\mu\,\rho\alpha\,\cosh\phi\big)\,.
\eea
The exact S-matrix for the perturbative bosonic and fermionic excitations is
related to  the above one by $k \to -k$ (see also section \ref{neq1ssgsmat})
\be\begin{split}
\la{sshcom} S_{sg}(\q,-\De(k))&\otimes S_{\mc{N}_1}(\q,-\De(k))\,,
\\\De(k)&=\fr{\pi}{k+\fr{1}{2}}\,.
\end{split}\ee
Unlike the $\mathcal{N}=1$ supersymmetric sine-Gordon case this theory does not
have a degenerate vacuum and thus has no soliton solutions. The perturbative
excitations are the only physical excitations and thus \eqref{sshcom} is the
complete S-matrix for this theory.

The pole structure of the S-matrix is consistent with this. Looking for poles
and and zeros at $\q=i\De$, the $S$ and $U$ channels have neither while the $T$
and $V$ channels have a simple zero. The lack of poles implies hat there are
no bound states of the perturbative excitations and is an  evidence for the
absence of any additional sectors.\,\foot{For reference, it is useful to note
that $S_{sg}(\q,-\De)$ has a zero at $\q=i\De$, whilst
$S_{\mc{N}_1}(\q,-\De)$ has a pole in the $S$ and $U$ channels and neither a
pole nor a zero in the $T$ and $V$ channels.}

\subsection{$\mc{N}=2$ supersymmetric sine-Gordon  \la{comp}}

In section \ref{neq2ssgsmat} the S-matrix for the perturbative excitations of
$\mathcal{N}=2$ supersymmetric sine-Gordon was interpreted \ci{ku2} as the
supersymmetrisation of the bosonic sine-Gordon S-matrix \eqref{sss}. One can
also interpret the same S-matrix as the supersymmetrisation of the bosonic
sinh-Gordon S-matrix. Indeed, rather than labelling the states as in
\eqref{decc} let us label them as follows  
\be\begin{split}
\ket{Y}=\ket{\Phi_{00}}\,,&\hs{20pt}\ket{\z}=\ket{\Phi_{01}}\,,
\\\ket{Z}=\ket{\Phi_{11}}\,,&\hs{20pt}\ket{\ch}=\ket{\Phi_{10}}\,,
\end{split}\ee
with the index $0$ being  bosonic and $1$ fermionic. Instead of factoring out
the S-matrix for the perturbative excitation of the sine-Gordon model we may 
factor out the S-matrix for the perturbative excitation of the sinh-Gordon
model. This corresponds to replacing $k\ra-k$ and $\De\ra -\De$ in
\eqref{bosfact} and \eqref{kde22}. The $\mathcal{N}=2$ supersymmetric
sine-Gordon S-matrix then factorises as follows
\be\la{facsom}\begin{split}
S_{SG}(\q,-\De)\otimes
S_{\mc{N}_1}&(\q,-\De)\otimes_{_G}
S_{\mc{N}_1}(\q,-\De)\,,
\\
\De=&\fr{\pi}{k}\,.
\end{split}\ee
The two ways of writing this S-matrix are consistent as they have the same
poles and zeroes. This can be seen using the results quoted in appendices
\ref{smatcom} and \ref{susysinh}.

While it is possible to factorise the S-matrix as in \eqref{facsom} this is
not the best way for interpreting it. The reason is that $\mc{N}=2$
supersymmetric sine-Gordon model  has solitonic excitations which, like in
bosonic sine-Gordon case, play the r\^{o}le of the elementary excitations in
this theory \ci{ku2}. To construct the S-matrix for these excitations we can
generalise the discussion in appendix \ref{smatcom} for sine-Gordon model, but 
not for the sinh-Gordon as it has no such excitations. The complete S-matrix
for $\mc{N}=2$ supersymmetric sine-Gordon theory takes the following schematic
form ($\De=\fr{\pi}{k}$) \ci{ku2}
\be\begin{split}
\trm{soliton-soliton:}&\hs{30pt}S_{SG}(\q,\De)\otimes
S^{(2)}_{RSG}(\q,\De)\otimes_{_G}
S^{(2)}_{RSG}(\q,\De)\,,
\\\trm{soliton-breather:}&\hs{30pt}S^{(n)}_{SG}(\q,\De)\otimes
S^{(n)}_{RSG}(\q,\De)\otimes_{_G}
S^{(n)}_{RSG}(\q,\De)\,,
\\\trm{breather-breather:}&\hs{30pt}S^{(n,m)}_{SG}(\q,\De)\otimes
S^{(n,m)}_{RSG}(\q,\De)\otimes_{_G}
S^{(n,m)}_{RSG}(\q,\De)\,.
\end{split}\ee
Identifying the lowest mass $n=m=1$ breather with the perturbative excitation
and using \eqref{ssg11}, \eqref{ide2} one finds agreement with \eqref{sss}.


\renewcommand{\theequation}{B.\arabic{equation}}
\setcounter{equation}{0}
\section{Comments on symmetries of \adss{3} superstring and reduced
theory\la{ads3s3pohl}}

In this appendix we discuss symmetries of the \adss{3} reduced and superstring
theories. The usual way of constructing the GS superstring sigma model is to
start with the coset \ci{ads3xs3stringtheory,gt2}
\be\la{coset33}
\fr{[PSU(1,1|2)]^2}{SU(1,1) \x SU(2)}\,.
\ee
The numerator group $[PSU(1,1|2)]^2$ has a bosonic subgroup $[SU(1,1)]^2 \x
[SU(2)]^2$ that can be extended by four central elements to $[U(1,1)]^2 \x
[U(2)]^2$. The coset \eqref{coset33} can then be rewritten as 
\be\la{coset}
\fr{[U(1,1|2)]^2}{U(1,1) \x U(2) \x [U(1)]^2}\,.
\ee
Following \ci{gt2} we use the following parametrisation for $[U(1,1|2)]^2$
\be\la{param3333}\left(\ba{cccc}
a_1& \al & 0 & 0
\\\td{\al} & b_1 &0&0
\\0&0&a_2& \bet
\\0&0& \td{\bet} &b_2\ea\right)\,.\ee
where $a_1$, $a_2$ are $2 \x 2$ $U(1,1)$ matrices and  $b_1$, $b_2$ are $2 \x 2$
$U(2)$ matrices.  $\al$ and $\bet$ are $2 \x 2$ complex fermionic matrices, as 
are $\td{\al}$ and $\td{\bet}$, which contain the same degrees of freedom as
$\al$ and $\bet$ respectively.

The corresponding superalgebra $\hat{\mf{f}}=[\mf{u}(1,1|2)]^2$ admits a
$\mbb{Z}_4$ decomposition
\be
\hat{\mf{f}}
=\hat{\mf{f}}_0 + \hat{\mf{f}}_1 +
 \hat{\mf{f}}_2 +  \hat{\mf{f}}_3\,, \ \ \ \ \ \ 
\com{\hat{\mf{f}}_i}{\hat{\mf{f}}_j}\subset
\hat{\mf{f}}_{i+j\trm{\,mod\,} 4}\, ,
\ee
where even/odd subspaces are bosonic/fermionic. The $\mbb{Z}_4$ decomposition
relevant for construction of the reduced \adss{3} theory is discussed in
\ci{gt2}; its key property is that it mixes the two copies of $[U(1,1|2)]^2$.

Two of the four central elements discussed above live in $\hat{\mf{f}}_0$ and
two in the $\hat{\mf{f}}_2$ subspace.\,\foot{This is different from the case 
of the \adss{2} and \adss{5} theories, where the numerator supergroup for is
of the form $PSU(n,n|2n)$, $n=1,\,2$. Compared to $U(n,n|2n)$ the two central
elements have been projected out (the supertrace and the trace). These both live
in $\hat{\mf{f}}_2$ when considering the $\mbb{Z}_4$ decomposition of
$U(n,n|2n)$.} In the \adss{2} and \adss{5} theories all central elements in
$\hat{\mf{f}}_2$ are ``manually'' projected out. We may do the same for the
\adss{3} case. For the symmetry analysis of the reduced theory we shall not
project out the central elements in $\hat{\mf{f}}_0$. Then the corresponding
supercoset we shall consider is 
\be\la{cosetfinhalf}
\fr{PS([U(1,1|2)]^2)}{U(1,1) \x U(2)}\,.
\ee
Here $P$ and $S$ in the numerator supergroup correspond to the following
constraints on the entries of \eqref{param3333}
\be\begin{split}
\Tr(a_1)+\Tr(b_1)&+\Tr(a_2)+\Tr(b_2)=0\,,
\\\Tr(a_1)-\Tr(b_1)&+\Tr(a_2)-\Tr(b_2)=0\,.
\end{split}\ee
Choosing the same element $T\in \mf{a}$ ($\mf{a}$ is the maximal abelian
subalgebra of  $\hat{\mf{f}}_2$) as in \ci{gt2}, we find that the subalgebra
$\mf{h}\subset \hat{\mf{f}}_0$  defined by $\com{\mf{h}}{T}=0$ is
$[\mf{u}(1)]^4$.

One of the four $\mf{u}(1)$s takes the following form
\be\left(\ba{cc}
\mbb{I}_4 & 0
\\0 &-\mbb{I}_4\ea\right)\,.
\ee
As the structure of the numerator supergroup \eqref{param3333} is block
diagonal,  any ``symmetry'' arising from this generator will have a trivial
action on all the physical fields in both the superstring and reduced theories. 
In the reduced \adss{3} theory the bosonic $H$  symmetry is therefore $[U(1)]^3$
rather than  $[U(1)]^2$  that one might predict by considering  just the coset
\eqref{coset33}. The bosonic symmetries of the reduced theory thus consist of
a global $[U(1)]^3$ and a gauged $[U(1)]^3$.

It is worth noting that the Lagrangian of the worldsheet superstring is not
altered by this discussion. All we have done is to  include the central elements
in the numerator supergroup and simultaneously divide by them in the denominator
group. Therefore, the $G$-gauge symmetry can always be used to remove them
giving back the original supercoset \eqref{coset33}. The addition of these two
central elements does not affect the construction of the reduced theory. They
both live in the algebra $\mf{h}$, which is gauged in the reduced theory (i.e. 
the central elements do not alter the degrees of freedom count).

Starting with the coset \eqref{cosetfinhalf} one may carry out the Pohlmeyer
reduction procedure by first gauge-fixing these central elements and then
proceed  as in \ci{gt2}. Alternatively, if the central elements are included, 
then there are two extra degrees of freedom in the group field $g$ and $A_\pm$,
but also two extra gauge symmetries.\,\foot{One of the $U(1)$ gauge symmetries
has trivial action on the group field $g$, containing the physical bosonic
excitations, and on  the fermions; however it will have a non-trivial action on
the gauge field.} Considering the $A_+=0$ gauge, integrating out $A_-$ and 
using the resulting constraint equation to eliminate $\xi$ one finds that 
the additional central elements have no effect on the resulting Lagrangian
\eqref{lag13}, i.e. we get again (the quartic expansion of) the Lagrangian
constructed in \ci{gt2}.


\renewcommand{\theequation}{C.\arabic{equation}}
\setcounter{equation}{0}
\section{Counterterms from functional determinants\la{olcfd}}

In \ci{dvm1} the one-loop S-matrix for the complex sine-Gordon model was
computed using a Lagrangian describing a single complex scalar and the 
Yang-Baxter equation was found to be  violated at this order. The authors
proposed that preserving the symmetry (integrability) should be the guiding
principle for quantization of the theory and found that there exists a local 
counterterm which restored the validity of YBE at one-loop order. 

It was subsequently suggested \ci{dh} that understanding the complex sine-Gordon
theory as arising from a gauged WZW \ci{bakps} should explain the origin of this
quantum counterterm. This was demonstrated explicitly in \ci{ht2} where several 
methods of deriving the one-loop quantum counterterm for the complex sine-Gordon
theory as defined by the gauged WZW plus integrable potential action (cf. the
bosonic part of \rf{gwzw}) in the path integral were presented, each giving 
precisely the corrections required to restore YBE and match the expansion of
the exact S-matrix of \ci{dh}. As was shown in \ci{ht2}, the quantum
counterterms may be understood as arising from the functional determinant
appearing after integrating out the gauge field $A_\pm$.

One of the methods discussed in \ci{ht2} was based on fixing the gauge $A_+=0$
and integrating over the gauge field $A_-$ to get a delta function constraint  
$\delta(g^{-1}\dpl g |_{\mf{h}})\,$ in the path integral. This constraint is
then used to eliminate the remaining unphysical degree of freedom $\xi$ in
\eqref{paramg} from the classical action. In addition, this delta function 
factor leads to a functional determinant of the form\,\foot{Here $\mc{B}_+$ is
a field dependent matrix built out of commutators acting on the vector space
$\mf{h}$. In particular, we require that  $\mc{B}_+v \in\mf{h}$  for $v \in
\mf{h}$. At leading order $\mc{B}_+$ is quadratic in fields. For further details
see \ci{ht2}.}
\be
(\det \mc{O}_+)^{-1} \,, \hs{20pt} \ \ \ \ 
\mc{O}_+ v \equiv \dpl v + \mc{B}_+ v \ , 
\hs{10pt}\ \ \ 
\trm{\ \ \ } \hs{5pt} v \in \mf{h}\,.
\ee
Introducing an orthonormal basis $\{T_i\}$ for $\mf{h}$ and defining
\be\la{matb}
\mc{B}_+ T_i = B_{+\,ij}T_j\,,
\ee
one can compute the contribution of this determinant (e.g., via standard anomaly
argument). This leads to the following one-loop correction to the Lagrangian
\be  \Delta L= 
-\fr{1}{8\pi}B_{+\,ij}\fr{\dm}{\dpl}B_{+\,ji} + \ord{B^3}\,.\la{uu}
\ee
For the purpose of computing the one-loop two-particle S-matrix higher order 
$\ord{B^3}$ corrections here can be ignored as $B$ is quadratic in the physical
field $X$ in \eqref{paramg}.

\subsection{Bosonic theories}

Let us  briefly review the structure of the resulting corrections for the
complex sine-Gordon and the $G/H= SO(N+1)/SO(N)$ generalised sine-Gordon 
theories \ci{ht2} using the parametrisation of $g$ given in \eqref{paramg}.
For the complex sine-Gordon case ($G/H=SU(2)/U(1)$) 
\be 
\mc{B}_+v = -\fr{1}{4}\Big(\com{X}{\com{\dpl X}{v}}
                          +\com{\dpl X}{\com{X}{v}}\Big)\,.
\ee 
Here $H=U(1)$ so the indices $i,j$ take only a single value and the matrix
$B_{+\,ij}$ \eqref{matb} may be denoted simply $B_+$. Also, $B_+$ is a total
derivative ($B_+= \del_+  b $) of a local function of $X$ and thus the
correction to the Lagrangian can be written as (we set $B_- = \del_- b = 
\fr{\del_-}{\del_+} B_+ $)\,\foot{As $H=U(1)$ is abelian there are no
$\ord{B^3}$ corrections in \rf{uu} this case.}
\be \la{csgcorrect}
\Delta L= -\fr{1}{8\pi}B_+B_- \ . 
\ee
Expanding the field $X= X_m T_m$ where $\{T_m\}$ is an orthonormal basis for
$\mf{m}$ (the coset part of the algebra of $G$) and recalling that we should
rescale the fields by $\sqrt{\fr{4\pi}{k}}$, \eqref{csgcorrect} precisely
matches the result for the local counterterm found by the second method in 
\ci{ht2}.      

For the $G/H = SO(N+1)/SO(N)$ theory we may split $\mc{B}_+$ into a symmetric
and antisymmetric parts\,\foot{To compare to \ci{ht2} note that $B_{+\,ij}=
-\fr{1}{2} f_{mpi}f_{npj}   X_m \dpl X_n \equiv -\fr{1}{2} V_{mnij} X_m \dpl
X_n$, where $f_{mpi}$ are the  structure constants of the algebra $\mf{g}$
decomposed as $\mf{g} = \mf{m}\oplus \mf{h}$.}
\be \la{oppe}
\begin{split}
\mc{B}_+v &= \mc{B}^s_+v +\mc{B}^a_+v\,,
\\\mc{B}^s_+v =-\fr{1}{4}(\com{X}{\com{\dpl X}{v}}&+\com{\dpl
X}{\com{X}{v}})\,,\ \ \ \ \ 
\hs{20pt}\mc{B}^a_+v =\fr{1}{4}\com{\com{X}{\dpl X}}{v}\,.
\end{split}
\ee
As for the complex sine-Gordon case  the symmetric part $B^s_{+\,ij}$
\eqref{matb} can be written as a total derivative of a local  bilinear
expression in $X$. The antisymmetric part satisfies the following
identity\,\foot{Here the differential operator $\fr{\dm}{\dpl}$ acts of course
only on the fields contained in $\mc{B}_+$ and not on $v$.} 
\be\begin{split}
\big(\fr{\dm}{\dpl}\mc{B}^a_+\big)v=\fr{\dm}{\dpl}\com{\com{X}{\dpl X}}{v}
 = -\com{\com{X}{\dm  X}}{v} +\fr{2}{\dpl}\com{\com{X}{\dpl \dm X}}{v}
\end{split}\ee
For the purpose of computing the one-loop S-matrix the second term can be
ignored as it vanishes on the linearised equation of motion for $X$
\be
\dpl \dm X +\mu^2 X =0\,.
\ee
We can therefore replace
\be\begin{split}
\big(\fr{\dm}{\dpl}\mc{B}^a_+\big)v
=\fr{\dm}{\dpl}\com{\com{X}{\dpl X}}{v} \approx-\com{\com{X}{\dm  X}}{v}
\equiv -\mc{B}^a_- v\,,
\end{split}\ee
and thus the  correction to the Lagrangian can be written
\be\begin{split}
\Delta L= -\fr{1}{8\pi}(B^s_{+\,ij}B^s_{-\,ji} - B^a_{+\,ij}B^a_{-\,ji})
+ \ord{B^3}\,.
\end{split}\ee

\subsection{Reduced  \adss{n}   theories}

The method of finding  counterterms that worked for the complex sine-Gordon
model \ci{ht2}, does not give the required counterterm for the reduced \adss{3}
theory discussed in  section \ref{ybe33}. Also, the counterterm obtained in the
same  way as above in the case of the reduced \adss{5} theory would break the
group factorisation  property of the one-loop S-matrix in section \ref{secfac}.

In this subsection we will postulate a single functional determinant based on
the group structure and fields of the theory that gives the required result in
all 3 cases: non-trivial  one-loop  correction  in the \adss{3} case and no
corrections in the \adss{2} and the \adss{5} cases.\,\foot{No counterterm is
needed in the reduced \adss{2} theory case to match the exact S-matrix of
\ci{ku1,ku2,sw,ahn}.}

Using integrability as our guiding principle the counterterm \eqref{counterterm}
is required in the reduced \adss{3} theory. Below we shall show that it may
originate from a particular functional determinant similarly to the bosonic
case discussed above. The corresponding operator acts on the superalgebra
$\hat{\mf{h}}$ (see section \ref{prgr}) whose bosonic subalgebra is $\mf{h}$. If
this determinant were to arise in a similar fashion to the bosonic case, there
should be analogues of the unphysical degrees of freedom $\xi$  taking values in
the fermionic subspaces of $\hat{\mf{h}}$. Comparing to the supersymmetric
gauged WZW theory written  in components or in superfield forms \ci{susygwzw,t1}
this  suggests that there may be an alternative formulation of the action 
\eqref{gwzw} that treats fermions and bosons on a more equal footing and thus 
requires extra fermionic components of the gauge field.

To define the first-order differential operator whose determinant produces the
required contribution let us first recall some group theory of the Pohlmeyer
reduction of the \adss{n} ($n=2,\,3,\,5$) superstring theories (see \ci{gt1}
and  section \ref{prgr}; for explicit bases and parametrisations see also 
\ci{gt2,rt,ht1,hiti}). Here the constant matrix $T$ that defines the potential
of the reduced theory is normalised as $T^2 =-\fr{1}{4}I\,.$ We take an
orthonormal basis for the bosonic generators of the superalgebra $\hat{\mf{f}}$.
Further we take  a basis for the fermionic generators of $\hat{\mf{f}}^\perp_1$
and $\hat{\mf{f}}^\perp_3$, i.e.  $T^{R^\perp}_i$ and $T^{L^\perp}_i$, 
respectively, such that
\be\la{baspef}\begin{split}
\STr(T^{R^\perp}_i T^{L^\perp}_j)=\de_{ij}\,,\hs{30pt}&
\STr(T^{L^\perp}_i T^{R^\perp}_j)=-\de_{ij}\,,
\\\STr(T^{R^\perp}_i T^{R^\perp}_j)=0\,,\hs{30pt}&
\STr(T^{L^\perp}_i T^{L^\perp}_i)=0\,,
\end{split}\ee
where the supertrace is defined as in \ci{gt1}.\,\foot{The same indices $i,j$
are used here for both fermionic subspaces as well as the bosonic algebra
$\mf{h} = \hat{\mf{f}}^\perp_0$ of the previous subsection. This is just a
notational convenience and is not meant to indicate that there are the same
number of generators in each of these spaces.} It will be useful to denote a
basis for the superalgebra $\hat{\mf{h}}$ (a fermionic extension of $\mf{h}$
defined in section \ref{prgr}) as ${T_I}$. This includes an orthonormal basis
for $\mf{h}$ along with the bases for $\hat{\mf{f}}^\perp_1$ and
$\hat{\mf{f}}^\perp_3$ described above. We also define the metric 
\be\la{etaij}
\hat{\eta}_{IJ}=\STr(T_I T_J)\,, \ \ \ \ \ \ \ 
\hat{\eta}_{IJ}\hat{\eta}^{JK}=\de_I^J\,,\hs{20pt}\hat{\eta}^{IJ}\hat{\eta}_{JK}
=\de_J^I\,,
\ee
so that $T^I=\hat{\eta}^{IJ}T_J\,,\hs{2pt} T_I = \hat{\eta}_{IJ}T^J$. This
metric is not as simple as in bosonic case due to the more involved supertraces
over the fermionic generators \eqref{baspef}. For the parallel fermionic
subspaces $\hat{\mf{f}}^\parallel_1$ and $\hat{\mf{f}}^\parallel_3$ we take
similar bases as for the perpendicular fermionic subspaces \eqref{baspef},
\ci{ht1}. We also have the following useful identities
\be\begin{split}
T^{L^\parallel_{1,2}}_m=2T T^{R^\parallel_{1,2}}_m\,, \ \ \ \ \ \ \ \ \ 
\acom{T}{T^{L^\parallel_{1,2}}}=\acom{T}{T^{R^\parallel_{1,2}}} =0 \,.
\end{split}
\ee
The physical excitations \eqref{spl} live in the subspaces, $X \in
\mf{f}_0^\parallel$, $\psr \in \mf{f}_1^\parallel$ and $\psr \in
\mf{f}_3^\parallel$. In the reduced \adss{n} theories each of these subspaces
can be split into two halves, both transforming in the same representation of a
subgroup of $H$ (see \ci{ht1} for more details). We take the parametrisation of
the physical fields
\be\begin{split}
X = Y_m T_m^A + Z_m T_m^S\,,
\ \ \ \ \ \ \psr = \sqrt{i} \zer{}_m T^{R^\parallel_1}_m+\sqrt{i} \chr{}_m
T^{R^\parallel_2}_m\,,\ \ \ \ \ 
\psl = \sqrt{i} \zel{}_m T^{L^\parallel_1}_m+\sqrt{i} \chl{}_m
T^{L^\parallel_2}_m\,.
\end{split}\ee
As is suggested by the notation the fields $Y_m$ parametrise the part of the
reduced theory corresponding to the bosonic $AdS_n$ space and similarly the
fields $Z_m$, the bosonic $S^n$ space.

The functional determinant we propose to consider is ($v \in \hat{\mf{h}}$)
\be\begin{split}\la{fermdet}
(\det \hat{\mc{O}}_+)^{-1} \,, \hs{20pt} \ \ 
\hat{\mc{O}}_+ v \equiv \dpl v + \hat{\mc{B}}_+ v \ , \ \  \ \ \ \ \
\hat{\mc{B}}_+ v = 
 \big[ {\com{X}{\dpl X}-\com{\psr}{2T \psr}}, \ {v} \big]  \ .
\end{split}\ee
Compared to the operator arising in the bosonic case \ci{ht2} there is not 
a part symmetric in $X$ and fermions are included ($\hat{\mc{O}}_+$ acts on a
superalgebra). The action of $\hat{\mc{B}}_+$ on the basis ${T_I}$ for
$\hat{\mf{h}}$ is defined as
\be
\hat{\mc{B}}_+ T_I = \hat{B}_{+\,I}^{\;\;\;\;\,J}T_J\,.
\ee
The linearised equations of motion for $X$ and $\psi$ are (see \eqref{spl})
\be
\dpl \dm X +\mu^2 X =0\,,\hs{10pt} 2 T \dpl\psl +\mu\psr =0  
\,,\hs{10pt} 2 T \dm\psr +\mu\psl =0  \,.
\ee
As in the bosonic example we can make use of these equations of motion to
replace
\be\begin{split}
\big(\fr{\dm}{\dpl}\hat{\mc{B}}_+\big) v &\approx-\big[\com{X}{\dm
X}-\com{2\psl T}{\psl}, \ {v}\big] \equiv -\hat{\mc{B}}_- v\ . 
\end{split}\ee
The one-loop correction to the Lagrangian coming from this determinant can then
be written as\,\foot{Here we used  that $\hat{\eta}^{IL}\hat{\eta}_{KL} =
(-1)^{[I]} \de^I_K$: it equals to $1$ if $I=K$ is a bosonic index and to $-1$ if
$I=K$ is a fermionic index. $\hat{\eta}$ was  defined in \eqref{etaij}.}
\be
\Delta L= \fr{1}{8\pi} \sum_{I,J} (-1)^{[I]} \hat{B}_{+\,I}^{\;\;\;\;\,J}
\hat{B}_{-\,J}^{\;\;\;\;\,I}
+ \ord{\hat{B}^3} \ .
\ee
For the reduced \adss{2} and \adss{5} theories this correction term vanishes as
required. In the case of the \adss{3} theory it gives, remarkably, the
non-trivial counterterm \eqref{counterterm} we postulated above to satisfy
the YBE.\,\foot{Recall that the physical fields should be rescaled by
$\sqrt{\fr{4\pi}{k}}$.}

We should emphasize again that the existence of a single universal expression
for the one-loop counterterm is rather  non-trivial. Its path integral origin  
remains, however, to be understood.\,\foot{One possibility to include fermions
$\psr,\psl$ in the determinant is by a rotation of them by the bosonic field
$g$. Similar rotations were used in the construction of the reduced theory
\ci{gt1,gt2}. However, just considering such  rotations of fermions in the
Lagrangian \eqref{gwzw} will not produce  \eqref{fermdet}: it would be necessary
in addition to have extra unphysical degrees of freedom living in the fermionic
part of $\hat{\mf{h}}$.} 


\renewcommand{\theequation}{D.\arabic{equation}}
\setcounter{equation}{0}
\section{Factorized  S-matrix of  reduced \adss{3} theory\la{33expansion}}

In this appendix we present the explicit form of the factorised S-matrix
\eqref{33fact}, \eqref{Sparam33} and rewrite it in terms of fields transforming
as a vector of the same $SO(2)$ group to enable comparison with the S-matrix of
section \ref{secollsg33}. This  single $SO(2)$ is obtained  by identifying the
two $SO(2)$s with indices $a$ and $\al$ and also the two $SO(2)$s with indices
$\adt$ and $\agdt$. Naively this gives an $[SO(2)]^2$ symmetry but the actions
of these $SO(2)$s coincide, leaving the  single $SO(2)$. After identifying the
pairs of $SO(2)$s the following rules can be used to translate to the single
$SO(2)$ notation 
\be\begin{split}
\mbb{I}\otimes\mbb{I} \ra \de_{mp}\de_{nq}\,,
&\hs{20pt}\mbb{I}\otimes \mbb{K}\ra
 \e_{mp}\e_{nq}\,,
\\\mbb{K}\otimes \mbb{I}\ra \e_{mp}\e_{nq}\,,
&\hs{20pt}\mbb{K}\otimes\mbb{K}\ra\de_{mp}\de_{nq}\,,
\\(\mbb{I})_{abcd}=\de_{ac}\de_{bd}\,,
&\hs{20pt}(\mbb{K})_{abcd}=\e_{ac}\e_{bd}\,,
\end{split}\ee
where the first entry in the tensor product corresponds to undotted indices and
the second entry to dotted indices.

The S-matrix has the following structure:

\footnotesize
{\allowdisplaybreaks
\tbf{Boson-Boson}
\begin{align*}
\Sc\ket{Y_m(p_1)Y_n(p_2)}
= & \big((L_1^2+L_2^2)\de_{mp}\de_{nq}
    +L_1L_2\e_{mp}\e_{nq}\big)
    \ket{Y_p(p_1)Y_q(p_2)}
\\& +\big(-2L_9^2(\de_{mn}\de_{pq}+
                  \e_{mn}\e_{pq})\big)\ket{Z_p(p_1)Z_q(p_2)}
\\& +\big((L_1+L_2)L_9(\de_{mn}\de_{pq}+\e_{mn}\e_{pq})
            \big)\ket{\z_p(p_1)\z_q(p_2)}
\\&  +\big((L_1+L_2)L_9(\de_{mn}\de_{pq}+\e_{mn}\e_{pq})
            \big)\ket{\ch_p(p_1)\ch_q(p_2)}
\\\Sc\ket{Z_m(p_1)Z_n(p_2)}
= & \big((L_3^2+L_4^2)\de_{mp}\de_{nq}
    +L_3L_4\e_{mp}\e_{nq}\big)\ket{Z_p(p_1)Z_q(p_2)}
\\&    +\big(-(L_3+L_4)L_{10}(\de_{mn}\de_{pq}+\e_{mn}\e_{pq}) 
\big)\ket{\ch_p(p_1)\ch_q(p_2)}
\\& +\big(-2L_{10}^2(\de_{mn}\de_{pq}+
                  \e_{mn}\e_{pq})\big)\ket{Y_p(p_1)Y_q(p_2)}
\\&
+\big(-(L_3+L_4)L_{10}(\de_{mn}\de_{pq}+\e_{mn}\e_{pq}) 
\big)\ket{\z_p(p_1)\z_q(p_2) }
\\\Sc\ket{Y_m(p_1)Z_n(p_2)}
= & \big((L_5^2+L_6^2)\de_{mp}\de_{nq} +2L_5
L_6 \e_{mp}\e_{nq}\big)\ket{Y_p(p_1)Z_q(p_2)}
\\& +\big(2L_{11}^2(\de_{mq}\de_{pn} +
\e_{mq}\e_{pn} \big)\ket{Z_p(p_1)Y_q(p_2)}
\\& +\big(-(L_5-L_6)L_{11}(-\de_{mn}\de_{pq}+\de_{mp}\de_{nq} +
\de_{mq}\de_{np})\big)
        \ket{\z_p(p_1)\ch_q(p_2)}
\\& +\big((L_5-L_6)L_{11}(-\de_{mn}\de_{pq}+\de_{mp}\de_{nq} +
\de_{mq}\de_{np})\big)
         \ket{\ch_p(p_1)\z_q(p_2)}
\\\Sc\ket{Z_m(p_1)Y_n(p_2)}
= & \big((L_7^2+L_8^2)\de_{mp}\de_{nq} +2L_7
L_8 \e_{mp}\e_{nq}\big)\ket{Z_p(p_1)Y_q(p_2)}
\\& +\big(2L_{12}^2(\de_{mq}\de_{pn} +
\e_{mq}\e_{pn} \big)\ket{Y_p(p_1)Z_q(p_2)}
\\& +\big((L_{7}-L_8)L_{12}(-\de_{mn}\de_{pq}+\de_{mp}\de_{nq} +
\de_{mq}\de_{np})\big)
        \ket{\ch_p(p_1)\z_q(p_2)}
\\&
+\big(-(L_7-L_8)L_{12}(-\de_{mn}\de_{pq}+\de_{mp}\de_{nq} +
\de_{mq}\de_{np})\big)
         \ket{\z_p(p_1)\ch_q(p_2)}
\end{align*}
\tbf{Boson-Fermion}
\begin{align*}
\Sc\ket{Y_m(p_1)\z_n(p_2)}
= & \big((L_1L_5+L_2L_6)\de_{mp}\de_{nq}
          +(L_1L_6+L_2L_5)\e_{mp}\e_{nq})\big)
        \ket{Y_p(p_1)\z_q(p_2)}
\\& +\big((L_1-L_2)L_{11}(\de_{mq}\de_{pn}+\e_{mq}\e_{pn}\big)
\ket{\z_p(p_1)Y_q(p_2)}
\\&    +\big((L_5+L_6)L_9(-\de_{mq}\de_{np}
        +\de_{mp}\de_{nq}+\de_{mn}\de_{pq})\big)
      \ket{\ch_p(p_1)Z_q(p_2)}
\\\Sc\ket{\z_m(p_1)Y_n(p_2)}
= & \big((L_1L_7+L_2L_8)\de_{mp}\de_{nq}
          +(L_1L_8+L_2L_7)\e_{mp}\e_{nq})\big)
        \ket{\z_p(p_1)Y_q(p_2)}
\\& +\big((L_1-L_2)L_{12}(\de_{mq}\de_{pn}+\e_{mq}\e_{pn})\big)
\ket{Y_p(p_1)\z_q(p_2)}
\\&    +\big(-(L_7+L_8)L_9(-\de_{mq}\de_{np}
        +\de_{mp}\de_{nq}+\de_{mn}\de_{pq})\big)
      \ket{Z_p(p_1)\ch_q(p_2)}
\\\Sc\ket{Y_m(p_1)\ch_n(p_2)}
= &  \big((L_1L_5+L_2L_6)\de_{mp}\de_{nq}
          +(L_1L_6+L_2L_5)\e_{mp}\e_{nq})\big)\ket{Y_p(p_1)\ch_q(p_2)}
\\&   
+\big((L_1-L_2)L_{11}(\de_{mq}\de_{pn}+\e_{mq}\e_{pn})\big)
\ket{\ch_p(p_1)Y_q(p_2)}
\\&    +\big(-(L_5+L_6)L_9(-\de_{mq}\de_{np}
        +\de_{mp}\de_{nq}+\de_{mn}\de_{pq})\big)
\ket{\z_p(p_1)Z_q(p_2)}
\\\Sc\ket{\ch_m(p_1)Y_n(p_2)}
= &  \big((L_1L_7+L_2L_8)\de_{mp}\de_{nq}
          +(L_1L_8+L_2L_7)\e_{mp}\e_{nq})\big)\ket{\ch_p(p_1)Y_q(p_2)}
\\&   
+\big((L_1-L_2)L_{12}(\de_{mq}\de_{pn}+\e_{mq}\e_{pn})\big)
\ket{Y_p(p_1)\ch_q(p_2)}
\\&    +\big((L_7+L_8)L_9(-\de_{mq}\de_{np}
        +\de_{mp}\de_{nq}+\de_{mn}\de_{pq})\big)
\ket{Z_p(p_1)\z_q(p_2)}
\\\Sc\ket{Z_m(p_1)\z_n(p_2)}
= & \big((L_3L_7+L_4L_8)\de_{mp}\de_{nq}
          +(L_3L_8+L_4L_7)\e_{mp}\e_{nq})\big)\ket{Z_p(p_1)\z_q(p_2)}
\\& +\big(-(L_3-L_4)L_{12}(\de_{mq}\de_{pn}+\e_{mq}\e_{pn})\big)
      \ket{\z_p(p_1)Z_q(p_2)}
\\& +\big((L_7+L_8)L_{10}(-\de_{mq}\de_{np}
        +\de_{mp}\de_{nq}+\de_{mn}\de_{pq})\big)
\ket{\ch_p(p_1)Y_q(p_2)}
\\\Sc\ket{\z_m(p_1)Z_n(p_2)}
= & \big((L_3L_5+L_4L_6)\de_{mp}\de_{nq}
          +(L_3L_6+L_4L_5)\e_{mp}\e_{nq})\big)\ket{\z_p(p_1)Z_q(p_2)}
\\& +\big(-(L_3-L_4)L_{11}(\de_{mq}\de_{pn}+\e_{mq}\e_{pn})\big)
      \ket{Z_p(p_1)\z_q(p_2)}
\\& +\big(-(L_5+L_6)L_{10}(-\de_{mq}\de_{np}
        +\de_{mp}\de_{nq}+\de_{mn}\de_{pq})\big)\ket{
Y_p(p_1)\ch_q(p_2)}
\\\Sc\ket{Z_m(p_1)\ch_n(p_2)}
= & \big((L_3L_7+L_4L_8)\de_{mp}\de_{nq}
          +(L_3L_8+L_4L_7)\e_{mp}\e_{nq})\big)\ket{Z_p(p_1)\ch_q(p_2)}
\\& +\big(-(L_3-L_4)L_{12}(\de_{mq}\de_{pn}+\e_{mq}\e_{pn})\big)
\ket{\ch_p(p_1)Z_q(p_2)}
\\&+\big(-(L_7+L_8)L_{10}(-\de_{mq}\de_{np}
        +\de_{mp}\de_{nq}+\de_{mn}\de_{pq})\big)\ket{\z_p(p_1)Y_q(p_2)}
\\\Sc\ket{\ch_m(p_1)Z_n(p_2)}
= & \big((L_3L_5+L_4L_6)\de_{mp}\de_{nq}
          +(L_3L_6+L_4L_5)\e_{mp}\e_{nq})\big)\ket{\ch_p(p_1)Z_q(p_2)}
\\& +\big(-(L_3-L_4)L_{11}(\de_{mq}\de_{pn}+\e_{mq}\e_{pn})\big) 
\ket{Z_p(p_1)\ch_q(p_2)}
\\& +\big((L_5+L_6)L_{10}(-\de_{mq}\de_{np}
        +\de_{mp}\de_{nq}+\de_{mn}\de_{pq})\big)
\ket{Y_p(p_1)\z_q(p_2)}
\end{align*}
\tbf{Fermion-Fermion}
\begin{align*}
\Sc\ket{\z_m(p_1)\z_n(p_2)}
= & \big((L_1L_3+L_2L_4)\de_{mp}\de_{nq}
          +(L_1L_4+L_2L_3)\e_{mp}\e_{nq}\big)
        \ket{\z_p(p_1)\z_q(p_2)}
\\&+
\big(2L_9L_{10}(\de_{mn}\de_{pq}+
              \e_{mn}\e_{pq})\big)\ket{\ch_p(p_1)\ch_q(p_2)}
\\& +\big((L_1+L_2)L_{10}(\de_{mn}\de_{pq} +
                     \e_{mn}\e_{pq})\big)\ket{Y_p(p_1)Y_q(p_2)}
\\& +\big(-(L_3+L_4)L_{9}(\de_{mn}\de_{pq} +
                     \e_{mn}\e_{pq})\big)\ket{Z_p(p_1)Z_q(p_2)}
\\\Sc\ket{\ch_m(p_1)\ch_n(p_2)}
= & \big((L_1L_3+L_2L_4)\de_{mp}\de_{nq}
          +(L_1L_4+L_2L_3)\e_{mp}\e_{nq}\big)
                        \ket{\ch_p(p_1)\ch_q(p_2)}
\\& +
\big(2L_9L_{10}(\de_{mn}\de_{pq}+
              \e_{mn}\e_{pq})\big)\ket{\z_p(p_1)\z_q(p_2)}
\\& +\big(-(L_3+L_4)L_{9}(\de_{mn}\de_{pq} +
                     \e_{mn}\e_{pq})\big)\ket{Z_p(p_1)Z_q(p_2)}
\\& +\big((L_1+L_2)L_{10}(\de_{mn}\de_{pq} +
                     \e_{mn}\e_{pq})\big)\ket{Y_p(p_1)Y_q(p_2)}
\\\Sc\ket{\z_m(p_1)\ch_n(p_2)}
= & \big((L_5 L_7 +L_6L_8)\de_{mp}\de_{nq}
       +(L_5 L_8+L_6L_7)\e_{mp}\e_{nq}\big)\ket{\z_p(p_1)\ch_q(p_2)}
\\& +\big(-L_{11}L_{12}(\de_{mq}\de_{pn}+\e_{mq}\e_{pn}\big)
\ket{\ch_p(p_1)\z_q(p_2) }
\\& + \big(-(L_5-L_6)L_{12}(-\de_{mn}\de_{pq} +
     \de_{mp}\de_{nq}+\de_{mq}\de_{np})\big)\ket{Y_p(p_1)Z_q(p_2)}
\\&
+\big(-(L_7-L_8)L_{11}(-\de_{mn}\de_{pq} +
     \de_{mp}\de_{nq}+\de_{mq}\de_{np})\big)\ket{Z_p(p_1)Y_q(p_2)}
\\\Sc\ket{\ch_m(p_1)\z_n(p_2)}
= & \big((L_5 L_7 +L_6L_8)\de_{mp}\de_{nq}
     +(L_5 L_8+L_6L_7)\e_{mp}\e_{nq}\big)\ket{\ch_p(p_1)\z_q(p_2)}
\\& +\big(-L_{11}L_{12}(\de_{mq}\de_{pn}+\e_{mq}\e_{pn}\big)
\ket{\z_p(p_1)\ch_q(p_2)}
\\& +\big((L_7-L_8)L_{11}(-\de_{mn}\de_{pq} +
     \de_{mp}\de_{nq}+\de_{mq}\de_{np})\big)\ket{Z_p(p_1)Y_q(p_2)}
\\& +\big((L_5-L_6)L_{12}(-\de_{mn}\de_{pq} +
     \de_{mp}\de_{nq}+\de_{mq}\de_{np})\big)\ket{Y_p(p_1)Z_q(p_2)}
\end{align*}
}
\normalsize


\renewcommand{\theequation}{E.\arabic{equation}}
\setcounter{equation}{0}
\section{Factorized  S-matrix of  reduced \adss{5} theory\la{55expansion}}

Similarly to the previous appendix, here we present the factorised S-matrix
\eqref{55fact}, \eqref{Sparam} in terms of states transforming as a vector of
the same $SO(4)$ group to enable comparison with the one-loop S-matrix of
section \ref{secollsg55}.

Following the discussion of symmetries in section \ref{secsym55}, the
$[SU(2)]^4$ is related to the single $SO(4)$ of sections \ref{seclag55} and
\ref{secollsg55} by identifying the two $SU(2)$s with indices $a$ and $\al$ and
the two $SU(2)$s with indices $\adt$ and $\agdt$. The resulting $[SU(2)]^2$ is 
(locally) the same as $SO(4)$.

After identifying the pairs of $SU(2)$s the following rules can be used to
translate to the $SO(4)$ notation 
\be\begin{split}
\mbb{I}\otimes\mbb{I} \ra \de_{mp}\de_{nq}\,,
&\hs{20pt}
\mbb{P}\otimes \mbb{I}+\mbb{I}\otimes\mbb{P}\ra
  \de_{mp}\de_{nq}+\de_{mq}\de_{np}-\de_{mn}\de_{pq}\,,
\\\mbb{P}\otimes \mbb{P}\ra \de_{mq}\de_{np}\,,
&\hs{20pt}\mbb{P}\otimes\mbb{I}-\mbb{I}\otimes\mbb{P}\ra
  \e_{mnpq}\,,
\\(\mbb{I})_{ab}^{cd}=\de_{a}^{c}\de_{b}^{d}\,,
&\hs{20pt}(\mbb{P})_{ab}^{cd}=\de_{a}^{d}\de_{b}^{c}\,,
\end{split}\ee
where the first entry in the tensor product corresponds to undotted indices and
the second to dotted ones.

The S-matrix has the following structure: 

\footnotesize
\tbf{Boson-Boson}
{\allowdisplaybreaks 
\begin{align*}
\Sc\ket{Y_m(p_1)Y_n(p_2)}
= & \big((K_1(K_1+K_2))\de_{mp}\de_{nq}
    -K_1K_2\de_{mn}\de_{pq}
    +K_2(K_1+K_2)\de_{mq}\de_{np}\big)
    \ket{Y_p(p_1)Y_q(p_2)}
\\& +\big(-K_5^2\de_{mn}\de_{pq}\big)\ket{Z_p(p_1)Z_q(p_2)}
\\& +\big(-\fr{1}{2}(K_1+K_2)K_5(\de_{mn}\de_{pq}+
     \de_{mp}\de_{nq}-\de_{mq}\de_{np}+\e_{mnpq})
    +K_2 K_5 \de_{mn}\de_{pq}\big)\ket{\z_p(p_1)\z_q(p_2)}
\\&  +\big(-\fr{1}{2}(K_1+K_2)K_5(\de_{mn}\de_{pq}+
     \de_{mp}\de_{nq}-\de_{mq}\de_{np}-\e_{mnpq})
    +K_2 K_5 \de_{mn}\de_{pq}\big)\ket{\ch_p(p_1)\ch_q(p_2)}
\\
\Sc\ket{Z_m(p_1)Z_n(p_2)}
= & \big((K_3(K_3+K_4))\de_{mp}\de_{nq}
    -K_3K_4\de_{mn}\de_{pq}
    +K_4(K_3+K_4)\de_{mq}\de_{np}\big)\ket{Z_p(p_1)Z_q(p_2)}
\\ & + \big(-K_6^2\de_{mn}\de_{pq}\big)\ket{Y_p(p_1)Y_q(p_2)}
\\&  +\big(\fr{1}{2}(K_3+K_4)K_6(\de_{mn}\de_{pq}+
     \de_{mp}\de_{nq}-\de_{mq}\de_{np}+\e_{mnpq})
    -K_4 K_6 \de_{mn}\de_{pq}\big)\ket{\ch_p(p_1)\ch_q(p_2)}
\\ & +\big(\fr{1}{2}(K_3+K_4)K_6(\de_{mn}\de_{pq}+
     \de_{mp}\de_{nq}-\de_{mq}\de_{np}-\e_{mnpq})
    -K_4 K_6 \de_{mn}\de_{pq}\big)\ket{\z_p(p_1)\z_q(p_2)}
\\
\Sc\ket{Y_m(p_1)Z_n(p_2)}
= & \big(K_9^2\de_{mp}\de_{nq}\big)\ket{Y_p(p_1)Z_q(p_2)}
\\& +\big(K_7^2\de_{mq}\de_{np}\big)\ket{Z_p(p_1)Y_q(p_2)}
\\& +\big(-\fr{1}{2}K_7K_9(-\de_{mn}\de_{pq}+
     \de_{mp}\de_{nq}+\de_{mq}\de_{np}-\e_{mnpq})\big)
        \ket{\z_p(p_1)\ch_q(p_2)}
\\& +\big(\fr{1}{2}K_7K_9(-\de_{mn}\de_{pq}+\de_{mp}\de_{nq}
         +\de_{mq}\de_{np} +\e_{mnpq})\big)
         \ket{\ch_p(p_1)\z_q(p_2)}
\\
\Sc\ket{Z_m(p_1)Y_n(p_2)}
= & \big(K_{10}^2\de_{mp}\de_{nq}\big)\ket{Z_p(p_1)Y_q(p_2)}
\\& +\big(K_8^2\de_{mq}\de_{np}\big)\ket{Y_p(p_1)Z_q(p_2)}
\\& +\big(\fr{1}{2}K_8K_{10}(-\de_{mn}\de_{pq}+
     \de_{mp}\de_{nq}+\de_{mq}\de_{np}-\e_{mnpq})\big)
        \ket{\ch_p(p_1)\z_q(p_2)}
\\&
+\big(-\fr{1}{2}K_8K_{10}(-\de_{mn}\de_{pq}+\de_{mp}\de_{nq}
         +\de_{mq}\de_{np} +\e_{mnpq})\big)
         \ket{\z_p(p_1)\ch_q(p_2)}
\end{align*}
\tbf{Boson-Fermion}
\begin{align*}
\Sc\ket{Y_m(p_1)\z_n(p_2)}
= & \big((K_1+\fr{1}{2}K_2)K_9\de_{mp}\de_{nq}
         +\fr{1}{2}K_2K_9(-\de_{mn}\de_{pq}+\de_{mq}\de_{np}
          +\e_{mnpq})\big)
        \ket{Y_p(p_1)\z_q(p_2)}
\\& +\big(\fr{1}{2}K_1K_7(\de_{mq}\de_{np}-\de_{mn}\de_{pq}
       +\de_{mp}\de_{nq}
-\e_{mnpq})+K_2K_7\de_{mq}\de_{np}\big)
\ket{\z_p(p_1)Y_q(p_2)}
\\&
+\big(-\fr{1}{2}K_5K_7(-\de_{mp}\de_{nq}+\de_{mn}\de_{pq}
       +\de_{mq}\de_{np}+\e_{mnpq})\big)
      \ket{Z_p(p_1)\ch_q(p_2)}
\\&    +\big(-\fr{1}{2}K_5K_9(-\de_{mq}\de_{np}
        +\de_{mn}\de_{pq}+\de_{mp}\de_{nq}-\e_{mnpq})\big)
      \ket{\ch_p(p_1)Z_q(p_2)}
\\
\Sc\ket{\z_m(p_1)Y_n(p_2)}
= & \big((K_1+\fr{1}{2}K_2)K_{10}\de_{mp}\de_{nq}
+\fr{1}{2}K_2K_{10}(-\de_{mn}\de_{pq}+\de_{mq}\de_{np}
          +\e_{mnpq})\big)
        \ket{\z_p(p_1)Y_q(p_2)}
\\& +\big(\fr{1}{2}K_1K_8(\de_{mq}\de_{np}-\de_{mn}\de_{pq}
       +\de_{mp}\de_{nq}
-\e_{mnpq})+K_2K_8\de_{mq}\de_{np}\big)\ket{
Y_p(p_1)\z_q(p_2)}
\\& +\big(\fr{1}{2}K_5K_8(-\de_{mp}\de_{nq}+\de_{mn}\de_{pq}
       +\de_{mq}\de_{np}+\e_{mnpq})\big)
      \ket{\ch_p(p_1)Z_q(p_2)}
\\&    +\big(\fr{1}{2}K_5K_{10}(-\de_{mq}\de_{np}
        +\de_{mn}\de_{pq}+\de_{mp}\de_{nq}-\e_{mnpq})\big)
      \ket{Z_p(p_1)\ch_q(p_2)}
\\
\Sc\ket{Y_m(p_1)\ch_n(p_2)}
= &  \big((K_1+\fr{1}{2}K_2)K_9\de_{mp}\de_{nq}
         +\fr{1}{2}K_2K_9(-\de_{mn}\de_{pq}+\de_{mq}\de_{np}
          -\e_{mnpq})\big)\ket{Y_p(p_1)\ch_q(p_2)}
\\&   
+\big(\fr{1}{2}K_1K_7(\de_{mq}\de_{np}-\de_{mn}\de_{pq}
       +\de_{mp}\de_{nq}
+\e_{mnpq})+K_2K_7\de_{mq}\de_{np}\big)\ket{
\ch_p(p_1)Y_q(p_2)}
\\& +\big(\fr{1}{2}K_5K_7(-\de_{mp}\de_{nq}+\de_{mn}\de_{pq}
+\de_{mq}\de_{np}-\e_{mnpq})\big)\ket{Z_p(p_1)\z_q(p_2)}
\\&    +\big(\fr{1}{2}K_5K_9(-\de_{mq}\de_{np}
+\de_{mn}\de_{pq}+\de_{mp}\de_{nq}+\e_{mnpq})\big)
\ket{\z_p(p_1)Z_q(p_2)}
\\
\Sc\ket{\ch_m(p_1)Y_n(p_2)}
= &  \big((K_1+\fr{1}{2}K_2)K_{10}\de_{mp}\de_{nq}
+\fr{1}{2}K_2K_{10}(-\de_{mn}\de_{pq}+\de_{mq}\de_{np}
          -\e_{mnpq})\big)\ket{\ch_p(p_1)Y_q(p_2)}
\\&   
+\big(\fr{1}{2}K_1K_8(\de_{mq}\de_{np}-\de_{mn}\de_{pq}
       +\de_{mp}\de_{nq}
+\e_{mnpq})+K_2K_8\de_{mq}\de_{np}\big)\ket{
Y_p(p_1)\ch_q(p_2)}
\\&
+\big(-\fr{1}{2}K_5K_8(-\de_{mp}\de_{nq}+\de_{mn}\de_{pq}
+\de_{mq}\de_{np}-\e_{mnpq})\big)\ket{\z_p(p_1)Z_q(p_2)}
\\&    +\big(-\fr{1}{2}K_5K_{10}(-\de_{mq}\de_{np}
+\de_{mn}\de_{pq}+\de_{mp}\de_{nq}+\e_{mnpq})\big)
\ket{Z_p(p_1)\z_q(p_2)}
\\
\Sc\ket{Z_m(p_1)\z_n(p_2)}
= & \big((K_3+\fr{1}{2}K_4)K_{10}\de_{mp}\de_{nq}+
    \fr{1}{2}K_4 K_{10}(-\de_{mn}\de_{pq}+\de_{mq}\de_{np}
      -\e_{mnpq})\big)\ket{Z_p(p_1)\z_q(p_2)}
\\& +\big(-\fr{1}{2}K_3 K_8(\de_{mq}\de_{np}
     -\de_{mn}\de_{pq}+\de_{mp}\de_{nq}+\e_{mnpq})-K_4 K_8
     \de_{mq}\de_{np}\big)
      \ket{\z_p(p_1)Z_q(p_2)}
\\& +\big(\fr{1}{2}K_6
K_8(-\de_{mp}\de_{nq}+\de_{mn}\de_{pq}
     +\de_{mq}\de_{np}-\e_{mnpq}\big)
      \ket{Y_p(p_1)\ch_q(p_2)}
\\& +\big(-\fr{1}{2}K_6 K_{10}(-\de_{mq}\de_{np}
+\de_{mn}\de_{pq}+\de_{mp}\de_{nq}+\e_{mnpq})\big)\ket{
\ch_p(p_1)Y_q(p_2)}
\\
\Sc\ket{\z_m(p_1)Z_n(p_2)}
= & \big((K_3+\fr{1}{2}K_4)K_9\de_{mp}\de_{nq}+
    \fr{1}{2}K_4 K_9(-\de_{mn}\de_{pq}+\de_{mq}\de_{np}
      -\e_{mnpq})\big)\ket{\z_p(p_1)Z_q(p_2)}
\\& +\big(-\fr{1}{2}K_3 K_7(\de_{mq}\de_{np}
     -\de_{mn}\de_{pq}+\de_{mp}\de_{nq}+\e_{mnpq})-K_4 K_7
     \de_{mq}\de_{np}\big)
      \ket{Z_p(p_1)\z_q(p_2)}
\\& +\big(-\fr{1}{2}K_6
K_7(-\de_{mp}\de_{nq}+\de_{mn}\de_{pq}
     +\de_{mq}\de_{np}-\e_{mnpq}\big)
      \ket{\ch_p(p_1)Y_q(p_2)}
\\& +\big(\fr{1}{2}K_6 K_9(-\de_{mq}\de_{np}
+\de_{mn}\de_{pq}+\de_{mp}\de_{nq}+\e_{mnpq})\big)\ket{
Y_p(p_1)\ch_q(p_2)}
\\
\Sc\ket{Z_m(p_1)\ch_n(p_2)}
= & \big((K_3+\fr{1}{2}K_4)K_{10}\de_{mp}\de_{nq}+
    \fr{1}{2}K_4 K_{10}(-\de_{mn}\de_{pq}+\de_{mq}\de_{np}
      +\e_{mnpq})\big)\ket{Z_p(p_1)\ch_q(p_2)}
\\& +\big(-\fr{1}{2}K_3 K_8(\de_{mq}\de_{np}
     -\de_{mn}\de_{pq}+\de_{mp}\de_{nq}-\e_{mnpq})-K_4 K_8
     \de_{mq}\de_{np}\big)\ket{\ch_p(p_1)Z_q(p_2)}
\\& +\big(-\fr{1}{2}K_6
K_8(-\de_{mp}\de_{nq}+\de_{mn}\de_{pq}
     +\de_{mq}\de_{pn}+\e_{mnpq}\big)\ket{Y_p(p_1)\z_q(p_2)}
\\& +\big(\fr{1}{2}K_6 K_{10}(-\de_{mq}\de_{np}
+\de_{mn}\de_{pq}+\de_{mp}\de_{nq}-\e_{mnpq})\big)\ket{
\z_p(p_1)Y_q(p_2)}
\\
\Sc\ket{\ch_m(p_1)Z_n(p_2)}
= & \big((K_3+\fr{1}{2}K_4)K_9\de_{mp}\de_{nq}+
    \fr{1}{2}K_4 K_9(-\de_{mn}\de_{pq}+\de_{mq}\de_{np}
      +\e_{mnpq})\big)\ket{\ch_p(p_1)Z_q(p_2)}
\\& +\big(-\fr{1}{2}K_3 K_7(\de_{mq}\de_{np}
     -\de_{mn}\de_{pq}+\de_{mp}\de_{nq}-\e_{mnpq})-K_4 K_7
     \de_{mq}\de_{np}\big)\ket{Z_p(p_1)\ch_q(p_2)}
\\& +\big(\fr{1}{2}K_6
K_7(-\de_{mp}\de_{nq}+\de_{mn}\de_{pq}
     +\de_{mq}\de_{np}+\e_{mnpq}\big)\ket{\z_p(p_1)Y_q(p_2)}
\\& +\big(-\fr{1}{2}K_6 K_9(-\de_{mq}\de_{np}
+\de_{mn}\de_{pq}+\de_{mp}\de_{nq}-\e_{mnpq})\big)
\ket{Y_p(p_1)\z_q(p_2)}
\end{align*}
\tbf{Fermion-Fermion}
\begin{align*}
\Sc\ket{\z_m(p_1)\z_n(p_2)}
= & \big(-\fr{1}{2}(K_1K_4-K_2K_3)\e_{mnpq}
         +(K_1 K_3+\fr{1}{2}(K_1K_4+K_2K_3))\de_{mp}\de_{nq}
     \\&\qquad    -\fr{1}{2}(K_1K_4+K_2K_3)\de_{mn}\de_{pq}
     +(K_2
      K_4+\fr{1}{2}(K_1K_4+K_2K_3))\de_{mq}\de_{np}\big)
        \ket{\z_p(p_1)\z_q(p_2)}
\\&+
\big(K_5K_6\de_{mn}\de_{pq}\big)\ket{\ch_p(p_1)\ch_q(p_2)}
\\& +
\big(-\fr{1}{2}(K_1+K_2)K_6(\de_{mn}\de_{pq}
     +\de_{mp}\de_{nq}-\de_{mq}\de_{np}+\e_{mnpq})
     +K_2K_6 \de_{mn}\de_{pq}\big)\ket{Y_p(p_1)Y_q(p_2)}
\\& +\big(\fr{1}{2}(K_3+K_4)K_5(\de_{mn}\de_{pq}
     +\de_{mp}\de_{nq}-\de_{mq}\de_{np}-\e_{mnpq})
     -K_4K_5 \de_{mn}\de_{pq}\big)\ket{Z_p(p_1)Z_q(p_2)}
\\
\Sc\ket{\ch_m(p_1)\ch_n(p_2)}
= & \big(\fr{1}{2}(K_1K_4-K_2K_3)\e_{mnpq}
         +(K_1 K_3+\fr{1}{2}(K_1K_4+K_2K_3))\de_{mp}\de_{nq}
     \\&\qquad    -\fr{1}{2}(K_1K_4+K_2K_3)\de_{mn}\de_{pq}
     +(K_2
K_4+\fr{1}{2}(K_1K_4+K_2K_3))\de_{mq}\de_{np}\big)
\ket{\ch_p(p_1)\ch_q(p_2)}
\\& +
\big(K_5K_6\de_{mn}\de_{pq}\big)\ket{\z_p(p_1)\z_q(p_2)}
\\& +\big(\fr{1}{2}(K_3+K_4)K_5(\de_{mn}\de_{pq}
     +\de_{mp}\de_{nq}-\de_{mq}\de_{np}+\e_{mnpq})
     -K_4K_5 \de_{mn}\de_{pq}\big)\ket{Z_p(p_1)Z_q(p_2)}
\\& +\big(-\fr{1}{2}(K_1+K_2)K_6(\de_{mn}\de_{pq}
     +\de_{mp}\de_{nq}-\de_{mq}\de_{np}-\e_{mnpq})
     +K_2K_6 \de_{mn}\de_{pq}\big)\ket{Y_p(p_1)Y_q(p_2)}
\\
\Sc\ket{\z_m(p_1)\ch_n(p_2)}
= & \big(K_9
K_{10}\de_{mp}\de_{nq}\big)\ket{\z_p(p_1)\ch_q(p_2)}
\\& +\big(-K_7
K_8\de_{mq}\de_{np}\big)\ket{\ch_p(p_1)\z_q(p_2)}
\\ & +
\big(-\fr{1}{2}K_8K_9(-\de_{mn}\de_{pq}+\de_{mp}\de_{nq}
+\de_{mq}\de_{np}-\e_{mnpq})\big)\ket{Y_p(p_1)Z_q(p_2)}
\\&
+\big(-\fr{1}{2}K_7K_{10}(-\de_{mn}\de_{pq}+\de_{mp}\de_{nq}
+\de_{mq}\de_{np}+\e_{mnpq})\big)\ket{Z_p(p_1)Y_q(p_2)}
\\
\Sc\ket{\ch_m(p_1)\z_n(p_2)}
= & \big(K_9
K_{10}\de_{mp}\de_{nq}\big)\ket{\ch_p(p_1)\z_q(p_2)}
\\& +\big(-K_7
K_8\de_{mq}\de_{np}\big)\ket{\z_p(p_1)\ch_q(p_2)}
\\ & +
\big(\fr{1}{2}K_7K_{10}(-\de_{mn}\de_{pq}+\de_{mp}\de_{nq}
+\de_{mq}\de_{np}-\e_{mnpq})\big)\ket{Z_p(p_1)Y_q(p_2)}
\\& +\big(\fr{1}{2}K_8K_9(-\de_{mn}\de_{pq}+\de_{mp}\de_{nq}
+\de_{mq}\de_{np}+\e_{mnpq})\big)\ket{Y_p(p_1)Z_q(p_2)}
\end{align*}
}
\normalsize


\renewcommand{\theequation}{F.\arabic{equation}}
\setcounter{equation}{0}
\section{Trigonometric relativistic limit of  quantum-deformed $\mf{psu}(2|2)
\ltimes\mbb{R}^3$\  R-matrix\la{qdefl}}

In \ci{bk} the fundamental R-matrix of the quantum deformation of the centrally
extended superalgebra $\mf{psu}(2|2) \ltimes \mbb{R}^3$ was constructed. This 
R-matrix depends on various parameters: the global algebra parameters,
$\al,\,g$; the quantum group deformation parameter $q$; and the spectral
parameters $x^+_i,\,x^-_i,\,\g_i$ ($i=1,2$), where $x^+_i$ and $x^-_i$ are
related by a constraint equation. There is a natural choice for the parameters
$\g_i$ (which effectively control the normalisation of the fermions) that is
given in \ci{bk}. We take $\g_i$ to be given by this choice, rescaled by a
factor of $\sqrt[4]{-g^2}$.

In this section we generalise the limit of \ci{bcl} that leads to a
trigonometric, relativistic, q-deformed, classical r-matrix giving the exact
trigonometric, relativistic, q-deformed R-matrix. The classical limit
investigated in \ci{bcl} corresponds to expanding the quantum deformation
parameter, 
\be
q=1+\fr{h}{2g}+\ord{g^{-2}}\,.
\ee
$g^{-1}$ is  playing the r\^{o}le of $\hbar$, i.e. it is small but finite. The
relativistic  trigonometric limit that is relevant for the reduced \adss{5}
theory is $h\ra \infty$ (discussed  in section 6.5 of
\ci{bcl}).

To generalise it we set 
\be
h \propto \fr{g}{k}\,,
\ee 
and take $k^{-1}$ to be the parameter playing the r\^{o}le of $\hbar$.
Assuming $q$ (or equivalently $k$) is finite, the strict $h \ra \infty$ limit
corresponds to the strict $g \ra \infty$ limit in the new variables $(g,\,k)$.

All dependence on $\al$ in the R-matrix then drops out (we just set $\al$  equal
to one). The spectral parameters $x^+_i,x^-_i$ are reinterpreted in terms of
rapidities, and the quantum deformation parameter $q$, is parametrised in terms
of the coupling $k$ as 
\be
q=1+\sum_{n=1}^{\infty}\fr{a_n}{k^n}\,.
\ee
This limit is essentially the same as the one described in section 6.5 of
\ci{bcl}, rewritten in a way that allows us to consider it to higher orders than
just leading (tree-level) one. The fixing of the remaining parameters is done to
match the resulting R-matrix with our  one-loop perturbation theory result 
for the perturbative  S-matrix as closely as possible.

In the strict $g\ra \infty$ limit the constraint equation relating $x^\pm_i$ 
reduces to 
\be
(x^+_i)^2=q^2 (x^-_i)^2\,.
\ee
To solve it we set 
\be
x^\pm_i=-iq^{\pm\fr{1}{2}}\ e^{-\vt_i}\,.
\ee
As suggested by this ansatz, the variables $\vt_i$ are identified with the
rapidities. The matching to the one-loop S-matrix as closely as possible
suggests the following expansion of $q$ to one-loop order
\be 
q=1-\fr{i\pi}{k}-\fr{\pi^2}{2k^2}+\ord{k^{-3}}\,.
\ee
This prompts us to make a conjecture that the exact form of $q$ should be
\be\la{qexp2}
q=\exp\Big(-\fr{i\pi}{k}\Big)\,.
\ee
For convenience we give some of the quantities that appear in the R-matrix of
\ci{bk} in this limit and parametrisation,
\be\la{qcu}
\g_i=\fr{1}{\sqrt{2\cos\fr{\pi}{2k}} } \  e^{-\fr{\vt_i}{2} -\fr{i\pi}{4k}}\,,
\hs{20pt}q^{C_i}=U_i=1\,.
\ee
The R-matrix is then parametrised by ten functions $J_i$,\,\foot{The ten
functions $J_I$ are related to the ten functions $A,B,\ldots,K$ of \ci{bk}
as follows
\be\no
(J_1,J_2,J_3,J_4,J_5,J_6,J_7,J_8,J_9,J_{10})
=(\fr{A-B}{2},\fr{A+B}{2},-\fr{D-E}{2},-\fr{D+E}{2},-\fr{C}{2},
\fr{F}{2},H,K,G,L)\,.
\ee
We have also renamed $\psi_{1,2}\ra \psi_{3,4}$.}
\be\la{rbk1}\begin{split}
\mathcal{R}\ket{\phi_1\phi_1}=&\big(J_1+J_2\big)\ket{\phi_1\phi_1}
\\\mathcal{R}\ket{\phi_1\phi_2}=&J_1\sec\fr{\pi}{k}\ket{\phi_1\phi_2}  
                +\big(J_2-iJ_1\tan\fr{\pi}{k}\big)\ket{\phi_2\phi_1}
                                -J_5\sec\fr{\pi}{k}\ket{\psi_3\psi_4}
+J_5(1+i\tan\fr{\pi}{k})\ket{\psi_4\psi_3}
\\\mathcal{R}\ket{\phi_2\phi_1}=&J_1\sec\fr{\pi}{k}\ket{\phi_2\phi_1}
+\big(J_2+iJ_1\tan\fr{\pi}{k}\big)\ket{\phi_1\phi_2}
                                -J_5\sec\fr{\pi}{k}\ket{\psi_4\psi_3}
+J_5(1-i\tan\fr{\pi}{k})\ket{\psi_3\psi_4}
\\\mathcal{R}\ket{\phi_2\phi_2}=&\big(J_1+J_2\big)\ket{\phi_2\phi_2}
\\\mathcal{R}\ket{\psi_3\psi_3}=&\big(J_3+J_4\big)\ket{\psi_3\psi_3}
\\\mathcal{R}\ket{\psi_3\psi_4}=&J_3\sec\fr{\pi}{k}\ket{\psi_3\psi_4}
+\big(J_4-iJ_3\tan\fr{\pi}{k}\big)\ket{\psi_4\psi_3}
                                -J_6\sec\fr{\pi}{k}\ket{\phi_1\phi_2}
+J_6(1+i\tan\fr{\pi}{k})\ket{\phi_2\phi_1}
\\\mathcal{R}\ket{\psi_4\psi_3}=&J_3\sec\fr{\pi}{k}\ket{\psi_4\psi_3}
+\big(J_4+iJ_3\tan\fr{\pi}{k}\big)\ket{\psi_3\psi_4}
                                -J_6\sec\fr{\pi}{k}\ket{\phi_2\phi_1}
+J_6(1-i\tan\fr{\pi}{k})\ket{\phi_1\phi_2}
\\\mathcal{R}\ket{\psi_4\psi_4}=&\big(J_3+J_4\big)\ket{\psi_4\psi_4}
\\\mathcal{R}\ket{\phi_a\psi_\bet}=&J_7\;\de_a^d\de_\bet^\g\ket{
\psi_\g\phi_d}
+J_9\;\de_a^c\de_\bet^\de\ket{\phi_c\psi_\de}
\\\mathcal{R}\ket{\psi_\al\phi_b}=&
                    J_8\;\de_\al^\de \de_b^c\ket{\phi_c\psi_\de}
                   +J_{10}\;\de_\al^\g \de_b^d\ket{\psi_\g\phi_d}
\end{split}\ee 
As in sections \ref{secfac} and \ref{bkcomp} our index notation in as follows
\be\begin{split}
a = (1,\,2)\,,&\hs{10pt}
\al = (3,\,4)\,,\hs{10pt}
A = (a,\,\al)\,,\hs{20pt}\trm{with the fermionic grading} \hs{10pt}
\bll a\brr =0\,,\hs{10pt}\bll \al\brr =1\,.
\end{split}\ee
In the trigonometric relativistic limit the functions $J_i$ are given in
\eqref{jfuncex}. The phase factor $R_0$ of \ci{bk} is related to $P_0$ by
\be\la{p0}
R_0(\q,k)=P_0(\q,k)\cosech\fr{\q}{2}\sinh\Big(\fr{\q}{2}+\fr{i\pi}{2k}
\Big)\,.
\ee

The requirement of unitarity of the R-matrix is \ci{bk}
\be\la{unit}
\mc{R}_{12}\mc{R}_{21}=\mbb{I}\otimes \mbb{I}\,.
\ee
In terms of the rapidities the interchange of $1$ and $2$ sends $\q \equiv
\vt_1-\vt_2 \ra-\q$. The R-matrix satisfies \eqref{unit} so long as the phase
factor satisfies
\be
R_0(\q,k)R_0(-\q,k)=1\,.
\ee
This implies the following constraint on the phase $P_0$ in \eqref{p0}
\be\la{p0int1}
P_0(\q,k)P_0(-\q,k)=
\fr{\sinh^2\fr{\q}{2}}{\sinh^2\fr{\q}{2}+\sin^2\fr{\pi}{2k}}\,.
\ee
The quantum-deformed R-matrix also has a crossing symmetry subject to a
constraint on the phase factor. This crossing symmetry is given by \ci{bk}
\be\la{crossss}
(\mc{C}^{-1}\otimes \mbb{I})\mc{R}_{\bar{1}2}^{\trm{ST}\otimes\mbb{I}}
(\mbb{I}\otimes \mc{C})\mc{R}_{12}=\mbb{I}\otimes \mbb{I}\,.
\ee
$\trm{ST}$ is supertransposition and the action of charge conjugation $\mc{C}$
on one-particle states is
\be\begin{split}
\mc{C}\ket{\phi_1}=-q^{\fr{1}{2}}\ket{\phi_2}\,,\hs{20pt}
\mc{C}\ket{\psi_3}=-q^{\fr{1}{2}}\ket{\psi_4}\,,&\hs{20pt}
\mc{C}\ket{\phi_2}=q^{-\fr{1}{2}}\ket{\phi_1}\,,\hs{20pt}
\mc{C}\ket{\psi_4}=q^{-\fr{1}{2}}\ket{\psi_3}\,.
\end{split}\ee
In the $g \ra \infty$ limit the crossed spectral parameters are given by
\be\ba{cc}
\bar{x}^{\pm}_i=-x^{\pm}_i\,,&\bar{\g_i}=-i\g_i\,,\hs{20pt}
q^{\bar{C}_i}=\bar{U}_i=1\,.
\ea\ee
The crossing symmetry relates the R-matrices $\mc{R}_{12}$ and
$\mc{R}_{\bar{1}{2}}$. Considering the  ``Lorentz invariant'' combinations of
the spectral parameters, e.g.,  $x_1/x_2$, and comparing to $\bar{x}_1/x_2$, we
see that these are related by $\q \equiv\vt_1-\vt_2 \ra \q+i\pi$. 
The R-matrix satisfies the crossing relation \eqref{crossss} if the phase
factor satisfies 
\be\la{dfdf}
R_0(\q,k)\ R_0(\q+i\pi,k)=
\fr{\cosh\Big(\fr{\q}{2}+\fr{i\pi}{2k} \Big)}
{\sinh\Big(\fr{\q}{2}-\fr{i\pi}{2k} \Big)}
\tanh\fr{\q}{2}\;\,.
\ee
Combining with the unitarity relation, \eqref{p0int1} then implies the following
constraint on the phase $P_0$ \eqref{p0}
\be
P_0(i\pi-\q,k)=P_0(\q,k)\,.
\ee
The crossing symmetry also implies the relations \eqref{crossj} between the
functions $J_i$.

The conjugation relations \eqref{congj} hold as long as the phase factor
satisfies 
\be
R_0(\q,k)=R_0^*(-\q,k)\,\hs{20pt} \Ra \hs{20pt} P_0(\q,k)=P_0^*(-\q,k)\,.
\ee
These relations are equivalent to the matrix unitarity relations of \ci{bk},
\be\la{unitmat}
(\mc{R}_{12})^\dagger\mc{R}_{12}=\mbb{I}\otimes \mbb{I}\,.
\ee


\renewcommand{\theequation}{G.\arabic{equation}}
\setcounter{equation}{0}
\section{One-loop S-matrix of $SO(N+1)/SO(N)$ generalized sine-Gordon models 
\la{bosonic}}

Here we shall discuss the one-loop S-matrix of the bosonic $G/H=SO(N+1)/SO(N)$
gauged WZW model deformed by an integrable potential as in \rf{gwzw}. This  
one-loop S-matrix, including the determinant corrections from integrating out
the unphysical fields, was computed in \ci{ht2}. Here we shall make some
additional comments and clarify the structure of the result.  

In the $N=4$ case the determinant corrections result in the S-matrix factorising
under $\so(4) \cong \su(2) \oplus \su(2)$ at one-loop \ci{ht2}. Below we will
see that,  like in  the reduced \adss{5} theory, one can find a quantum-deformed
S-matrix similar to the one-loop factorised S-matrix that satisfies the
Yang-Baxter equation.

The Lagrangian found  by fixing the $A_+=0$ gauge and eliminating $A_-$ and
$\xi$ is given by taking \eqref{lag15} with $Y=\z=\ch=0$ and $m,n,p,q$ now
being $SO(N)$ vector indices
\bea
\Lag_{b} & = & 
\no         
             \fr{1}{2}\dpl X_m \dm X_m
             - \fr{\mu^2}{2} X_m X_m
\\       & & + \fr{\pi}{k}\Big[
         \fr{1}{3}X_m X_m \dpl X_n \dm X_n
             - \fr{1}{3}X_m \dpl X_m X_n \dm X_n
             + \fr{\mu^2}{6}X_m X_m X_n X_n\Big]
+\ord{k^{-2}}\,.
\eea
The one-loop S-matrix for this theory \ci{ht2} is given by the usual Feynman
diagram  contribution  plus the contribution of the determinant obtained upon
integrating out $A_-$ and $\xi$\,\foot{This determinant contribution is present
only for $N\geq 2$ (for $N=1$ or the sine-Gordon model the group $H$ is
trivial).}
\bea
&&\Sc\ket{X_m(p_1) X_n(p_2)}=\Big(  S_1(\q,k)\de_{mn} \de_{pq}
                + S_2(\q,k)\de_{mp}\de_{nq} 
                + S_3(\q,k)\de_{mq}\de_{np}\Big)\ket{X_p(p_1)X_q(p_2)} \ ,  \\
&& \ \ \ \ \ \ \ \ \ \ \ \ \ \ \    S_i = \bar S_i + \De S_i \ , 
\eea
\be\begin{split}
\bar S_3(\q,k)=&\bar S_1(i\pi-\q,k)=\fr{i\pi}{k}\coth\q
+\fr{i\pi}{2k^2}(\cosech\q-\coth\q)
\\&\hs{80pt}-\fr{\pi^2}{k^2}
\coth\q\cosech\q+\fr{i\pi}{2k^2}(N-2)\q\coth^2\q
+\ord{k^{-3}}
\\ \bar S_2(\q,k)=&1+\fr{i\pi}{k}\cosech\q-\fr{\pi^2}{k^2}
(\fr{1}{2}+\cosech^2\q)+\fr{i\pi}{2k^2}(N-2)\cosech\q
+\ord{k^{-3}}
\end{split}\ee
\be\begin{split}
\De S_3(\q,k)=\De S_1(i\pi-\q,k)=&-\fr{i\pi}{2k^2}
(\cosech\q-\coth\q)-(N-2)\fr{i\pi}{k^2}\coth\q
\\\De S_2(\q,k)=&-(N-2)\fr{i\pi}{k^2}\cosech\q
\end{split}\ee
The correction $\De S_k$ coming from the determinant (or the corresponding
one-loop counterterm which it produces) splits into two parts. The part not
proportional to $(N-2)$ is required to maintain some of the consequences of
integrability: in the abelian $H$ case, $N=2$, the corresponding counterterm 
contribution restores the satisfaction of the Yang-Baxter equation at one-loop
and agreement with the exact S-matrix of \ci{dh}. Also, in the $N=4$ case the
counterterm restores the group factorisation of the S-matrix under 
$\mf{so}(4)\cong\su(2)\oplus\su(2)$. However, as in the reduced \adss{5} theory,
in the non-abelian case, $N>2$ the addition of the counterterms does not
restore the validity of YBE. 

The part with coefficient $N-2$ is proportional to the tree-level S-matrix and
may be interpreted as being due to a shift in the coupling $k$ by the dual
Coxeter number of $H=SO(N)$,\  $c_H=N-2$. This is a recurring feature of these
theories. In general, there are two  shifts --  $k \to k+\fr{c_H}{2}$ and $k
\to k+c_H$ that play a r\^{o}le, hence we define the following shifted
couplings, 
\be
\kt=k+\fr{c_H}{2}\,,\hs{25pt}\kh=k+c_H\,.
\ee
It is useful to extract the phase factor
\be\la{bosphase}
P_B(\q,k)=1+\fr{i\pi}{\td{k}}\cosech\q
-\fr{\pi^2}{2\td{k}^2}(\cosech\q)^2+\ord{\fr{1}{\td{k}^3}}\,.
\ee
This phase factor satisfies the unitarity and crossing symmetry relations,
\be\la{boscross}
P_B(\q,k)\ P_B(-\q,k)=1+\ord{\fr{1}{\td{k}^3}}\,,
\hs{30pt}P_B(\q,k)=P_B(i\pi-\q,k)\,.
\ee
Then the total  S-matrix coefficients are given by 
\be\la{smatboss}\begin{split}
S_i =   P_B \hat{S}_i  \ , \ \ \  \ 
\hat{S}_2(\q,k)=&1-\fr{\pi^2}{\hat{k}^2}\coth^2\q
+\ord{\fr{1}{\hat{k}^3}}\,, \\
\hat{S}_3(\q,k)=\hat{S}_1(i\pi-\q,k)
=&\fr{i\pi}{\hat{k}}\coth\q
+\fr{i\pi}{2\hat{k}^2}(N-2)\q\coth^2\q
+\ord{\fr{1}{\hat{k}^3}}\,.
\end{split}
\ee
Note that while in  the phase factor $k$ enters as $\kt$, in the $\hat{S}_i$  it
enters as $\kh$.

\subsection{$N=1$: sine-Gordon model}

In the $N=1$ case the index $m$ only takes a single value and there is only a
single amplitude\,\foot{Taking $N=1$ and summing the three determinant
contributions gives $ \De S_1(\q,k)+\De S_2(\q,k)+\De S_3(\q,k)=0\,.$ This is
expected as for $N=1$ the group $H$ is trivial and thus there is no functional
determinant contribution.}
\be\begin{split}
S(\q,k)&=P_B(\q,1)(\hat{S}_1(\q,1)+\hat{S}_2(\q,1)+\hat{S}_3(\q,1))
\\&=1+\fr{i\pi}{k}\Big(1+\fr{1}{2k}\Big)\cosech\q-\fr{\pi^2}{2k^2}
\cosech^2\q+\ord{k^{-3}}\,.
\end{split}\ee
This agrees with the expansion of the exact S-matrix for the perturbative
excitations of the sine-Gordon model 
\be
S_{SG}^{(1,1)}(\q,\De(k))=
\fr{\sinh\q+i\sin\De(k)}{\sinh\q-i\sin\De(k)}\,, \ \ \ \ \ \ \ \ \ 
\De(k)=\fr{\pi}{k-\fr{1}{2}}\,,
\ee
where\,\footnote{\la{foot2}In footnote \ref{sine} the shift in $k$ for bosonic
the sine-Gordon model was given as $k \ra k-1$. Here $k$ has been rescaled by
$2$ as we are considering a truncation of the $G/H=SO(N+1)/SO(N)$ theory. As $G$
is abelian in this case there is no quantization of $k$ and thus this rescaling
is arbitrary.} for $N=1$, $\hat{S}_1(\q,k)+\hat{S}_2(\q,k)+\hat{S}_3(\q,k)=1\,,$
and thus  all the information is contained in the phase factor $P_B(\q,k)$
\eqref{bosphase}.

\subsection{$N=2$: complex sine-Gordon model\la{csg}}

In the $N=2$ case $H$ is abelian. Its dual Coxeter number vanishes and the
coupling $k$ is  unshifted, $\kt=\kh=k$. Usually for the complex sine-Gordon 
model  one would take the coset $G/H=SU(2)/U(1)$ rather than $G/H=SO(3)/SO(2)$.
For the perturbative S-matrix the only difference amounts to a rescaling $k$
by $2$. This is a consequence of the dual Coxeter number of $SU(2)$ being twice
that of $SO(3)$.

The determinant contribution is non-trivial and as $H$ is abelian we expect
the corrections to restore the satisfaction of YBE. This can be seen easily by
noting that for $N=2$ the reflection coefficient in the corrected S-matrix
vanishes, 
\be
R(\q,k)=\hat{S}_1(\q,k)+\hat{S}_3(\q,k)=0+\ord{k^{-3}}\,.
\ee
If the reflection coefficient vanishes and we have a crossing symmetry then
there is only one independent amplitude. The S-matrix can then be encoded in
a single function 
\be 
P_B(\q,k)\Big(\hat{S}_2(\q,k)+\hat{S}_3(\q,k)\Big)=1+\fr{i\pi}{k}\coth\fr{\q}{2}
-\fr{\pi^2}{2k^2}\coth^2\fr{\q}{2}+\ord{k^{-3}}\,.
\ee
This then agrees with the exact result derived  based on assumption of exact 
integrability \ci{dh}.\,\foot{$k$ has been rescaled by a factor of $2$ relative
to the one in \ci{dh}, see above.}

\subsection{$N=4$: group factorization\la{so4fact}}

For $H=SO(4)$ the field $X_m$ transforms in a vector representation of $H$. As
we have the isomorphism $\so(4) = \su(2)\oplus \su(2)$, with the vector
representation of $SO(4)$ equivalent to the bifundamental of $SU(2) \x SU(2)$ we
can rewrite the S-matrix using the $SU(2)$ indices
\be 
\hat{S}_{mn}^{pq}(\q,k)\ \sim\  \hat{S}_{a\al,b\bet}^{c\g,d\de}
(\q,k)\,.
\ee
Due to the integrability of the theory this S-matrix should factorise into the
tensor product of two $SU(2)$ S-matrices,
\be
\hat{S}_{a\al,b\bet}^{c\g,d\de}
(\q,k)=\hat{S}_{ab}^{cd}(\q,k)\ \hat{S}_{\al\bet}^{\g\de}(\q,k)\,.
\ee
This is indeed  the case for the S-matrix \eqref{smatboss} with $N=4$
\be\begin{split}\la{s1s2}
\hat{S}_{ab}^{cd}(\q,k)=\hat{s}_1(\q,k)&\de_a^c\de_b^d
+\hat{s}_2(\q,k)\de_a^d\de_b^c\,,
\\
\hat{s}_1(\q,k)=&1-\fr{i\pi}{2\kh}\coth\q 
 +\fr{i\pi}{8\kh^2}(5i\pi -4\q)\coth^2\q +\ord{\fr{1}{\kh^3}}
\\\hat{s}_2(\q,k)=&\fr{i\pi}{\kh}\coth\q
   -\fr{i\pi}{2\kh^2}(i\pi -2\q)\coth^2\q +\ord{\fr{1}{\kh^3}}
\end{split}\ee
These functions satisfy the crossing symmetry relations
$\hat{s}_1(i\pi-\q)=\hat{s}_1(\q,k)+\hat{s}_2(\q,k)\,,\hs{5pt}
\hat{s}_2(i\pi-\q)=-\hat{s}_2(\q,k)\,.$
The S-matrix \eqref{s1s2} does not satisfy the Yang-Baxter equation \ci{ht2}.
Motivated by the discussion of the  reduced \adss{5} theory in section
\ref{seclag55}, we may  consider a quantum-deformed $SU(2)$ S-matrix taking the
following form (with the quantum deformation parameter
$q=\exp\big(-\fr{i\pi}{\kh}\big)$)
\be\begin{split}\la{rsu2}
\mathcal{S}\ket{\phi_1\phi_1}&=(r_1+r_2)\ket{\phi_1\phi_1}
\\ \mathcal{S}\ket{\phi_1\phi_2}&=r_1\sec\fr{\pi}{\kh}\ket{\phi_1\phi_2}
                              +\big(r_2-i r_1
\tan\fr{\pi}{\kh}\big)\ket{\phi_2\phi_1}
\\\mathcal{S}\ket{\phi_2\phi_1}&=r_1\sec\fr{\pi}{\kh}\ket{\phi_2\phi_1}
                              +\big(r_2+i r_1
\tan\fr{\pi}{\kh}\big)\ket{\phi_1\phi_2}
\\\mathcal{S}\ket{\phi_2\phi_2}&=(r_1+r_2)\ket{\phi_2\phi_2} \ . 
\end{split}\ee
The quantum-deformed crossing symmetry relations are
\be\begin{split}\la{crosssu2}
r_1(i\pi-\q)=\cos\fr{\pi}{\kh}\big[ r_1(\q,k)+r_2(\q,k)\big]\,,\hs{30pt}
r_2(i\pi-\q)=-\cos\fr{\pi}{\kh}\big[ r_2(\q,k)-\tan^2\fr{\pi}{\kh}\ r_1(\q,k)
\big]\,.
\end{split}\ee
In analogy with the reduced \adss{5} theory one may find  the ``closest''
functions to \eqref{s1s2} satisfying the crossing relations \eqref{crosssu2} 
such that the quantum-deformed S-matrix \eqref{rsu2} is consistent with YBE:
\be\begin{split}\la{r1r2}
r_1(\q,k)=&P_0(\q,k)(1-\fr{i\pi}{2\kh}\coth\q 
 +\fr{i\pi}{8\kh^2}(5i\pi\coth^2\q -4\q\cosech^2\q) 
 +\ord{\fr{1}{\kh^3}})\,,
\\r_2(\q,k)=&P_0(\q,k)(\fr{i\pi}{\kh}\coth\q
   -\fr{i\pi}{2\kh^2}(i\pi\coth^2\q -2\q\cosech^2\q)
+\ord{\fr{1}{\kh^3}})\,.
\end{split}\ee
Similarly to the reduced \adss{5} theory these functions agree precisely with
\eqref{rsu2} at tree-level with a modification of the undressed $\q$ terms at
one-loop

\subsection{Symmetries}
The classical bosonic $SO(N+1)/SO(N)$ theories arise as the Pohlmeyer reductions
of the string theory on $\mbb{R}_t \x S^{N+1}$, with $S^{N+1} =
{SO(N+2)}/{SO(N+1)}$. Their symmetries fall into two classes, depending on
$H=SO(N)$ being abelian or non-abelian.

In the case of abelian $H$ the symmetry group is given by
\be
\mf{iso}(1,1) \oplus \mf{h} \oplus \mf{h}^{(g)}\,,
\ee
where the superscript $(g)$ denotes the gauge symmetry. The fields on which the
global part of the gauge symmetry has a linear action are field redefinitions of
the fields on which the global $H$ symmetry has a linear action. Therefore, the
physical symmetry acting on on-shell states is
\be\la{adsbossymmb}
\mf{iso}(1,1) \oplus \mf{h}\,.
\ee
In the abelian $H$ case the perturbative S-matrix (including the determinant
corrections arising from gauge-fixing etc.) satisfies the Yang-Baxter equation.

In the case of non-abelian $H$ the symmetry group is given by
\be
\mf{iso}(1,1) \oplus \mf{h}^{(g)}\,,
\ee
i.e. there is no additional global $H$ symmetry. The perturbative S-matrix
computed following \ci{ht1,ht2} has a manifest symmetry given by the global part
of the gauge group $H$. However, this S-matrix does not satisfy the Yang-Baxter
equation already at the tree level.

As in the examples discussed in  section \ref{sumofsym}, we may expect the
physical symmetry of these theories to be given by 
\be
U_q(\mf{iso}(1,1) \oplus \mf{h})\,.
\ee
In the case of the abelian $H$ this quantum deformation should have no effect as
the perturbative computation agrees with integrability results and satisfies the
YBE \ci{dvm1,dh,ht2}. In the non-abelian $H$ case (section \ref{so4fact}) there
are quantum-deformed S-matrices (e.g., \eqref{rsu2}, \eqref{crosssu2}) closely
related to the perturbative  S-matrix  that  satisfies the Yang-Baxter equation
and have a quantum-deformed $H$ symmetry.


\end{document}